\documentclass[fleqn]{2020SCGE}
\usepackage[]{xcolor}

\usepackage[T1]{fontenc}

\begin{document}

\ensubject{subject}

\ArticleType{Article}
\SpecialTopic{SPECIAL TOPIC: }
\Year{2020}
\Month{January}
\Vol{63}
\No{1}
\DOI{??}
\ArtNo{000000}
\ReceiveDate{? ?, 2023}
\AcceptDate{? ?, 2023}

\title{Science objectives of the Einstein Probe mission}{Science objectives of the Einstein Probe mission}


\author[1,2]{Weimin Yuan}{{wmy@nao.cas.cn}}
\author[3]{Lixin Dai}{}
\author[4]{Hua Feng}{}
\author[1,2]{Chichuan Jin}{}
\author[5,6]{Peter Jonker}{}
\author[7]{Erik Kuulkers}{}
\author[1]{\\Yuan Liu}{}
\author[8]{Kirpal Nandra}{}
\author[9]{Paul O’Brien}{}
\author[10]{Luigi Piro}{}
\author[8]{Arne Rau}{}
\author[11,12]{Nanda Rea}{}
\author[8]{\\Jeremy Sanders}{}
\author[4]{Lian Tao}{}
\author[13]{Junfeng Wang}{}
\author[14,15]{Xuefeng Wu}{}
\author[16,17]{Bing Zhang}{}
\author[4,2]{\\Shuangnan Zhang}{}
\author[18]{Shunke Ai}{}
\author[8]{Johannes Buchner}{}
\author[8]{Esra Bulbul}{}
\author[19]{Hechao Chen}{}
\author[20,2]{\\Minghua Chen}{}
\author[4]{Yong Chen}{}
\author[4]{Yu-Peng Chen}{}
\author[21]{Alexis Coleiro}{}
\author[11,12]{Francesco Coti Zelati}{}
\author[22]{\\Zigao Dai}{}
\author[18]{Xilong Fan}{}
\author[1,2]{Zhou Fan}{}
\author[8]{Susanne Friedrich}{}
\author[23,24]{He Gao}{}
\author[13]{Chong Ge}{}
\author[4]{\\Mingyu Ge}{}
\author[14]{Jinjun Geng}{}
\author[25,26]{Giancarlo Ghirlanda}{}
\author[10,27]{Giulia Gianfagna}{}
\author[1,2]{Lijun Gou}{}
\author[28]{\\Sébastien Guillot}{}
\author[29,30]{Xian Hou}{}
\author[1]{Jingwei Hu}{}
\author[31,32]{Yongfeng Huang}{}
\author[33]{Long Ji}{}
\author[4,2]{Shumei Jia}{}
\author[34]{\\S. Komossa}{}
\author[35]{Albert K. H. Kong}{}
\author[1]{Lin Lan}{}
\author[23,24]{An Li}{}
\author[13]{Ang Li}{}
\author[4]{Chengkui Li}{}
\author[1]{Dongyue Li}{}
\author[22,15]{\\Jian Li}{}
\author[36]{Zhaosheng Li}{}
\author[1,2]{Zhixing Ling}{}
\author[8]{Ang Liu}{}
\author[20,2]{Jinzhong Liu}{}
\author[37,38]{Liangduan Liu}{}
\author[8]{\\Zhu Liu}{}
\author[39]{Jiawei Luo}{}
\author[4,2]{Ruican Ma}{}
\author[40]{Pierre Maggi}{}
\author[8]{Chandreyee Maitra}{}
\author[11,12]{Alessio Marino}{}
\author[3]{\\Stephen Chi-Yung Ng}{}
\author[1]{Haiwu Pan}{}
\author[8]{Surangkhana Rukdee}{}
\author[2,41,42]{Roberto Soria}{}
\author[1]{Hui Sun}{}
\author[33]{\\Pak-Hin Thomas Tam}{}
\author[10]{Aishwarya Linesh Thakur}{}
\author[43]{Hui Tian}{}
\author[44,10]{Eleonora Troja}{}
\author[18]{Wei Wang}{}
\author[31]{\\Xiangyu Wang}{}
\author[1]{Yanan Wang}{}
\author[14,15]{Junjie Wei}{}
\author[1]{Sixiang Wen}{}
\author[13]{Jianfeng Wu}{}
\author[20]{Ting Wu}{}
\author[14,15]{\\Di Xiao}{}
\author[1]{Dong Xu}{}
\author[45,46]{Renxin Xu}{}
\author[4]{Yanjun Xu}{}
\author[43]{Yu Xu}{}
\author[1,2]{Haonan Yang}{}
\author[18]{Bei You}{}
\author[24]{\\Heng Yu}{}
\author[37,38]{Yunwei Yu}{}
\author[31,32,14]{Binbin Zhang}{}
\author[1,2]{Chen Zhang}{}
\author[29,30]{Guobao Zhang}{}
\author[4]{\\Liang Zhang}{}
\author[1]{Wenda Zhang}{}
\author[20,2]{Yu Zhang}{}
\author[31]{Ping Zhou}{}
\author[31,32]{Zecheng Zou}{}

\AuthorMark{W. Yuan, et al. }

\AuthorCitation{Yuan W., Dai L., Feng H., et al}


\address[1]{National Astronomical Observatories, Chinese Academy of Sciences, Beijing, 100101, People’s Republic of China}
\address[2]{School of Astronomy and Space Sciences, University of Chinese Academy of Sciences, Beijing, 100049, People’s Republic of China}
\address[3]{Department of Physics, University of Hong Kong, Pokfulam Road, Hong Kong, People’s Republic of China}
\address[4]{Key Laboratory of Particle Astrophysics, Institute of High Energy Physics, Chinese Academy of Sciences, \\Beijing 100049, People’s Republic of China} 
\address[5]{Department of Astrophysics/IMAPP, Radboud University, Nijmegen, 6500 GL, the Netherlands}
\address[6]{SRON, Netherlands Institute for Space Research, Leiden, 2333 CA, the Netherlands} 
\address[7]{ESA/ESTEC, Noordwijk, 2201 AZ, The Netherlands}
\address[8]{Max Planck Institute for Extraterrestrial Physics, Garching, 85748, Germany}
\address[9]{School of Physics and Astronomy, University of Leicester, Leicester, LE1 7RH, UK}
\address[10]{INAF - Istituto di Astrofisica e Planetologia Spaziali, Rome, 00133, Italy} 
\address[11]{Institute of Space Sciences (ICE), CSIC, Campus UAB, Barcelona, E-08193, Spain} 
\address[12]{Institut d’Estudis Espacials de Catalunya (IEEC), Barcelona, E-08034, Spain} 
\address[13]{Department of Astronomy, Xiamen University, Xiamen, Fujian 361005, People’s Republic of China} 
\address[14]{Purple Mountain Observatory, Chinese Academy of Sciences, Nanjing 210023, People’s Republic of China} 
\address[15]{School of Astronomy and Space Sciences, University of Science and Technology of China, Hefei 230026, People’s Republic of China} 
\address[16]{Nevada Center for Astrophysics, University of Nevada Las Vegas, NV 89154, USA} 
\address[17]{Department of Physics and Astronomy, University of Nevada Las Vegas, NV 89154, USA} 
\address[18]{Department of Astronomy, School of Physics and Technology, Wuhan University, Wuhan 430072, People’s Republic of China} 
\address[19]{School of Physics and Astronomy, Yunnan University, Kunming 650500, People’s Republic of China}
\address[20]{Xinjiang Astronomical Observatory, Chinese Academy of Sciences, Urumqi 830011, People’s Republic of China} 
\address[21]{Université Paris Cité, CNRS, Astroparticule et Cosmologie, Paris, F-75013, France} 
\address[22]{Department of Astronomy, University of Science and Technology of China, Hefei 230026, People’s Republic of China} 
\address[23]{Institute for Frontier in Astronomy and Astrophysics, Beĳing Normal University, Beĳing 102206, People’s Republic of China} 
\address[24]{Department of Astronomy, Beĳing Normal University, Beĳing 100875, People’s Republic of China} 
\address[25]{INAF – Osservatorio Astronomico di Brera, Merate (LC), I-23807, Italy} 
\address[26]{INFN – sezione di Milano-Bicocca, Milano (MI), I-20126, Italy} 
\address[27]{Dipartimento di Fisica, Università di Roma "Sapienza", Roma, I-00185, Italy} 
\address[28]{Institut de Recherche en Astrophysique et Planétologie, UPS-OMP, CNRS, CNES, Toulouse Cedex 4, F-31028, France} 
\address[29]{Yunnan Observatories, Chinese Academy of Sciences, Kunming 650216, People’s Republic of China} 
\address[30]{Key Laboratory for the Structure and Evolution of Celestial Objects, Chinese Academy of Sciences, \\Kunming 650216, People’s Republic of China} 
\address[31]{School of Astronomy and Space Science, Nanjing University, Nanjing 210023, People's Republic of China} 
\address[32]{Key Laboratory of Modern Astronomy and Astrophysics (Nanjing University), Ministry of Education, \\Nanjing 210023, People’s Republic of China} 
\address[33]{School of Physics and Astronomy, Sun Yat-Sen University, Zhuhai 519082, People’s Republic of China}
\address[34]{Max-Planck-Institut f\"ur Radioastronomie, Bonn, 53121, Germany}
\address[35]{Institute for Cosmic Ray Research, The University of Tokyo, Kashiwa City, Chiba 277-8582, Japan} 
\address[36]{Key Laboratory of Stars and Interstellar Medium, Xiangtan University, Xiangtan, Hunan 411105, People’s Republic of China} 
\address[37]{Institute of Astrophysics, Central China Normal University, Wuhan 430079, People’s Republic of China} 
\address[38]{Key Laboratory of Quark and Lepton Physics (Central China Normal University), Ministry of Education, \\Wuhan 430079, People’s Republic of China} 
\address[39]{College of Physics and Hebei Key Laboratory of Photophysics Research and Application, Hebei Normal University, \\Shijiazhuang, Hebei 050024, People’s Republic of China} 
\address[40]{Observatoire Astronomique de Strasbourg, Université de Strasbourg, CNRS, Strasbourg, F-67000, France}  
\address[41]{INAF – Osservatorio Astrofisico di Torino, Pino Torinese, I-10025, Italy} 
\address[42]{Sydney Institute for Astronomy, School of Physics A28, The University of Sydney, NSW 2006, Australia} 
\address[43]{School of Earth and Space Sciences, Peking University, Beijing, 100871, People's Republic of China} 
\address[44]{Department of Physics, University of Rome “Tor Vergata”, Rome, I-00133, Italy} 
\address[45]{School of Physics and State Key Laboratory of Nuclear Physics and Technology, Peking University, \\Beijing 100871, People's Republic of China} 
\address[46]{Kavli Institute for Astronomy and Astrophysics, Peking University, Beijing 100871, People’s Republic of China} 


\abstract{ 
The Einstein Probe (EP) is an interdisciplinary mission of time-domain and X-ray astronomy. Equipped with a wide-field lobster-eye X-ray focusing imager, EP will discover cosmic X-ray transients and monitor the X-ray variability of known sources in 0.5--4 keV, at a combination of detecting sensitivity and cadence that is not accessible to the previous and current wide-field monitoring missions. EP can perform quick characterisation of transients or outbursts with a Wolter-I X-ray telescope onboard. In this paper, the science objectives of the Einstein Probe mission are presented. EP is expected to enlarge the sample of previously known or predicted but rare types of transients with a wide range of timescales. Among them, fast extragalactic transients will be surveyed systematically in soft X-rays, which include $\gamma$-ray bursts and their variants, supernova shock breakouts, and the predicted X-ray transients associated with binary neutron star mergers. EP will detect X-ray tidal disruption events and outbursts from active galactic nuclei, possibly at an early phase of the flares for some. EP will monitor the variability and outbursts of X-rays from white dwarfs, neutron stars and black holes in our and neighbouring galaxies at flux levels fainter than those detectable by the current instruments, and is expected to discover new objects. A large sample of stellar X-ray flares will also be detected and characterised. In the era of multi-messenger astronomy, EP has the potential of detecting the possible X-ray counterparts of gravitational wave events, neutrino sources, and ultra-high energy $\gamma$-ray and cosmic ray sources. EP is expected to help advance the studies of extreme objects and phenomena revealed in the dynamic X-ray universe, and their underlying physical processes. Besides EP's strength in time-domain science, its follow-up telescope, with excellent performance, will also enable advances in many areas of X-ray astronomy.
}



\maketitle


\begin{multicols}{2}
\section{Introduction}\label{sec:1}

The X-ray sky is a fascinating and dynamic realm of astrophysics, where events and sources can vary dramatically, or appear, evolve, and disappear over timescales ranging from milliseconds to years.
The well-known examples range from stellar flares in the neighbourhood of the Solar system, to variability and outbursts involving neutron stars and black holes (BHs) in our Galaxy and other galaxies nearby, and to spectacular $\gamma$-ray bursts (GRBs) out to the edge of the universe. Their X-ray radiation may change in brightness from a few to tens to hundred times, and to many orders of magnitudes, and their variability timescales also span many orders of magnitudes, from sub-seconds, to days and years. The origin and mechanism of these events involve a diversity of energy sources and physical processes, such as magnetic field, thermal nuclear reaction, gravitational energy, particle acceleration and plasma physics. 

As an interdisciplinary research of X-ray and time-domain astronomy, the studies of X-ray transient and variable sources have significantly advanced our understanding of the origin and evolution of some of the celestial objects and the physics in the extreme conditions in the universe. This has in part been achieved via the building, launching and operating of several generations of wide-field (i.e., hundreds to thousands square degrees) X-ray monitoring instruments in the past decades, including BATSE/CGRO  \cite{Barthelmy.1995}, ASM/RXTE \cite{Levine.1996}, INTEGRAL \cite{Winkler2003,Kuulkers2021}, the Neil-Gehrels Swift Observatory (Swift) \cite{Gehrels2004} and MAXI \cite{Matsuoka2009} that are currently operating. Meanwhile, discoveries of transients have also been made with narrow-field X-ray telescopes in surveying the sky, e.g. ROSAT \cite{Truemper1982} and eROSITA \cite{Predehl2021}, and serendipitous observations, e.g. Chandra \cite{Weisskopf2002}, XMM-Newton \cite{Jansen2001} and Swift/XRT \cite{Burrows2005}, and even during spacecraft slews from one pointing direction to another \cite{Saxton2008}.

In recent years, some transient and variable sources or phenomena have generated enormous interest among the time-domain and X-ray astronomy communities.  
These include, as examples, supernova (SN) shock breakouts (SBOs), tidal disruption events (TDEs) and quasi-periodic eruptions (QPEs) from massive black holes (MBHs), GRBs at redshifts around the reionization era, flares and stellar corona mass ejections far stronger than those seen on our own Sun. As the most remarkable event, the first detection of electromagnetic-wave emission associated with the gravitational-wave (GW) event GW 170817 \cite{Abbott17}, originating from a binary neutron star (BNS) merger, marked the true advent of multi-messenger astronomy. This motivated the search for possible X-ray signatures resulting from binary neutron star mergers that accompany gravitational-wave events \cite{Troja+17,Troja2020}. 
Another frontier of multi-messenger astronomy is concerned with the synergetic detection of transient neutrino events and electromagnetic-wave sources, where a few candidate cases that might have already been seen \cite{IceCube-a.2018, IceCube-b.2018, Icecube.2022, Icecube.2023}.      
These objects and phenomena provide new insights into some of the important questions in contemporary astrophysics, such as the properties of supernova progenitor stars; the ubiquity of MBH at the centre of galaxies and the lower end of their mass function; how mass falls onto black holes and relativistic jets are launched; how the coronal activities of host stars impact the atmosphere and habitability of exoplanets; what kind of remnants and what kind of X-ray radiation are produced in the process of binary neutron star mergers, and how to further constrain the neutron star equation of state. 
To help address the above questions, it is essential to characterise these important transient and variable types by detecting them in larger numbers and complementing their studies with high-quality follow-up observations at multi-wavelengths.

The Einstein Probe (EP) mission \cite{Yuan2022} is aimed at detecting such transients and phenomena in a systematic way by looking deeper beyond the reach of the current instruments in orbit while maintaining a similarly large instantaneous field of view. 
EP is a mission led by the Chinese Academy of Sciences (CAS) and a collaborative project participated by the European Space Agency (ESA), the Max-Planck Institute for Extraterrestrial Physics (MPE) in Germany, and the Centre National d’Etudes Spatiales (CNES) in France. 
EP was successfully launched into orbit from the Xichang Satellite Launch Center on January 9, 2024. Following the launch was the commissioning and calibration phase, during which the spacecraft and the instruments were tested and calibrated. From 2024 July the commissioning and calibration phase was successfully completed and the nominal operations phase has started. 

This paper discusses the science objectives of EP, starting with a summary of its scientific capabilities (Section\,\ref{sec:ep}).  As an interdisciplinary mission, EP covers a wide range of science topics concerned with time-domain and high-energy astrophysics.
By expanding the monitoring horizon beyond the Milky Way, EP enables systematic detections of extragalactic transients and variables fainter than typical GRBs. Among them, those with short timescales (from seconds to days) that are produced in processes involving stellar objects are discussed in Section\,\ref{sec:fegt}. In contrast, sources with longer timescales, mostly related to massive black holes at the center of galaxies, e.g. TDEs and active galactic nuclei (AGNs), are discussed in Section\,\ref{sec:smbh}. The scientific opportunities and potential of EP in the era of multi-messenger astronomy concerned with gravitational waves and neutrinos are discussed in Section\,\ref{sec:mma}. The studies of stellar-mass compact objects, including black holes, neutron stars (NSs) and white dwarfs (WDs), isolated or in binary systems, in our and nearby galaxies are presented in Section\,\ref{sec:sco}. 
Section\,\ref{sec:flarestar} discusses the science cases in the research of stellar flares.
The capabilities of EP as an X-ray observatory, as well as its unique features, allow advances in many areas of X-ray astronomy, which are discussed in Section\,\ref{sec:obssci}, focusing mainly on supernova remnants (SNRs) and clusters of galaxies. A brief summary is given in Section\,\ref{sec:sum}.

\section{Scientific capabilities and observing strategies} 
\subsection{Scientific capabilities}\label{sec:ep}

EP is equipped with two X-ray telescopes, both of the focusing optics type. 
The Wide-field X-ray Telescope (WXT) is designed to monitor the sky with an onboard triggering system. 
The Follow-up X-ray Telescope (FXT, \cite{Chen.2020}) is for quick onboard observations to characterize and precisely localize transients detected, and for observations of targets of opportunity (ToO) by command uplink.
Some of the specifications of WXT and FXT are listed in Table\,\ref{tab:EPinstrument}. 
These specifications and performances have been verified by the onground calibration of the instruments as well as the in-orbit calibration (results of the calibrations for WXT and FXT will be presented elsewhere). Some of the WXT performances, e.g. the point spread function (PSF) and limiting fluxes, have also been demonstrated by the EP/WXT pathfinder LEIA (Lobster-Eye Imager for Astronomy), launched in July 2022 \cite{Cheng.2024, ZhangC2022, Ling2023}.

\begin{table*}[tb]
\footnotesize
\caption{Specifications of the instruments}
\label{tab:EPinstrument}
\tabcolsep 35pt 
\begin{tabular*}{\textwidth}{ccc}
\toprule
Parameters & Wide-field X-ray Telescope & Follow-up X-ray Telescope  \\\hline
Number of modules & 12 & 2\\
Telescope optic & lobster-eye MPO & Wolter-I \\
Detector   & CMOS & pn-CCD \\
Field of view & $\ge$ 3600 sq.deg. & $\ge$ 60' (diameter) \\
Focal length (mm) & 375 & 1600 \\
Effective area @1.25\,keV (cm$^2$)& 2--3 & $\sim$300 (one unit)\\
Spatial resolution (1\,keV) &  5' (FWHM) & 20-24'' (HPD, on-axis) \\
Bandpass (keV) & 0.5--4 & 0.3--10 \\
Energy resolution (eV) & 122 @1.25\,keV  & 100 @1.5\,keV \\
 & $\sim$$8.9 \times10^{-10}$\,(27.6\,mCrab) @10\,s & \\
Limiting flux (ergs\,s$^{-1}$\,cm$^{-2}$) & $\sim$$1.2 \times10^{-10}$\,(3.9\,mCrab) @100\,s & $\sim$1$\times10^{-14}$ @10\,ks\\
 & $\sim$$2.6 \times10^{-11}$\,(0.8\,mCrab) @1\,ks & \\
 & & 50\,ms (full-frame) \\
Time resolution & 50\,ms & 2\,ms (partial window) \\
 & & 42\,$\mu$s (timing) \\
\bottomrule
\end{tabular*} 
\vspace{5mm}

Notes: The typical limiting fluxes in 0.5-4\,keV are derived for a point-like source from simulations, assuming a power-law spectrum with a photon index of 2 and a Galactic absorption column $3\times10^{20}$\,cm$^{-2}$.\\
MPO: micro-pore optics; FWHM: full width at half maximum; HPD: half-power diameter; mCrab: 1/1000 of the strength of the X-ray flux of the Crab nebular.  
\end{table*}

\begin{figure}[H]
\centering
\includegraphics[width=0.49\textwidth]{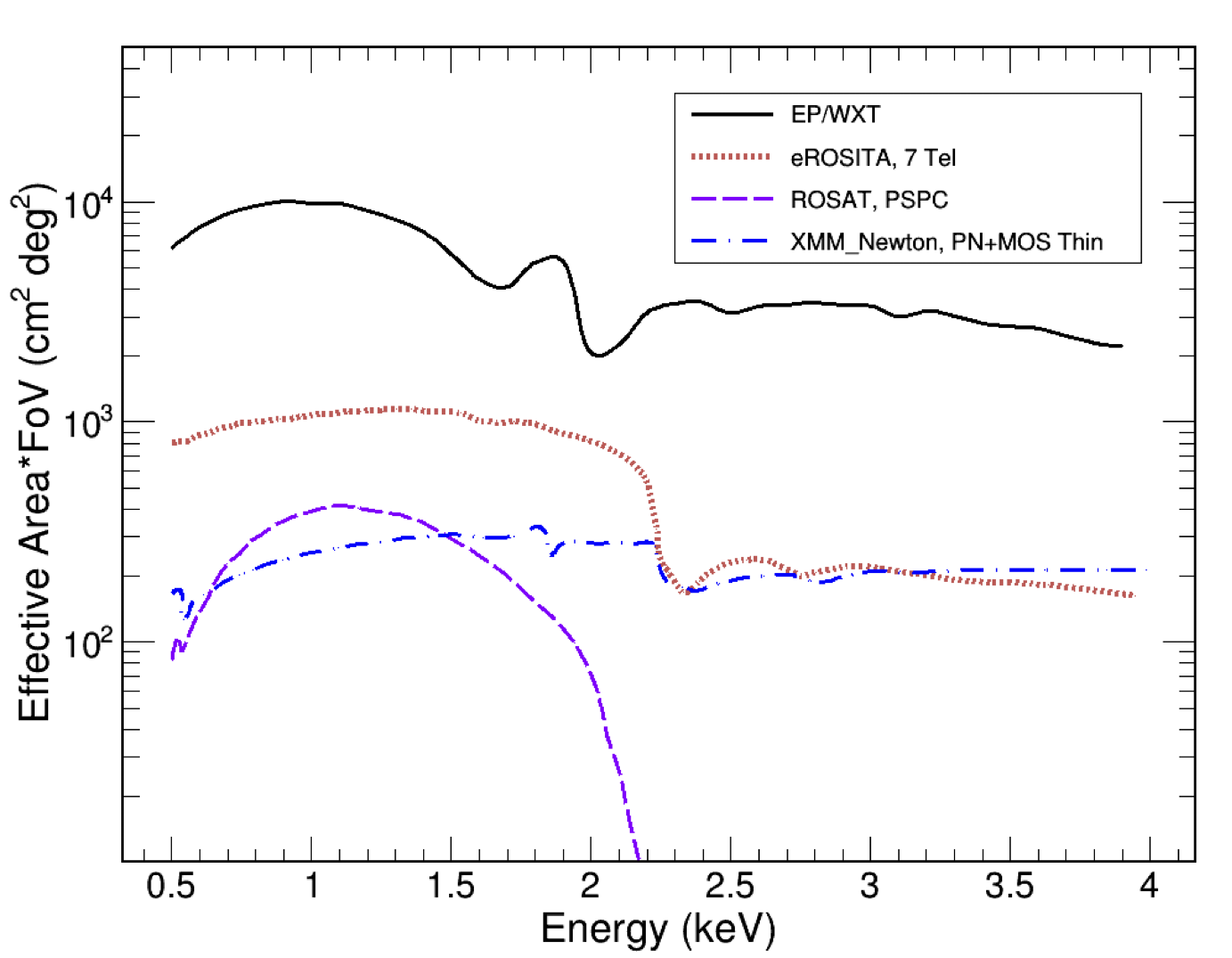}
\caption{Representative grasp (effective area times FoV) of WXT as a function of photon energy (black). The grasp parameters of several X-ray focusing telescopes are overplotted for a comparison (adapted from Figure 10 in  \cite{Yuan2022}).}
\label{fig:wxt_grasp}
\end{figure}

The WXT employs novel lobster-eye micro-pore optics (MPO), combined with a large array of total 48 back-illuminated scientific Complementary Metal-Oxide-Semiconductor (CMOS) \cite{Wu.2022, Wang.2022} detectors (each of $6\times6\,{\rm cm}^2 $), enabling a large instantaneous field of view (FoV) of over 3600 square degrees.
As a characteristic feature of lobster-eye optics, the imaging properties, e.g. PSF and effective area, are largely uniform across almost the entire FoV, i.e., free from strong vignetting effects as inherent in most other optics (e.g. Wolter-I mirrors). 
The PSF on the focal sphere has a characteristic cruciform shape and a bright central spot. The effective area for photons focused onto the central spot peaks around $2-3$~cm$^2$ at 1~keV, whilst the total area is $7-8$~cm$^2$ when all the photons falling onto both the central spot and two cross-arms are collected.
Although the effective area is small, its variations are minor across almost the FoV except at the edges.
These properties lead to the Grasp parameter (product of effective area and FoV), a figure of merit for the power of sky survey, $\sim$10,000\,cm$^2$\,deg$^2$ at 1\,keV, the largest in the soft X-ray band among all focusing X-ray telescopes (Figure \ref{fig:wxt_grasp}). 
The time resolution of WXT is 50 milliseconds, making WXT less susceptible to the pile-up effect for bright sources.

The nominal values of the detecting sensitivity for typical WXT observations are given in Table\,\ref{tab:EPinstrument}. They were estimated from simulations for point-like sources by making use of some of the instrumental parameters derived from the on-ground calibration (e.g.  PSF and effective area).
A background level of about 0.3~counts~s$^{-1}$\,cm$^{-2}$ (0.5--4 keV), which was predicted for the WXT detectors in orbit, is adopted (this value is within the range, though slightly lower than the average value, of the in-orbit measurements).  
It is clear that the WXT sensitivity improves up on those of the wide-field X-ray monitors currently in orbit by one order of magnitude or more. 
This is mainly enabled by the true focusing imaging capability of the lobster-eye mirror with several-arc-minute angular resolution.
Meanwhile, another advantageous factor is the soft X-ray bandpass of WXT, in which the number of photons is the largest for cosmic X-ray sources with a typical spectral shape.
Therefore, WXT is best suited to detect thermal sources with spectra peaking in soft X-rays, such as TDEs and SN-SBOs, whose effective temperatures are of the order of several tens of electron-Volt (eV).  

\begin{figure}[H]
\centering
\includegraphics[width=0.5\textwidth]{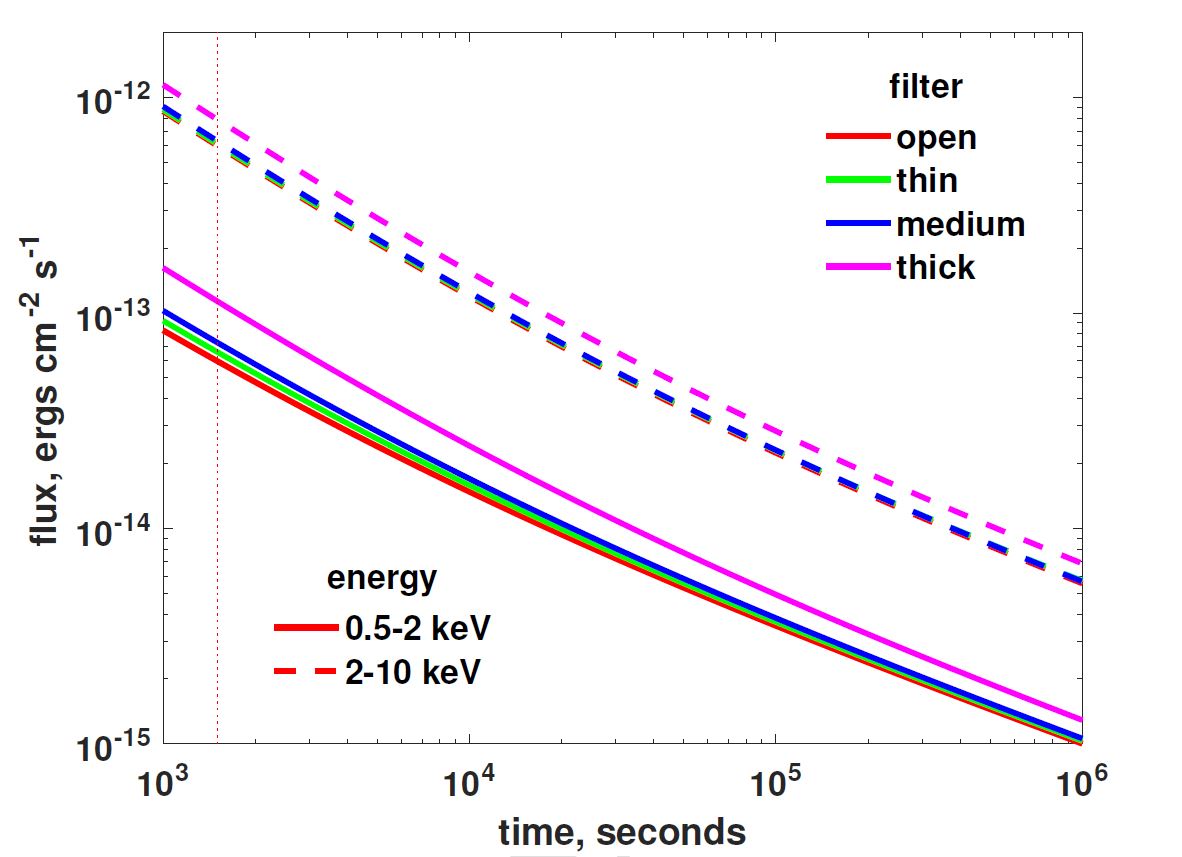}
\caption{Simulated sensitivity of the FXT for one telescope unit in the 0.5--2~keV and 2--10~keV bands for different filter settings (figure from \cite{zhangj22}).}
\label{fig:sec4_figure6}
\end{figure}

The FXT employs a set of twin co-aligned conventional Wolter-I telescopes equipped with pn-CCDs. Each of the mirror assemblies is of the same design as one of the seven  eROSITA telescopes \cite{Predehl2021}. 
The combined effective area, as measured in on-ground calibration, reaches $\sim$600~cm$^2$ at around 1.25~keV, which is excellent among dedicated follow-up telescopes.
The combined effective area is $94$~cm$^2$ at the Fe K line energies around 6\,keV, comparable to that of the Swift/XRT.
These values are in good agreement with those measured in in-orbit calibrations. 
The simulated detection sensitivity for point-like sources as a function of exposure is shown in Figure\,\ref{fig:sec4_figure6} \cite{zhangj22} for one FXT unit, reaching $\sim 10^{-14}$~ergs\,s$^{-1}$\,cm$^{-2}$ in 0.5--2 keV with a 25-minute exposure.
FXT can operate in three readout modes, full-frame, partial-window, and timing mode, to adapt to different observational requirements on the time resolution.
These modes render a time resolution of 50\,ms, 2\,ms, and 42\,$\mu$s, respectively, and a pipe-up flux limit for bright sources of 10\,mCrab, 200\,mCrab and 5\,Crab, respectively, for a pipe-up fraction $<10\%$). 

\subsection{Observing and follow-up strategies}

The EP satellite operates in a circular orbit of 593 km with an orbital period of 96.5 minutes and an inclination of 29 degrees. Its science observing modes include survey observations with WXT, follow-up observations with FXT triggered by on-board detection of transients and outbursts, and ToO observations via command uplink. 
During WXT survey, three pointed observations are carried out to cover three regions without much overlap in the night sky in each orbit. In this way most of the night sky is covered in 3 orbits. For a specific position in the night sky, the cadence ranges from several to about 10 visits per day. WXT data taken on the same sky regions in multiple orbits will be stacked to detect sources and transients that are fainter than those detectable in single snapshots. In half a year, the entire sky will be covered. 

During WXT observations, the data taken are processed by the on-board processing and triggering unit in real time to search for transients brighter than a preset threshold. 
Once a transient is detected, its basic information (e.g.  the coordinates, flux, spectral hardness ratio) is downlinked to the ground segment at a latency of a few minutes. 
There are two routes for this: the VHF network system of the CNES of France (built for the Chinese-French SVOM mission) and the Chinese \emph{Beidou} satellite navigation system. The alert data will be distributed to the entire astronomical community at the first instance to trigger multi-wavelength follow-up observations.
For normal ToO command uplink, the latency is about 3 hours; 
for time-critical ToOs, the latency is less than 10 minutes via the \emph{Beidou} system.

\begin{figure*}
\centering
\begin{tabular}{c}
\includegraphics[width=0.99\textwidth]{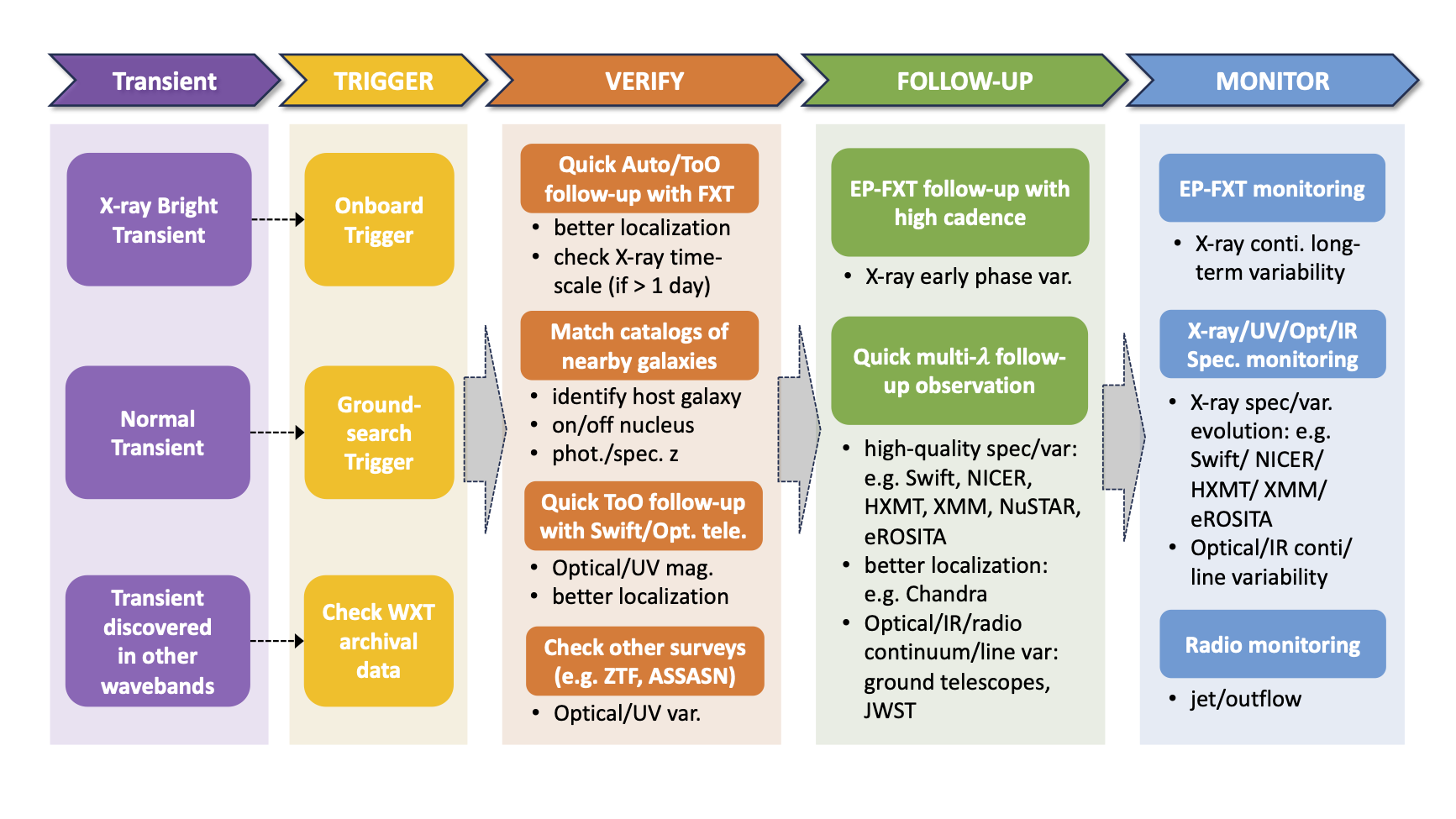}  \\
\end{tabular}
\caption{A basic decision-making process for the follow-up observations of EP transients.}
\label{fig-transient-flowchart}
\end{figure*}

In general, the following observing strategies are applied to transient candidates.
(1) Upon the detection of a new transient onboard, a follow-up observation with FXT is triggered automatically; it takes 3--5 minutes for the spacecraft to slew to point FXT to the target. 
Meanwhile, the alert data is downlinked to the EP science center. 
For transients of high scientific significance, further FXT monitoring with enhanced cadence is carried out as ToO observations. 
(2) Transients fainter than the preset on-board triggering threshold may also be detected from on-ground data processing, at a time latency of typically several hours to half a day depending on the data telemetry speed;
they may also trigger FXT observations as internal ToO observations. 
(3) For transients detected by other telescopes in other wavebands, X-ray follow-up observations can be performed with EP, with both FXT and WXT. For events with a poorly determined position, WXT may first provide improved source positions if an X-ray afterglow can be detected, which can in turn be used to guide the pointing of FXT.
(4) The observing strategies for multi-messenger events are specially designed and described in Section\,\ref{sec:obs_strategy_mm}.
A decision-making process for follow-up observations of transients is demonstrated in Figure\,\ref{fig-transient-flowchart}.

\section{Extragalactic Fast X-ray Transients}
\label{sec:fegt}



Extragalactic Fast X-ray Transients (FXRTs) with timescales from seconds to hours are of heterogeneous origin. 
A large fraction of the early detections were associated with GRBs as the prompt X-ray emission by several missions, e.g. Ginga, WATCH, BeppoSAX and HETE-2 (e.g. \cite{Strohmayer1998, Sazonov1998, Lamb2004}).

Long-lasting X-ray afterglows were discovered by the BeppoSAX mission \cite{Costa1997, Piro1998}. Fast X-ray transients without GRB counterparts were also found, sometimes referred to as X-ray flashes (XRFs) \cite{2001grba.conf...16H}, and suggested to be very soft, X-ray rich GRB. Other types of FXRTs have been discovered in recent years, mostly serendipitously by X-ray telescopes with a narrow FoV. A marked example is the X-ray outburst (XRO) 080109 accompanying SN 2008D detected with Swift/XRT \cite{Mazzali2008, Soderberg2008}. Early multi-wavelength (ultraviolet, optical and near-infrared) photometric data confirmed XRO~080109 originating from the supernova SBO process \cite{Modjaz2009}. Although real-time detection has not been the main science driver of these narrow-field X-ray telescopes, there have been recent attempts to enable serendipitous discovery and quick dissemination of extragalactic FXRTs with, e.g. Swift/XRT  \cite{evans2023}. Most of the efforts, however, have been devoted to mining the archival data of various X-ray missions, particularly of Chandra and XMM-Newton \cite{YangG2019, Pastor-Marazuela2020}.
These studies resulted in samples of FXRTs, e.g. those interpreted as SBO candidates from XMM-Newton data \cite{Alp2020, Novara2020}, and samples of FXRTs from Chandra \cite{Quirola-Vasquez2022,Quirola-Vasquez2023}.

 

Of particular interest are two FXRTs found in the Chandra Deep Field-South (CDF-S) survey, CDF-S XT1 and XT2 \cite{Bauer2017, XueYQ2019}. XT1 is characterized by a rise time of $\sim$100 seconds and a power-law temporal decay. It is consistent with a few theoretical scenarios including an ``orphan" X-ray afterglow from an off-axis GRB, a low-luminosity GRB at high redshift, or a beamed intermediate-mass black hole (IMBH)-WD TDE \cite{Bauer2017, PengZK2019, Sarin2021} (see Section\,\ref{sec:tde} for discussion on TDE).
In contrast, XT2 shows an X-ray plateau similar to those seen in some short GRB afterglows \cite{XueYQ2019}. Multi-wavelength constraints and host galaxy properties suggest that XT2 was most likely produced by wind emission from a massive, rapidly spinning newborn magnetar formed in a BNS merger \cite{ZhangB2013, XiaoD2019}. If confirmed, XT2 represents a new type of electromagnetic (EM) counterparts of GWs although its presumed GW signal is beyond the reach of the current operational GW detectors. Interestingly, XT1 can also be explained by such a model, with the line-of-sight different from that in the case of XT2 \cite{SunH2019}. 
The studies of XT1 and XT2 are timely in light of the milestone event GW\,170817, and have stimulated search for other similar candidate events \cite{LinDC2022}.

The aforementioned sources constitute the current samples of extragalactic FXRTs. Clearly they have heterogeneous origins. 
In the Chandra sample, FXRTs come in two categories: those ``nearby" ones apparently associated with galaxies in the local universe and ``distant" ones beyond 100 Mpc \cite{Quirola-Vasquez2022,Quirola-Vasquez2023}. Despite of some similar features in their light curves, the great differences in distance lead to a wide luminosity distribution spanning several orders of magnitude. Generally, the ``nearby" FXRTs have peak luminosities below a few $\times10^{42}\,\rm ergs\,s^{-1}$, falling into the realm of ultraluminous X-ray Sources and X-ray binaries, as well as soft $\gamma$-ray repeaters (SGRs) and anomalous X-ray pulsars (AXPs) that have comparable luminosities \cite{Gogus1999, Hurley1999, Terasawa2005, Woods2006}. For the distant sample, possible origins include TDEs involving a white dwarf and an intermediate-mass black hole with a possibly wide luminosity range of $\sim$$10^{42}-10^{48}\,\rm ergs\,s^{-1}$ \cite{Saxton2021}, SN SBOs with typical luminosities $\sim$$10^{42}-10^{44}\,\rm ergs\,s^{-1}$ \cite{Nakar2010, Waxman2017}, XRFs associated with low-luminosity GRBs (LL-GRBs) \cite{Campana2006, Pian2006, Starling2011}, ``orphan" X-ray afterglows from off-axis GRBs \cite{Bauer2017,Sarin2021} and magnetar wind emission as the aftermath of BNS mergers \cite{SunH2019, XiaoD2019, XueYQ2019}. 
However, the origins of most FXRTs in the current samples remain controversial, due to limited observational data, especially a lack of multi-wavelength counterparts except for just one single case (XRO 080109/SN 2008D).

With the large FoV, soft X-ray bandpass, and good sensitivity, EP-WXT is well suited to detect FXRTs of various types and expected to enlarge the samples substantially. The yearly detection rate for WXT is estimated to be from a dozen \cite{Quirola-Vasquez2023} to $\sim 100$ based on the results of LEIA (Li D.Y. in preparation), which is also confirmed with the first phase of EP operations. For some FXRTs, deep X-ray follow-ups can be performed with EP-FXT. EP's rapid dissemination of transient alerts to the whole community, a designed goal of the mission, will trigger follow-up observations in time, enabling search of multi-wavelength counterparts and redshift measurement. This will, as evidenced by Beppo-SAX, HETE-2, and Swift for the GRB research, transform the field. 


\subsection{$\gamma$-ray Bursts}

GRBs are the most violent explosive phenomena in the universe and are classified by their duration. Bursts with prompt emission longer than 2 seconds are named long GRBs (LGRBs), and those shorter than 2 seconds are short GRBs (SGRBs). After half a century of research, there has been significant progress in our understanding of GRBs. A great leap forward was achieved with the discovery of the counterparts of LGRBs led by BeppoSAX, and then of SGRBs by Swift and HETE-2, yet some mysteries remain regarding the progenitor star, jet property, and radiation mechanism \cite{zhang2018book}. 
With its large FoV, high sensitivity and unique soft X-ray bandpass, EP provides new impetus to the study of GRBs in the following aspects.

\subsubsection{High-redshift GRBs}  

As the brightest cosmic beacons in the universe, GRBs can be detected up to very high redshifts. Currently, the most distant spectroscopically confirmed burst is GRB 090423 at a redshift of $z=8.2$ \cite{2009Natur.461.1258S,2009Natur.461.1254T}.
GRBs are thus promising probes of the early universe, complementary to quasar and galaxy surveys. The study of high-$z$ GRBs can shed light on the dark energy and cosmological parameters, the cosmic star formation rate, the first stars (the so-called Population III stars), the cosmic reionization epoch, the metal enrichment history, and more (see \cite{2015JHEAp...7...35S,2015NewAR..67....1W} for reviews).

\begin{table}[H]
\footnotesize
\begin{threeparttable}
\caption{List of high-redshift ($z\geq6$) GRBs detected by Swift}\label{tab:high-zGRBs}
\doublerulesep 0.1pt \tabcolsep 10pt 
\begin{tabular}{ccc|ccc}
\toprule
\multicolumn{3}{c}{Spectroscopy}  & \multicolumn{3}{c}{Photometry} \\\hline
  GRB & $z$ & Refs. & GRB & $z$ & Refs. \\\hline
  050904 & 6.3 &  \cite{2006Natur.440..184K}  & 090429B & 9.4 & \cite{2011ApJ...736....7C} \\
  080913 & 6.7 &  \cite{2009ApJ...693.1610G}  & 120521C & 6.0 & \cite{2014ApJ...781....1L} \\
  090423 & 8.2 & \cite{2009Natur.461.1258S,2009Natur.461.1254T}   & 120923A & 8.5 & \cite{2013arXiv1307.6156T} \\
  130606A & 5.9 &  \cite{2013ApJ...774...26C,2013arXiv1312.5631C,2014PASJ...66...63T,2015A&A...580A.139H}  &         &     &  \\
  140515A & 6.3 & \cite{2014arXiv1405.7400C,2015A&A...581A..86M}   &         &     &  \\
  210905A & 6.3 & \cite{2023A&A...671A..84S}   &         &     &  \\
  240218A & 6.8 & \cite{2024GCN.35756....1S}   &         &     &  \\
\bottomrule
\end{tabular}
\end{threeparttable}
\end{table}

A major limitation is that, with current instruments, the detected high-$z$ ($z\geq6$) GRBs are rare, with only ten bursts identified after 20 years of Swift operations.
Among these, seven have spectroscopic redshifts and the other three have only photometric redshifts (Table\,\ref{tab:high-zGRBs}). On the other hand, population synthesis models predict that bursts at $z>6$ account for $\sim10\%$ of the whole population, strongly supporting that GRBs are effective in selecting high-$z$ objects \cite{2008MNRAS.385..189S,2015MNRAS.448.2514G}.
It is expected that the paucity of high-$z$ GRBs is mainly caused by the limited sensitivity and relatively high energy bandpass of the current GRB missions, as high-$z$ GRBs are faint and their peak emission is redshifted to low energies. Moreover, some high-$z$ GRBs might have been missed during the patchy follow-up observation campaigns \cite{2022NatAs...6.1101C}.

\begin{figure}[H]%
\centering
\includegraphics[width=0.49\textwidth]{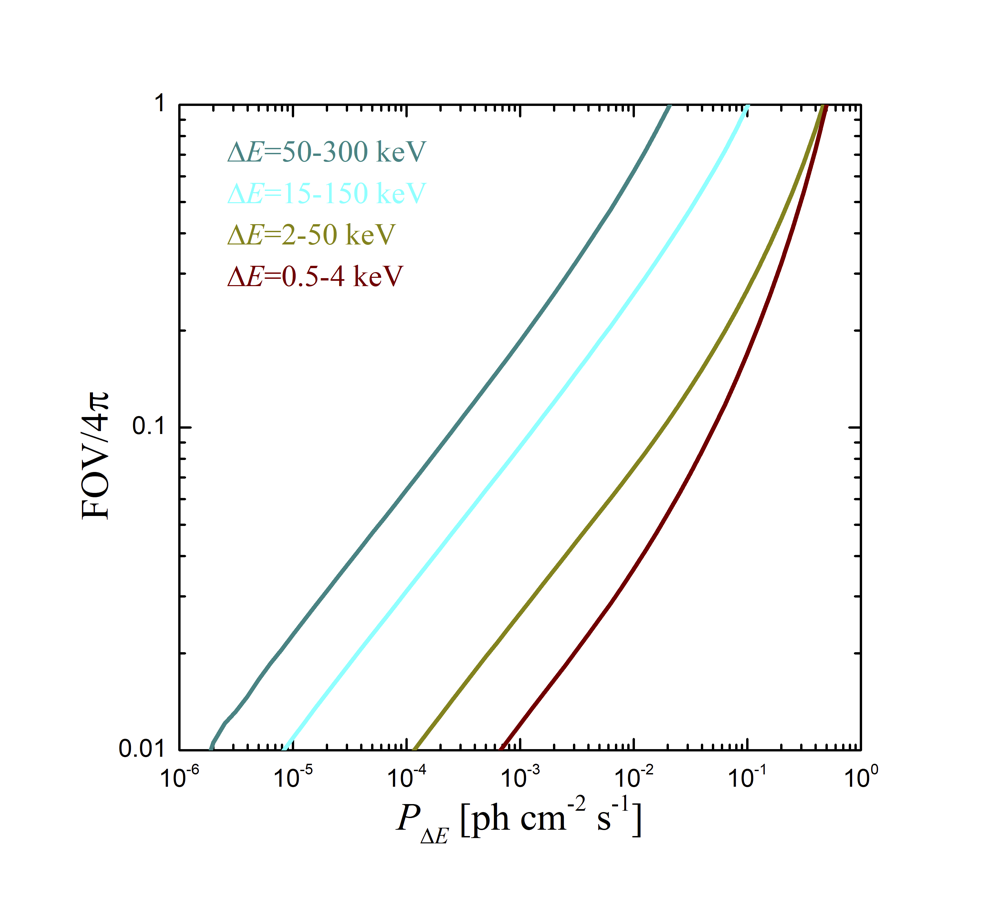}
\vskip-0.2in
\caption{Required sensitivity (in terms of peak flux $P_{\Delta E}$) and FoV,
operating in a given energy band $\Delta E$, to detect 10 GRBs $\rm yr^{-1}$ at $z>8$.
Different lines stand for different energy bands (see also in \cite{2015JHEAp...7...35S}).}
\label{fig:high-zGRB}
\end{figure}

In order to take full advantage of the potential of GRBs for probing the early Universe, a much larger sample of high-$z$ GRBs than the present one is needed. The detection of high-$z$ GRBs is therefore one of the important goals driving the design of the current and future space missions, including EP. Based on an observational tested population model, the detection rate of high-$z$ GRBs by a generic detector with a defined sensitivity and energy band was estimated \cite{2015MNRAS.448.2514G}. Following the procedure in \cite{2015MNRAS.448.2514G}, the sensitivity and FoV required for a mission to be able to detect 10 GRBs $\rm yr^{-1}$ at $z>8$ in various energy bands are shown in Figure\,\ref{fig:high-zGRB} (see also \cite{2015JHEAp...7...35S}).
For example, by adopting the Swift FoV of 1.4\,sr and the energy band of 15--150 keV, to detect 10 GRBs $\rm yr^{-1}$ at $z>8$ requires an improvement by 100 times over the current sensitivity of $\sim0.4$ ph $\rm cm^{-2}$ $\rm s^{-1}$. 
Whereas for an instrument with the same FoV and operating in the 0.5--4 keV energy band, a sensitivity level of $\sim0.06$ ph $\rm cm^{-2}$ $\rm s^{-1}$ is required.
Obviously, a better strategy for capturing a large number of high-$z$ GRBs is to have an instrument operating in preferably the soft X-ray band while retaining high sensitivity \cite{2015MNRAS.448.2514G,2015JHEAp...7...35S}. 

Compared to the other currently operating wide-field monitors, EP-WXT's higher sensitivity ($\sim 8.9\times 10^{-10}$ ergs\,$\rm s^{-1}$\,$\rm cm^{-2}$ or $\sim 0.45$ photons\,$\rm s^{-1}$\,$\rm cm^{-2}$ for an exposure of $10$\,s, assuming a power-law spectrum with a photon index -2 and a Galactic absorption column $3\times10^{20}$\,cm$^{-2}$) and soft X-ray bandpass (0.5--4 keV) make it a suitable instrument to detect high-$z$ GRBs \cite{2018SSPMA..48c9505W}. Based on the same population model, the high-$z$ GRB detection rate by EP is expected to be $\sim5$ events $\rm yr^{-1}$ at $z>6$. 
Assuming an efficiency of roughly 30\% for determining the redshift by ground-based spectroscopic follow-ups,  it is expected that some 1.5 GRBs will be found at $z>6$ by EP per year (there is an uncertainty of a factor of $\sim1.5$). 
Clearly, quick optical/IR follow-ups of the afterglows and the measurement of the redshifts are essential to identify high-$z$ GRBs detected by EP.   
It is worth noting that the first EP GRB with measured redshift during its commissioning phase, EP240315a, has a redshift of 4.859 \cite{ricci.2024, Gillanders2024arXiv, Levan2024arXiv, LiuY2024arXiv}.
After several years of operations, EP will demonstrate the long-anticipated prospects for detecting high-$z$ GRBs with lobster-eye telescopes.

\subsubsection{X-ray Flashes}
\begin{figure}[H]%
\centering
\includegraphics[width=0.49\textwidth]{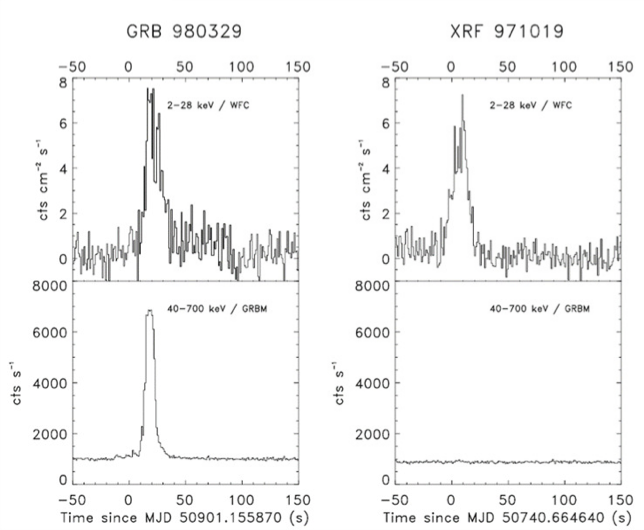}
\caption{Light curves of GRB 980329 and XRF 971019 in the X-ray (BeppoSAX/WFC) and $\gamma$-ray (BeppoSAX/GRBM) bands
plotted on the same scale displays the absence of $\gamma$-rays of XRFs. Adapted from \cite{2003AIPC..662..229H}.}
\label{fig:XRF}
\end{figure}

X-ray Flashes are an extension of classical GRBs to the softer and fainter regime, with similar duration and isotropic spatial distributions. Ever since XRFs were first confirmed by
the BeppoSAX satellite \cite{2001grba.conf...16H,2003AIPC..662..229H}, they have been
studied extensively by performing a population study of both the prompt and afterglow properties of BeppoSAX \cite{Dalessio06}, HETE-2 and Swift events \cite{2005ApJ...629..311S,2008ApJ...679..570S}.
An example of the light curves of a GRB and an XRF in the X-ray and $\gamma$-ray bands is shown in Figure \ref{fig:XRF}. It is seen that the XRF emission is concentrated in the X-ray band.
The spectral properties of XRFs are similar to those of GRBs, except that the values of
the spectral peak energy, the peak flux, and the energy fluence of XRFs are relatively lower \cite{2005ApJ...629..311S}. There is no clear boundary between XRFs and GRBs. Bursts in the
grey zone are sometimes termed ``X-ray rich'' GRBs (XRRs). 
XRFs, XRRs, and classical GRBs are usually defined as bursts with $\log[S(2-30\;{\rm keV})/S(30-400\;{\rm keV})]>0.0$,
$-0.5<\log[S(2-30\;{\rm keV})/S(30-400\;{\rm keV})]\leq0$, and $\log[S(2-30\;{\rm keV})/S(30-400\;{\rm keV})]\leq-0.5$,
respectively, where $S$ is the energy fluence in the corresponding energy band.

Up to now, several theoretical models have been proposed to explain XRFs. 
\begin{itemize}
  \item High-redshift GRBs: due to the redshift effect, the prompt emission of
  high-$z$ GRBs is expected to peak at softer energies, appearing as XRFs \cite{2001grba.conf...16H}.
  However, a systematic analysis of the redshift distribution of XRFs suggests that
  at least some of them are nearby events, thereby excluding the large distance hypothesis \cite{2007A&A...465L..13G}.
  \item  Off-axis GRBs: in this model, GRBs are observed on the axis of the jetted fireball  that produces the burst, while XRFs correspond to an off-axis geometry \cite{2002ApJ...571L..31Y,2004ApJ...606L..33Y}. It was shown that the combined properties of the prompt and afterglow emission of XRFs and XRRs are consistent with an off-axis scenario only if the jet is structured (either Gaussian or power-law shaped), while a top-hat jet is excluded \cite{Dalessio06}.
  \item  Structured jets: GRB jets are significantly structured in this model, 
  with an angle-dependent energy per solid angle. It was shown that a variable jet opening angle model can explain the
  observed properties of XRFs, XRRs, and GRBs in a unified picture \cite{2005ApJ...620..355L,Salafia15}.
  \item  Dirty fireballs: a naive hypothesis is that baryon-contaminated  fireballs with smaller Lorentz factors would not produce $\gamma$-ray emission 
  but give softer and fainter emission. Thus, these dirty fireballs 
  \cite{1999ApJ...513..656D,2002MNRAS.332..735H} have 
  been regarded as the origin of XRF-like events.
\end{itemize}

Thanks to the soft X-ray passband and high sensitivity, EP-WXT is expected to significantly enlarge the sample of XRFs. 
In X-rays, XRFs and high-$z$ GRBs may exhibit similar spectral hardness, and can only be distinguished by redshift measurements. The study of a large sample of XRFs will provide valuable insights into the emission mechanism of GRBs, the nature of GRB progenitors, cosmic starformation history, SN explosion, and so on.

\subsubsection{Low-luminosity GRBs}

Low-luminosity GRBs are fainter by four orders of magnitude ($10^{46} - 10^{48}$~erg~s$^{-1}$ \cite{Campana06,Soderberg06,Kaneko07}) 
than normal GRBs ($10^{50} - 10^{52}$~erg~s$^{-1}$),
but with a higher local event rate, making them crucial in understanding the sources of high-energy cosmic rays, neutrinos and gravitational waves \cite{LiangEW07}.
While normal GRBs are produced by a highly relativistic jet seen by an on-axis observer, LL-GRBs are interpreted as the shock breakout
of a mildly relativistic jet \cite{Nakar15}, poorly collimated jets, or off-axis viewing effect \cite{Urata15,Salafia15}.
The jet launched from the central engine may be shocked as it propagates through the extended star envelope \cite{Bromberg11}
or the merger ejecta \cite{Bromberg18,GengJJ19}, then the emission from the resulting cocoon material is expected to be thermal in the X-ray band.
According to prevailing magnetohydrodynamical simulations of GRB jets (e.g.  \cite{Gottlieb20}),
for an off-axis observer, faint thermal-like X-ray emission from the cocoon may be common in LL-GRBs.   

Although the local event rate of LL-GRBs is predicted to be high according to the extension of the luminosity function \cite{Pescalli15}, so far only a few LL-GRBs have been observed and investigated in detail.
On the one hand, the current $\gamma$-ray detectors are insensitive to them due to their low luminosity. On the other hand, it is argued that LL-GRBs have softer spectra of the prompt emission than those of normal GRBs.
Therefore, the good sensitivity and 0.5--4 keV soft X-ray bandpass (c.f. 20-300 keV for most GRB monitors) of EP-WXT are suitable to catch LL-GRBs more efficiently.
For a fraction of bursts detected by EP and GRB monitors (e.g. Swift, Fermi, INTEGRAL, GECAM and SVOM) simultaneously, their prompt emission properties could be well characterized, which would help unveil the physical origin of LL-GRBs. 

\subsubsection{Ultra-long GRBs}

\begin{figure}[H]%
\centering
\includegraphics[width=0.49\textwidth]{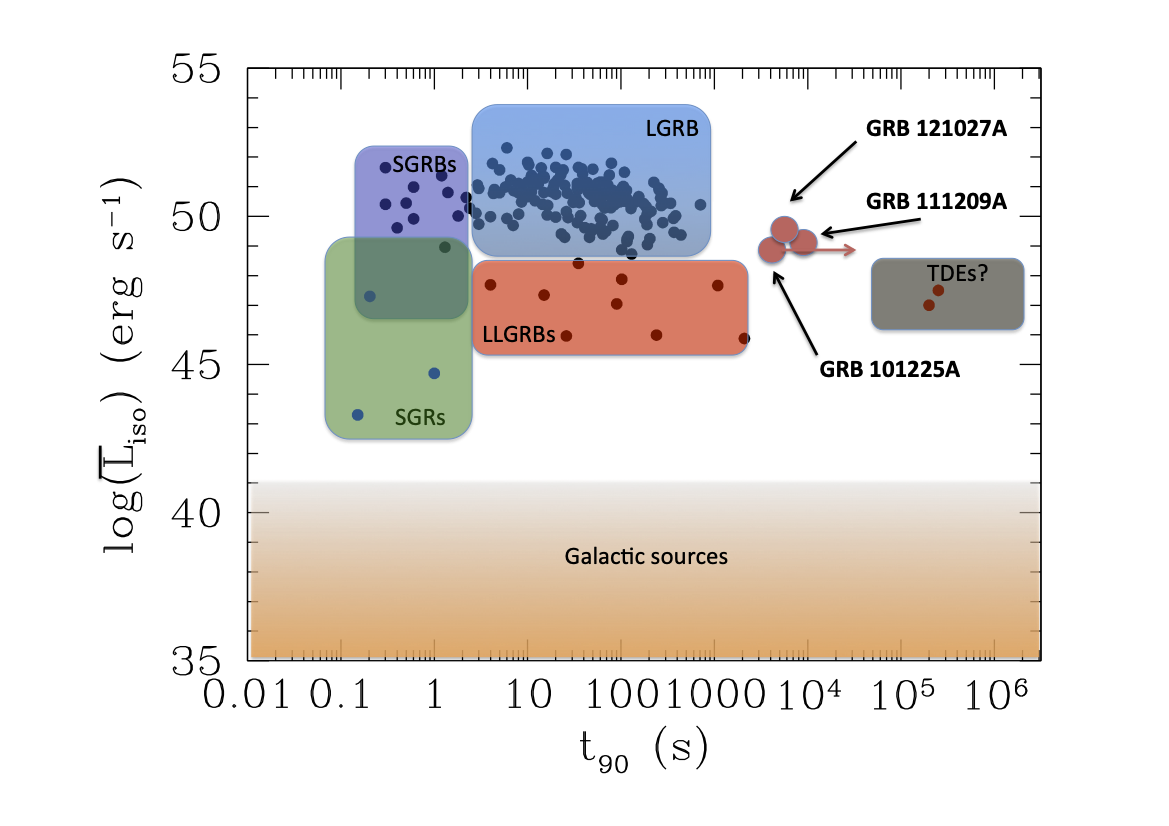}
\vskip-0.2in
\caption{Distribution of different classes of GRBs in the $T_{90}$ vs $L_{\rm iso}$ diagram, including short GRBs (SGRBs), long GRBs (LGRBs), low-luminosity GRBs (LL-GRBs), soft $\gamma$-ray repeaters (SGRs) and tidal disruption events (TDEs) (figure adapted from \cite{Levan14}).}
\label{fig:GRBs}
\end{figure}

Ultra-long GRBs (UL-GRBs) are another interesting subgroup whose prompt emission can last for more than several thousand seconds, as shown in Figure \ref{fig:GRBs}.
Their host galaxies are fainter and smaller than those of normal GRBs.
Several theories have been proposed to model the progenitor and radiation mechanism of UL-GRBs,
including the blue supergiant star (BSG) model \cite{Gendre13}, TDE by a massive black hole \cite{Levan14}, and a spin-down black hole model \cite{Nathanail15}. 
The UL-GRB 130925A exhibited an extremely soft X-ray component,  lasting for a day, that is attributed to a hot cocoon formed as the relativistic jet pierces through a very massive, low metallicity star progenitor \cite{Piro14}. The fallback of the stellar outer layers supports a long-duration jet, which entrains a large mass of baryons in the hot cocoon. On the other hand,
it is found that UL-GRB 111209A was accompanied by a bright SN explosion, 
named SN 2011kl. The magnitude of SN 2011kl was $M_{\rm bol} = -20.0$, which is between ordinary supernovae (SNe) ($M_{\rm bol} > -19.0$) 
and superluminous supernovae (SLSNe) ($M_{\rm bol} < -21.0$), and is three times or more brighter than those of typical SNe associated with GRBs \cite{Greiner15}. Moreover, the spectra of SN 2011kl are different from normal SNe,
but similar to SLSNe. Hence, SN 2011kl is suggested to be powered by a magnetar.
The association of UL-GRB 111209A with SN 2011kl does not support the progenitor model involving a blue giant star or a TDE of a massive black hole.

During the prompt phase of an UL-GRB, the X-ray lightcurve exhibits violent flares,
with an average luminosity of $10^{49}$~erg~s$^{-1}$. Assuming a half-opening angle of 10 degrees and an event rate of $1/3$ of normal GRBs, it is estimated that EP may detect several UL-GRBs per year. This will greatly enlarge the sample of UL-GRBs, providing crucial clues to their progenitors and radiation mechanism.

\subsubsection{Other features observed in GRBs}

In addition to the aforementioned types of GRBs, some GRBs may also exhibit special and interesting features. For instance, in a fraction of GRBs ($\sim10\%$ of LGRBs and $\sim3\%$ of SGRBs), some weak emission features preceding the main bursts separated by quiescent gaps have been detected, which are usually termed as precursors \cite{Koshut1995,Lazzati2005a,Burlon2008,Burlon2009,Troja2010,HuYD2014,LanL2018,ZhongSQ2019,WangJS2020}. From a theoretical point of view, the precursors may be explained by various models, being possibly associated with e.g. relativistic fireballs \cite{Paczynski1986,Meszaros2001}, progenitor-linking jet breakout from massive stars envelope \cite{Ramirez-Ruiz2002,ZhangWQ2003,Lazzati2005b,Nakar2017}, the magnetospheric interaction or the tidal crust cracking at inspiral prior to binary compact stars coalescence \cite{Hansen2001,Troja2010,Tsang2012,Palenzuela2013,WangJS2018,Dichiara2023}, or directly the central engine activity \cite{Ramirez-Ruiz2001,Bernardini2013}. Precursors observed in some GRBs exhibit a spectrum significantly softer than the main prompt pulse \cite{Piro05,Dichiara2023}. 
EP may detect the precursor emission more efficiently than the previous missions and may shed light on whether any faint X-ray emission is produced during the periods of the apparent quiescence.  
This is of great importance for testing models of the GRB precursor emission.

In observations, about $20\%$ of the SGRBs (originating from binary compact star mergers) detected with Swift have extended emission following their initial short, hard spike \cite{Sakamoto2011}. Compared to the short spike, the extended emission has relatively softer spectra and can extend to the soft X-ray band \cite{Sakamoto2011}. In addition, some SGRBs show an internal X-ray plateau followed by a very rapid decay, which is consistent with being the dimmer and softer extended emission \cite{Troja2007,Rowlinson2010,Rowlinson2013,LvHJ2014,LvHJ2015,LanL2021}. Both the extended emission and internal plateau are difficult to interpret within the framework of a BH as the central engine, whereas they are consistent with the existence of a rapidly spinning millisecond magnetar that survives for an extended period of time and collapses into a BH as it spins down sufficiently \cite{Troja2007,Rowlinson2010,Rowlinson2013,LvHJ2014,LvHJ2015}. 

EP is expected to detect SGRBs with extended radiation or internal plateau. They can be used to estimate the maximum mass of neutron stars \cite{GaoH2016,Piro2017}. Moreover, some recent studies propose that, if the sudden cessation of an observed flux after the internal plateau indicates the collapse of a magnetar into a BH, some detectable signals from this newborn BH should be expected. For instance, a small X-ray bump (shown in GRB 070110) or a second plateau (shown in GRBs 070802, 090111, and 120213A) would show up if the fallback accretion could activate the newborn BH and sufficient energy could be internally dissipated or transferred from the newborn BH to the GRB blast wave \cite{ChenW2017,ZhaoLT2020}. Quick EP-FXT follow-up observations of SGRBs (triggered by EP or other high-energy wide-field monitors) will enlarge the sample of SGRBs with a late flare or a second plateau, which will provide a crucial test on whether the internal plateau truly is powered by a magnetar that is about to collapse. 

In the X-ray afterglow emission of several LGRBs, such as GRB 111209A and GRB 121027A, a distinct feature known as the ``giant X-ray bump" was observed \cite{WuXF2013,Greiner2015,GaoH2016b}. Unlike typical X-ray flares, where the rising and decaying phases have comparable timescales \cite{LiangEW2006,Chincarini2010}, the giant X-ray bump shows a ``step-like" rise followed by a much shallower decay. It is believed to be produced by the reactivation of a long-lasting central engine activity, which releases energy and powers both the prompt emission and the giant X-ray bump. The giant X-ray bump could be driven by the accretion of a fraction of the progenitor's envelope to the central black hole \cite{WuXF2013,GaoH2016b}. The detection of SN 2011k associated with GRB 111209A, which exhibits a giant X-ray bump, suggests that additional energy might be injected into the SN ejecta \cite{Greiner2015}. One plausible explanation is the Blandford-Payne mechanism \cite{Blandford1982}, which can extract energy and angular momentum from the accretion disk through a baryon-rich wide wind/outflow along the field lines that extend from the disk surface to large distances \cite{GaoH2016b}. 

Recently, a comprehensive search in archival data has identified 19 additional candidates of the giant X-ray bumps, mostly detected by Swift \cite{ZhaoLT2021}. EP allows for the detection of this feature in new GRBs and may aid in confirming its association with an observed SN, which can help determine the origin of giant X-ray bumps. Furthermore, such a study may contribute to our understanding of the physical processes involved in core collapsing and the power source behind SNe.

\subsection{Supernovae}
Among the thousands of supernovae observed, only over 60 or so have been detected in X-rays \cite{Dwarkadas2016,Brunton2023}. A collection of the observed SN X-ray emission can be found in the SN X-Ray Database (SNaX) \footnote{\url{https://kronos.uchicago.edu/snax/}} \cite{Nisenoff2020}.

\subsubsection{Circumstellar Medium Interaction}
The primary X-ray sources of SNe can be produced by the interaction of the high-speed SN ejecta with the surrounding circumstellar medium (CSM) \cite{Chevalier2006}. 
This process can drive two shocks: a forward shock propagating into the CSM and a reverse shock moving back into the SN ejecta \cite{Chevalier1982}. On the one hand, the forward shock has a typical velocity around 10,000 km s$^{-1}$, which can lead to effecitve temperatures of up to $10^{9}$ K. On the other hand, the reverse shock moves at a speed of about 1000 km s$^{-1}$, which can heat up the SN ejecta to roughly 10$^{7}$ K.
Both the forward and reverse shocks are, therefore, extremely hot and can emit soft and hard X-rays through free-free bremsstrahlung of electrons \cite{Fransson1996,Chevalier2006}. This X-ray emission can further excite the unshocked CSM to produce coronal emission lines such as Fe VII and Fe X.

The luminosity of the bremsstrahlung X-ray emission can be expressed as \cite{Chevalier2006}
\begin{eqnarray}
L_{\mathrm{X}} \approx  3.0 \times 10^{39} \bar{g}_{\mathrm{ff}}&C_{n}&\left(\frac{\dot{M}}{10^{-5} M_{\odot} \mathrm{~yr}^{-1}}\right)^2\left(\frac{v_{\text {wind }}}{10 \mathrm{~km} \mathrm{~s}^{-1}}\right)^{-2} \nonumber\\
&\times&\left(\frac{t}{10 \mathrm{~days}}\right)^{-1} \mathrm{ergs}\,\rm{s}^{-1} \label{Lxinteraction}
\end{eqnarray}
where $\bar{g}_{\mathrm{ff}} $ is the free-free Gaunt factor, $C_n=1$ for the forward shock and $(n-3)(n-4)/4(n-2)$ for the reverse shock, which depends on the ejecta density index $n$, and $\dot{M}$ and $v_{\text {wind}}$ are the mass-loss rate and wind velocity of the progenitor, respectively. 
Equation (\ref{Lxinteraction}) shows that the X-ray luminosity due to the bremsstrahlung of electrons is proportional to the square of the CSM density. Therefore, it is expected that the X-ray emission of interacting SNe can help to determine the density of the CSM and, furthermore, the late evolution of the progenitors since the formation of the CSM is determined by the stellar mass losses. For a steady wind of a density profile $\rho_{\rm CSM}\propto r^{-2}$, the thermal X-ray luminosity is expected to decline as $L_X \propto t^{-1}$. In comparison, most SNe Ia are expected to lack X-ray emission because of a very low density of the CSM. One exception is the type Ia SN\,2012ca, which has been detected in X-rays, indicating an asymmetric medium with a high-density component  \cite{Bochenek2018}.

The observed light curves of the X-ray emission of SNe arising from interaction with CSM exhibit a large diversity. 
Generally, the X-ray emission can last for from a few days to a few tens of years, and thus SNe behave as slowly-evolving X-ray sources rather than fast transient phenomena. The strongest X-ray emission is usually found in SNe IIn, with typical peak luminosities being around $L_X \sim 10^{41}-10^{42}\,\mathrm{ergs}\,\mathrm{s}^{-1}$. 
The X-ray emission from SNe IIL and SNe IIP is relatively weak, with luminosities $L_X \lesssim 10^{40}\, \mathrm{ergs}\,\mathrm{s}^{-1}$. These light curves usually indicate a CSM density decaying more quickly than the steady wind case. Furthermore, in some cases, long-term X-ray observations can even reveal that the SN ejecta moves in an evolving CSM, which hints towards the fact that the environment can be altered by the progenitor star. 
EP could possibly reveal some detailed and unknown transient features from a generally smooth X-ray light curve. This would help us probe the detailed structures of the CSM and, hopefully, the complicated late-phase evolution of the progenitors.


\subsubsection{Supernova Shock Breakouts}

The supernova shock breakout is the initial electromagnetic signal from core-collapse SNe. It is prominently observed in the soft X-ray and ultraviolet wavebands, lasting from seconds to less than an hour. This is followed by ultraviolet and optical emission from the expanding cooling envelope, which occur over a period of several days \cite{Waxman2017}.

The specific characteristics of the SBO emission depend on the type of the progenitors. 
On the one hand, the time scale of SBO radiation is short and closely related to the radius of the progenitor star $R_{*}$. On the other hand, the total energy of SBO radiation can be estimated as the thermal energy in the shock front at the time of the breakout, which can be expressed as  
$E_{\rm SBO} \sim 2 \times 10^{48}\,E_{51}^{0.56}\,R_{*,500}^{1.74}$\,ergs and 
$E_{\rm SBO} \sim 8 \times 10^{46}\,E_{51}^{0.58}\,R_{*,50}^{1.68}$\,ergs for red supergiants (RSGs) and BSGs, respectively \cite{Nakar2010,Waxman2017}. Here $E_{51}$ is the explosion energy in units of $10^{51}$\,ergs and $R_{*,500}$ (or $R_{*,50}$) the radius of the progenitor in units of $500\,R_{\odot}$ (or $50\,R_{\odot}$). Based on the above estimation, a classical analytic model of SBO emission \cite{Matzner1999} can determine the post-shock radiation temperature for a RSG, which is the progenitor of SNe IIP, to be $T_{\rm SBO} \sim 5 \times 10^{5}\,E_{51}^{0.2}\,R_{*,500}^{-0.54}$\,K. 
Meanwhile, we can also have $T_{\rm SBO} \sim 1 \times 10^{6}\,E_{51}^{0.18}\,R_{*,50}^{-0.48}$\,K for a BSG of a typical radius of $R_{*} \simeq 50 \,R_{\odot}$. Finally, for SNe Ib/c, their progenitors are Wolf-Rayet stars of small radii ($R_{*}\lesssim10\,R_{\odot}$) and the consequent peak luminosity and duration of the SBO emission is $L_X \sim 10^{44}$ ergs\,s$^{-1}$ and around 10 seconds. Therefore, SBO observations are of great significance for deep understanding of the explosion of core-collapse SNe (CCSNe), since the duration, temperature, and luminosity of SBO emission are closely related to the radius of the progenitor stars, which can be used to directly constrain the stellar types of the progenitors.

Detecting SBO emission is challenging, however, given the rather short duration of the flares. In the ultraviolet survey with the GALEX satellite, two nearby galaxies (redshifts $\sim 0.2-0.3$) exhibited rapid brightening, which was later linked to SNe II-P that were discovered by ground-based optical telescopes. The observed behaviour of the rapid brightening can be attributed to a classical SBO event emerging from a RSG, with a timescale of around 2000 seconds \cite{Schawinski2008}.  
A thermal component found in the spectral analysis of GRB\,060218 can be attributed to a quasi-relativistic SBO originating from the dense stellar wind of its progenitor star \cite{Campana2006}. The main issue with explaining GRB\,060218 as a SBO is to explain, in a reasonable way, the required dense wind of the Wolf-Rayet progenitor prior to its explosion, which remains unsolved in late-stage stellar evolution theory.  
Another well-known SN SBO candidate observed in X-ray is XRO\,080109 (SN\,2008D), which was serendipitously discovered by Swift/XRT \cite{Soderberg2008}. The total energy of the radiation detected in 0.3-10\,keV was approximately $2\times 10^{46}$\,ergs. The peak luminosity was about $6\times10^{43}$\,ergs\,s$^{-1}$, and it took about 60 seconds to reach the peak. The characteristic timescale for the decline in luminosity is approximately 130 seconds. The spectrum can be fitted with either a blackbody of $\sim 0.7$\,keV or a power-law model. The peak luminosity and blackbody temperature of this event are comparable to the theoretical predictions of SBO from Ib/Ic, but the duration is too long. Therefore, it is also necessary to introduce the assumption of a dense stellar wind. Recently, 12 possible candidates of SN SBO were reported by a search from the XMM-Newton archival data \cite{Alp2020}. 

WXT is advantageous for detecting SBO events of CCSNe. 
Prompt follow-up observations with FXT and other facilities at multi-wavelengths may provide good sampling of the light curves, which are useful for the in-depth study of some of the SBOs and their subsequent early adiabatic cooling period. 
This will provide comprehensive constraints on the explosion models, physical processes, and the evolution of the progenitor stars of CCSNe.
The estimation of the detection rate with EP is rather uncertain \cite{Sun2022}, however, due to the largely unknown SBO event rates. 

\subsubsection{Superluminous Supernovae}\label{sect:SLSNe}

Superluminous supernovae are a unique type of SNe that are intrinsically bright. They can be a few hundred times brighter than normal SN events \cite{GalYam2012,GalYam2019}. Although SLSNe have been actively studied for nearly two decades, there are still a number of questions unanswered concerning the explosion mechanisms and progenitor stars. 

The energy sources powering SLSNe remain unclear yet, as the conventional models invoking the radioactive decay of heavy elements are greatly challenged by their unusually high luminosities. 
The extremely high mass of $^{56}$Ni could be produced during some unusual SN explosions that are triggered due to electron–positron pair-production instability \cite{Barkat1967,Rakavy1967}.  The slow evolution of the SN 2007bi light curve in its later phase, which aligns with the decay of radioactive $^{56}$Co, may indicate a pair-instability supernova (PISN) signature \cite{Gal-Yam2009}.
However, while a few SLSNe may indeed be explained by PISNe, the majority require alternative powerful energy sources instead of radioactive decay.
Interaction between the ejecta and the CSM around an exploding star may efficiently convert the kinetic energy of the ejected material into radiation through powerful shocks. Such a model is often used to explain luminous and superluminous SNe IIn \cite{Chevalier2011,Ginzburg2012,Chatzopoulos2012}.

Alternatively, a spinning-down magnetar is an appealing explanation for the extremely high energy requirements of SLSNe \cite{Kasen2010,Woosley2010,YuYW2017}. 
In this scenario, the X-ray emission can, in principle, be expected from the cooling of relativistic particles that are accelerated in the magnetar's wind nebula (MWN) \cite{Metzger2014}. The X-ray ionization breakout is a sensitive probe of the properties of a hidden magnetar central engine in SLSNe-I.  X-ray photons may escape without strongly affecting the ionization state of the ejecta. Therefore, the high-energy counterparts of SLSNe may provide valuable insight into the future modeling of MWNs if they are indeed responsible for these central engines.

\begin{figure}[H]%
\centering
\includegraphics[width=0.4\textwidth]{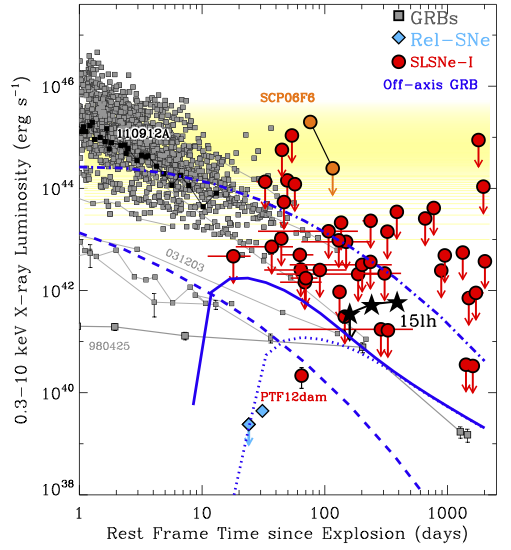}
\caption{X-ray emission from SLSNe-I (red circles), GRB X-ray afterglows (gray squares), relativistic SNe
(blue diamonds), and representative off-axis afterglow models (blue lines) from collimated outflows. Adapted from Ref. \cite{Margutti2018}.}
\label{fig:LC-Xay}
\end{figure}

Follow-up campaigns have been conducted on nearby SLSNe in the X-ray band using XMM-Newton, Chandra, and Swift. These campaigns spanned from a few days to approximately 2000 days after the alleged explosion epoch, giving upper limits on the X-ray luminosities as $L_X \lesssim 10^{41}$ ergs\,s$^{-1}$\cite{Levan2013,Bhirombhakdi2018,Margutti2018}. In most cases these observations placed stringent constraints on emission in the 0.3--10\,keV band, as shown in Figure \ref{fig:LC-Xay}. Only two SLSNe exhibited a signature of X-ray emission: SCP06F6 and PTF12dam. However, the faint X-ray source detected in PTF12dam is likely attributed to the star-forming activity in the host galaxy rather than the SLSN \cite{Margutti2018}.  

SCP06F6, a distant SLSN-I at $z=1.189$, holds the record of being the most luminous X-ray SN ever observed, with $L_{\rm X} \simeq 10^{45}$\,ergs\,s$^{-1}$ in the 0.2-2\,keV range, approximately 100 days after its initial discovery \cite{Levan2013}. Such highly luminous X-ray emission is rare among SLSNe-I.  As a possible explanation, the lack of significant X-rays in SLSNe may be caused by strong bound–free absorption of the X-rays by the SN ejecta \cite{Andreoni2022}.

More observations are needed to further characterize the X-ray emission from SLSNe, as well as to understand their evolution in a comprehensive way. 
EP can contribute to the research of this field by conducting sensitive observations for nearby SLSN-Is, providing possible new insight in distinguishing between plausible models of SLSNe. EP may also detect more X-ray flares similar to that of SCP06F6.

\subsubsection{Fast Blue Optical Transients}

Fast Blue Optical Transients (FBOTs) are a mysterious type of explosive events that were discovered through optical transient surveys in the past decade. FBOTs are characterized by their rapid rise to maximum brightness ($t_{\rm rise} < 10$ days), intense emission ($L_{\rm peak} > 10^{43}$ ergs s$^{-1}$), and blue colors  ($(g-r)< -0.2$ mag) \cite{Drout2014,Pursiainen2018,Ho2021-ZTF}. FBOTs pose challenges to traditional SN models, given their short timescales and high peak luminosities. 

In view of the similarity of FBOTs with SLSNe, models invoking shock interaction or a central engine have also been suggested to explain FBOTs. In the first class of models, the FBOT emission is commonly attributed to either the breakout emission of an outwards propagating shock caused by SN explosion \cite{Chevalier2011,Rest2018,Pellegrino2021} or a jet choked in a dense stellar envelope or stellar wind \cite{Gottlieb2022}. In the second class of models, the FBOT emission is generated by low-mass ejecta driven primarily by a central engine, i.e., a rapidly spinning magnetar \cite{Drout2014,YuYW2015,Pursiainen2018,LiuJF2022} or an accreting BH \cite{Kashiyama2015}. Specifically, low-mass ejecta plus a newborn magnetar system could originate from SN explosion of ultra-stripped progenitor stars \cite{Tauris2013,Tauris2015,Tauris2017,Suwa2015,Hotokezaka2017,DeK2018,Sawada2022}, including electron capture SNe \cite{Moriya2016,Mor2022}, accretion-induced collapses (AICs) of white dwarfs \cite{Kasliwal2010,Brooks2017,YuYW2015,YuYW2019a,Lyutikov2022}, and even mergers of binary NSs, binary WDs or NS-WD binaries\cite{YuYW2013,YuYW2015,YuYW2019b,Zenati2019}.

In the framework of these models, as discussed in Section \ref{sect:SLSNe}, some X-ray emission could also be expected to accompany the FBOT emission. For example, the X-ray emission produced by a MWN could finally escape from the low-mass ejecta \cite{YuYW2019a}. The magnetar wind-driven SBO can also provide a signature for the existence of a newborn magnetar \cite{LiSZ2016,WuSC2022}, which is more likely to be in the X-ray band than the normal SN SBO
emission.

\begin{figure}[H]%
\centering
\includegraphics[width=0.46\textwidth]{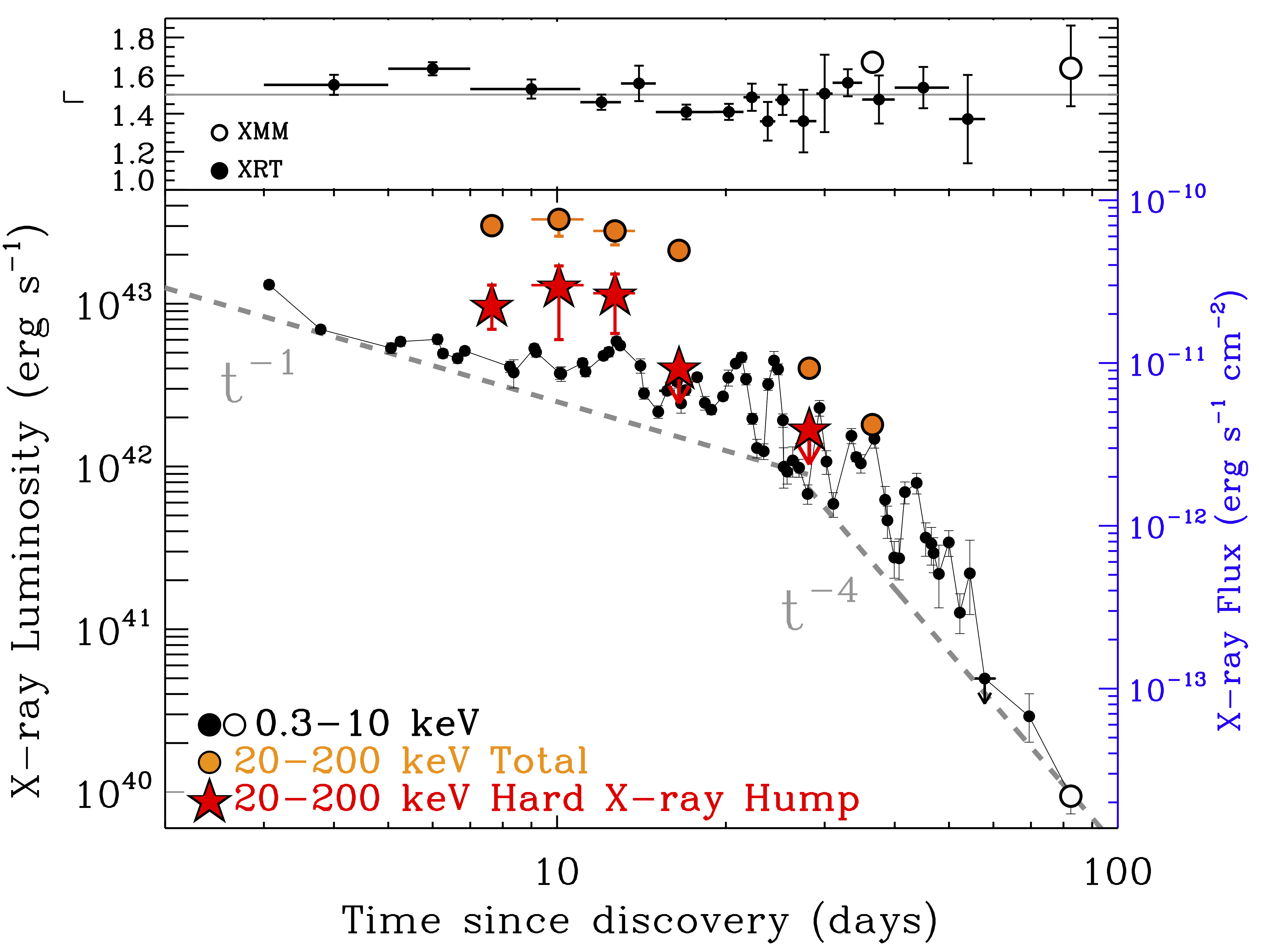}
\caption{Temporal evolution of AT 2018cow at soft (black, $0.3-10$ keV) and
hard (orange and red, $20-200$ keV) X-ray energies. Adapted from \cite{Margutti2019}.}
\label{fig:AT2018-Xay}
\end{figure}

AT2018cow stands out as the first FBOT with well-sequenced photometric and spectroscopic observations across a wide range of wavelengths, from $\gamma$-ray to radio \cite{Perley2019,Rivera2018,Margutti2019}. As shown in Figure \ref{fig:AT2018-Xay}, the X-ray luminosities of AT 2018cow are at the level of $L_X \sim 10^{43}$\,ergs\,s$^{-1}$, which are significantly higher than those seen in normal SNe and are similar to the values found in GRBs. There is a continuous occurrence of X-ray flares that have short periods of variability lasting only a few days. These flares are observed alongside a gradual decay, initially following an inverse proportion to time ($t^{-1}$). However, after approximately 25 days, the decay becomes steeper and follows an even faster decay ($t^{-4}$).  Interestingly, the X-ray emission from AT 2018cow provides suggestive evidence of one or more quasi-periodic oscillations (QPOs) \cite{Pasham2022,ZhangWJ2022b}. QPOs are commonly observed in accreting systems. If the high luminosity and the detection of QPO are indeed genuine, it may indicate that AT2018cow is powered by an accreting compact object.

Apart from AT2018cow, successful multi-wavelength observations of FBOTs are still rare. Only a few FBOTs were detected in X-rays \cite{Coppejans2020,Bright2022,YaoYH2022}, while the majority were not. This may be due to that the optimal time for X-ray observation had already passed when they were found from archival data.
X-ray observations of AT2020xnd revealed a luminosity consistent with that of AT2018cow at around 20-40 days \cite{Bright2022}. The X-ray luminosity of AT2020mrf, the highest among FBOTs \cite{YaoYH2022}, was detected by eROSITA approximately 35 days after the first optical detection. The X-ray emission at the 0.1-10 keV band had a luminosity of $L_X \sim 2 \times10^{43}$\,ergs\,s$^{-1}$. AT2020mrf stands out as the first multi-wavelength FBOT identified through X-ray survey, highlighting the potential of surveys in different wavelengths, such as EP in the soft X-ray band, in identifying these types of transients.

X-ray light curves can help determine the energy source of FBOTs, while broad-band X-ray spectroscopy can reveal how the material in the closest vicinity to the central engine evolves in its geometry \cite{Margutti2019}. The expected number of detections per year for events like AT 2018cow and AT 2022mrf by EP ranges from approximately 0.012 to 43, depending on the volumetric rate of FBOTs \cite{YaoYH2022}.

\section{Supermassive black holes}
\label{sec:smbh}


Supermassive black holes (SMBHs) with masses in the range of $M_{\rm BH} \sim 10^{5}-10^{10} M_\odot$ (those in the lower end of this mass range are sometimes referred to as "massive black holes" or MBHs in this paper) are thought to reside in the nuclei of most, if not all, massive galaxies. The tight correlations between the masses of SMBHs and the properties of their host galaxies \cite{Magorrian1998} suggest that SMBHs and galaxies have co-evolved throughout the cosmic history. This means that SMBHs have profound impacts on the formation and growth of their host galaxies by means of energetic feedback into their environments.  

SMBHs can accrete gaseous materials through disks. During the accretion process, they can produce luminous radiation and launch powerful outflows and even relativistic jets. About ten percent of SMBHs continuously accrete gas from their surroundings, manifesting themselves as AGNs. In the local universe, the nuclei of nearly $\sim 40$ percent of galaxies show certain levels of activities or low-luminosity AGN \cite{Ho1997}. SMBHs can also occasionally tidally disrupt stars and accrete stellar materials, causing them to temporarily brighten. Such events, termed TDEs, make it possible to probe the dormant SMBH population. Both AGNs and TDEs can produce luminous radiation outshining their host galaxies, and their emission spectra cover almost the entire electromagnetic spectrum from radio to $\gamma$-rays. Furthermore, winds and relativistic jets have both been observed from TDEs and AGNs. Last but not least, TDEs and AGNs are both multi-messenger emitters, which are possible major sources for high-energy cosmic rays and neutrinos.




One of the key science objectives of Einstein Probe is to search and study TDEs  \cite{Yuan.2016}.
WXT is expected to detect a sample of TDEs and flaring AGNs and trigger timely multi-wavelength follow-up observations.
For some of them, as well as for those discovered by other facilities, FXT can perform quick follow-up and long-term monitoring observations in X-rays (see Figure\,\ref{fig-transient-flowchart} for the follow-up strategy of EP transients). 
Such unique capabilities enable the exploration of the key evolution phases of various SMBH outbursts, such as the early phase of TDEs. With a carefully designed observational strategy, FXT can monitor the spectral and temporal evolution of a sample of known AGNs and TDEs at low flux levels for an extended period.
Complemented by follow-up and synergetic observations in multi-wavelengths, EP is expected to help address some of the key questions in the SMBH astrophysical research:

\begin{itemize}
    \item What are the demographics of the majority of SMBHs in quiescence?
    \item What are the occupation fractions of SMBH/MBHs at the centers of galaxies, especially at the lower-mass end?
    \item Are IMBHs truly rare and where are they located? How to find binary SMBHs? 
    \item How are accretion disks formed? How are outflows, including winds and relativistic jets, launched?
    \item How do the properties of an accretion flow change with the accretion state? 
    \item How are the X-ray coronae formed, maintained, and destroyed around black holes?
     \item How are high-energy particles accelerated near SMBHs?
\end{itemize}

\subsection{Tidal Disruption Events}
\label{sec:tde}
\subsubsection{Overview}

\noindent Stars in the central parsec of a galaxy are occasionally scattered to approach a central MBH. If the stellar orbit passes within the tidal disruption radius of the MBH, $R_T\approx R_\star(M_{\rm BH}/M_\star)^{1/3}$, the tidal force exerted by the MBH exceeds the self-gravity of the star, and the star will be torn apart \cite{Hills.1975, Rees.1988}. About half of the stellar debris remains bound to the MBH and falls back to its vicinity following a characteristic $t^{-5/3}$ power-law \cite{Phinney.1989, Evans.1989}, while the other half escapes on unbound orbits. 
The bound stellar debris compresses, collides, and eventually accretes onto the MBH, producing a very luminous flare. The flare is expected to rise quickly and then decay over several years, with its luminosity evolution following the debris mass fallback rate to some extent \cite{Komossa.2004}.

TDEs can be used as a powerful probe for studying quiescent MBHs, which are otherwise difficult to observe. The evolution of the light curve can be used to constrain the BH mass \cite{Guillochon.2013}, and the maximum BH mass allowing stars to be disrupted outside the BH event horizon depends on the BH spin \cite{Kesden.2012}. Therefore, TDEs can provide essential clues about the demographics of MBHs.
TDEs are also expected to provide alternative ways to detect IMBHs and binary SMBHs \cite{Liu-F.2009, Liu-F.2014, komossa.2015, Rossi.2021}, and to constrain the population and densities of stars in the innermost regions of galaxies \cite{Stone.2020}.

TDEs are ideal laboratories for studying BH disk formation, accretion and outflow physics \cite{komossa.2023}. Recent models show that TDE debris can assemble accretion disks through debris stream-stream collision caused by general relativistic apsidal precession or nozzle shocks produced by compression near the pericenter \cite{Bonnerot.2021}. The timescale of the disk formation is model-dependent, and currently different predictions were made by different models. Nonetheless, TDEs offer a great opportunity to study how gaseous materials form accretion disks around BHs within human-observable timescales. Once the accretion disk is formed, the gas accretion rate onto the MBH in typical TDEs can reach as high as two orders of magnitude above the Eddington accretion rate. This offers us an invaluable opportunity to study super-Eddington accretion and outflow physics around MBHs in the local universe \cite{Dai-L.2021}. The knowledge gained can be applied to understand how the most massive quasars at high redshifts have grown and provided feedback to their host galaxies.

Since the first detections of TDE in X-ray in the late 1990's \cite{bade.1996, komossa.1999, komossa.1999b}, about 100 candidates have been detected so far \cite{komossa.2015, saxton.2021, Gezari.2021}. The observed population exhibits a broad range of properties, including continuum and line emission \cite{komossa.2008, wang-t.2012, charalampopoulos.2022 }. Furthermore, it is found that in a few TDEs, relativistic jets have been launched from the transient accretion disks. We discuss a few characteristic types of TDEs or TDE-related phenomena and the prospects of studying them using EP below.

\subsubsection{Thermal X-ray TDEs}
Most of the TDE candidates discovered in the X-ray band so far belong to the class of thermal TDEs, characterised by dominance of thermal emission in their X-ray spectra near the maximum brightness. The thermal X-ray emission is thought to be produced from the innermost part of an optically thick accretion disk formed from the stellar debris. Its temperature is determined by the rate at which material is accreted and can often be well described by a black body with an effective temperature in the range of $kT\approx$ 50--100\,eV (or an equivalent steep power law with a slope of $\Gamma\approx$ 4--5)\cite{komossa.2002}. This notably differs from the typically harder AGN spectra and provides a powerful method of selecting TDE candidates by their soft X-ray spectra, as enabled by WXT's energy range. 

In some TDEs, a hard spectral component emerged several months to years after the onset of the event, potentially indicating the formation of a hot optically thin plasma, by which soft disk photons are inverse Compton scattered to produce the hard spectral tail \cite{komossa.1999, wevers.2019, yao.2022}. A prominent example is RXJ\,1242–1119, which showed a significant spectral hardening 
from $\Gamma$=5.1$\pm$0.9 at peak to $\Gamma$=2.5$\pm$0.2 nine years later \cite{Komossa.2004}. EP is able to search for these spectral transitions by monitoring TDEs with FXT during the whole mission lifetime.

\begin{figure*}[]
\centering
\begin{tabular}{cc}
\includegraphics[width=17cm]{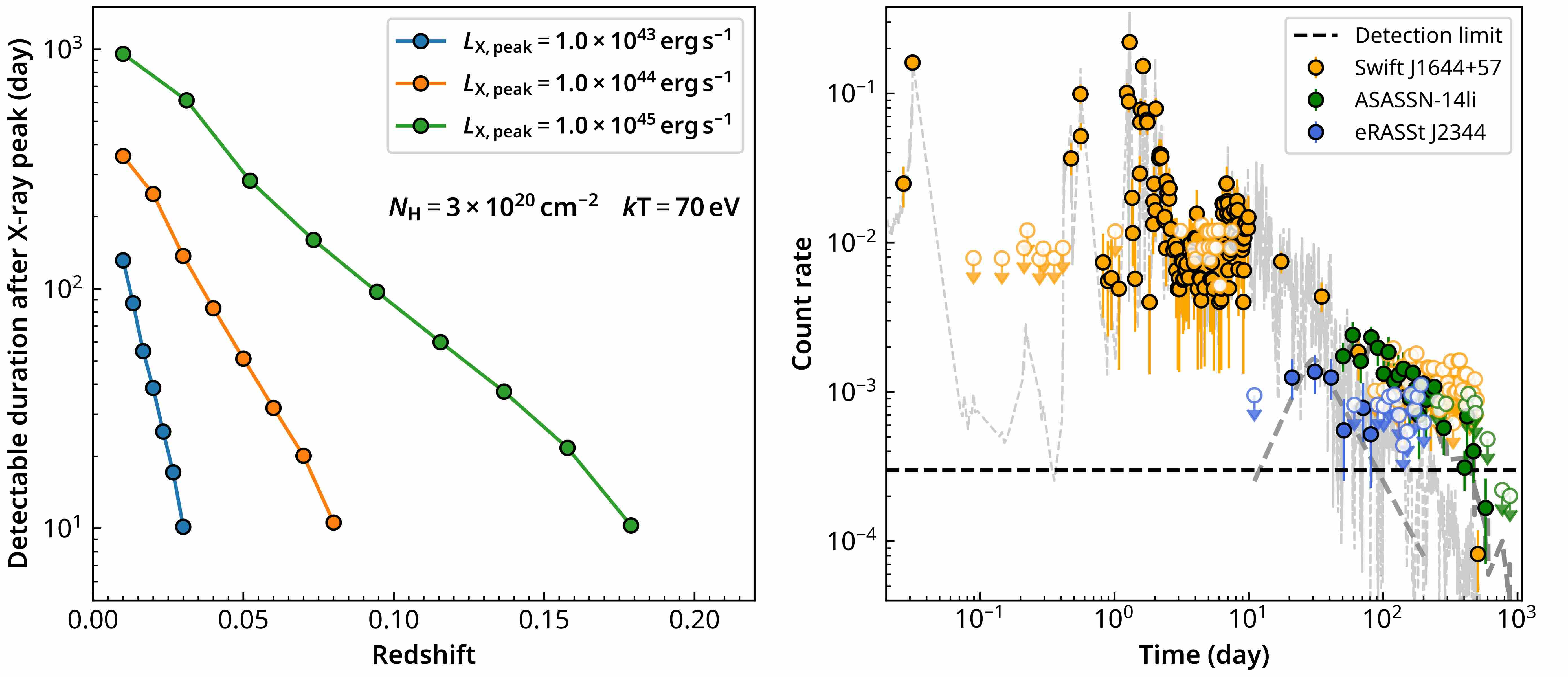}
\end{tabular}
\caption{{\it Left:} WXT detection window as function of maximum redshift for canonical thermal TDEs with peak luminosities of 10$^{43}$\,ergs\,s$^{-1}$, 10$^{44}$\,ergs\,s$^{-1}$, and 10$^{45}$\,ergs\,s$^{-1}$ (figure revised based on Figure 4 in \cite{2018SSPMA..48c9503L}). {\it Right:} Simulated WXT count rates using the observed data for the relativistic TDE Swift~J1644+57 \cite{Mangano.2016}, the nearby bright thermal TDE ASASSN-14li \cite{Bright.2018}, and the most luminous eROSITA detected TDE eRASSt~J2344 \cite{Homan.2023}. }
\label{fig:sec3_t_vs_z}
\end{figure*}

Thermal TDEs can reach peak luminosities of several 10$^{44}$ ergs s$^{-1}$ and remain at this level for a few weeks. 
Giving the sensitivity reached by stacking data taken over days to weeks, WXT is expected to find a sample of TDEs out to several hundred Mpc. 
Figure\,\ref{fig:sec3_t_vs_z} shows the period over which a typical thermal ($kT$=70\,eV) TDE can be detected by WXT out to a given redshift. For example, a TDE with $L_{\rm 0.5-4\,keV}$=10$^{44}$\,ergs\,s$^{-1}$ at $z$=0.05, comparable to RXJ\,1242-1119 mentioned above, is detectable by WXT for nearly two months. The monitoring observations with FXT can last for months to years.

Due to its lower energy threshold of 0.5\,keV, WXT is more sensitive to TDEs with slightly higher effective temperatures. This might favor events associated with less massive BHs, as the disk temperature is inversely related to BH mass. However, BH spin and additional spectral components can also lead to apparently harder spectra when being fit with simple spectral models.

\subsubsection{Relativistic TDEs}
Some TDEs have been found to launch powerful relativistic jets. If the jet is aligned with the line of sight from the observer, the emission in all wavelengths can be Doppler-boosted by several orders of magnitude, leading to a very bright (L$_{\rm X,iso}\sim$10$^{47-48}$ ergs s$^{-1}$) transient with significant short-time variability. In the last twenty years, four such relativistic TDEs have been discovered, three (Swift~J1644+57 \cite{Burrows.2011}; Swift~J2058+05 \cite{Cenko.2012}; Swift~J1112.2-8238 \cite{Brown.2015}) in the X-ray band and one (AT~2022cmc \cite{Pasham.2023}) in the optical band. 

Due to their 3--4 orders of magnitude higher peak brightness compared to thermal TDEs, relativistic TDEs can be discovered out to significantly higher distances. They thus provide a rare opportunity to study otherwise quiescent SMBHs at redshifts up to $z\sim$1 and beyond. Figure\,\ref{fig:sec3_t_vs_z} (right) shows the simulated light curve of Swift~J1644+57 at $z$=0.35 as if it would have been seen with the WXT. The event would have appeared as an exceptionally bright and variable X-ray source for several weeks, enabling even longer and more detailed FXT follow-up.  The observed X-ray spectra were found to be considerably harder than those of thermal TDEs, best fitted with power laws with slopes of $\Gamma$$\sim$1.3-1.5. 

The small number of relativistic TDEs discovered to date suggests that only a subset of TDEs produces outflows with sufficiently high velocities or that the outflows are very narrowly collimated. The finding of more jetted TDEs will help further constrain the jet production efficiency and understand the radiative mechanisms operating in super-Eddington jets.

\subsubsection{Intermediate-mass black hole TDEs}
\label{sec:IMBH-TDE}


The existence of intermediate-mass black holes ($M_{\rm BH} \sim 10^2$--$10^5 M_{\odot}$) \cite{greene20} is an essential prediction of the current scenarios of galaxy assembly and evolution. 
The mass function and occupation fraction of IMBHs, which are highly uncertain though \cite{Gallo.2019}, can provide robust constraints to the seeding mechanisms of the SMBHs in the early Universe \cite{Volonteri.2010, Greene.2012}.
They are also among the key science motivations for space-borne gravitational-wave detectors in the mHz band. IMBHs could either be fossil remnants from the early Universe or continue to form today in the dense core of massive star clusters. 
They are expected to be found also in the nuclei of dwarf and late-type spiral galaxies, 
or in the outskirts of more massive galaxies that have accreted and shredded a large population of satellite dwarfs. IMBHs in the latter scenario are more likely to manifest as X-ray transients, by accreting gas either from a donor star with a much lower mass (see Section\,\ref{sec:imbhs}), or as a consequence of full or partial TDEs. 

Indeed several IMBH-TDE candidates have been reported, e.g.\ \cite{Wevers.2017, Lin-D.2018, Angus.2022}, some of which produce strong X-rays. In fact, only IMBHs with a mass $\lesssim$10$^5 M_{\odot}$ can tidally disrupt a WD, producing a strong electromagnetic flare, instead of swallowing it entirely \cite{haas12, anninos18, Eracleous.2019, chen23}. The cores of massive globular clusters, likely the remnants of accreted satellites, are ideal environments for such events \cite{ye23}. IMBH-WD TDEs are a source of not only electromagnetic signals, but also gravitational waves (see Section\,\ref{sec:X-rayCounterpartsGWsources}). Such IMBHs are of particular interest among prospective EP targets. 


\subsubsection{TDEs in SMBH binaries}

SMBH binaries are a natural consequence of galaxy mergers \cite{Begelman.1980, Volonteri.2003}. When the binary separation is around milliparsec scale, the dynamics of the debris of a tidally disrupted star will be strongly affected, with the fallback pattern deviating from that in single-BH TDEs. Theoretical studies predict that gaps may appear in the observed light curves, and in certain cases the debris accretion onto the secondary MBH can be more significant than that onto the primary \cite{Liu-F.2009, Ricarte.2016, Coughlin.2017}. Interestingly, such signatures similar to the model predictions have been observed in a few TDEs \cite{Liu-F.2014, Shu-X.2020}. Furthermore, TDEs from SMBH binaries can also be used as precursors for the coalescence of the binary which produces low-frequency GWs. While most electromagnetic searches for binary SMBHs require that at least one, or both, SMBHs are AGNs, TDEs have the advantage that they trace the {\em {quiescent}} binary population.


\subsubsection{Rates and demographics of X-ray TDEs}

Statistical studies of TDEs can test models of TDE formation, probe nuclear stellar populations, and study BH demographics. The current best constraints on the rates of X-ray selected events have been obtained with ROSAT \cite{Donley.2002}, XMM-Newton \cite{Esquej.2008, Khabibulin.2014}, and SRG/eROSITA \cite{Sazonov.2021} of $\sim$1$\times$10$^{-5}$ to 2$\times$10$^{-4}$ TDEs per galaxy and per year, based on only a few tens of events. 
EP is expected to provide a statistically well-selected sample of new TDEs (cf. the previous, mostly heterogeneous samples), which will enable to further constrain the rates and thus to test and refine models that describe the conditions under which TDEs occur, including the effects of stellar dynamics, and binary interactions near the central SMBHs.

Currently the demographics of MBHs is mostly explored through observing the AGN population, such as measuring their masses through reverberation mapping and their spins through Fe K line profiles. 
As tracers of dormant MBHs, TDEs can allow us to detect quiescent MBHs with masses on the lower end ($M_{\rm BH}\lesssim10^8 M_\odot$) at the centers of galaxies.
A uniform sample of TDEs will provide an opportunity to study the demographics of BHs. 
TDEs offer an effective means for measuring the BH masses, from the timescale of the luminosity evolution \cite{Guillochon.2013} and fitting the X-ray spectra \cite{WenSixiang2020}, and can therefore be used to independently constrain the mass function of MBHs and occupation fraction. 
Moreover, the TDE rate in a galaxy is also linked to the BH mass as well as the stellar structure in its vicinity, enabling to probe the density, structure, and population of stars in the inner cusp of galaxies \cite{Stone.2016, Pfister.2021, Wong.2022}. 
Last but not least, TDEs may also offer a means to constrain the MBH spin, through quasi-periodic signals which likely originate near the BH event horizon \cite{Pasham.2019}, fitting the measured X-ray spectra \cite{WenSixiang2020, WenSixiang2022}, or in extreme cases when stars are disrupted by very massive BHs with $M_{\rm BH}>10^8 M_\odot$ \cite{Leloudas.2016}.

\subsection{Quasi-Periodic Eruptions}

Quasi-periodic eruptions are a new class of X-ray transients discovered at the centers of nearby low-mass galaxies, e.g.  with stellar masses $M_\star\approx(1-3)\times10^{9}\,M_\odot$ \cite{Arcodia.2021}. They are characterised by high-amplitude X-ray bursts with a quasi-periodic recurrence time of a few hours to one day and a duration of tens of minutes to a few hours. The first QPE was serendipitously discovered in the galaxy GSN 069 \cite{Miniutti.2019}. It shows repeating X-ray flares with a recurrence time of around 9 hours and a duration of about 1 hour. Another QPE was later identified in RE J1034 \cite{Giustini.2020} through a dedicated search in the archival XMM-Newton data. Two additional QPEs (eRO-QPE1 and eRO-QPE2) \cite{Arcodia.2021} were discovered through a blind search in the eROSITA all-sky survey data. In addition to these four securely classified QPEs, several QPE candidates have also been reported (XMMSL1 J024916.6-041244, \cite{Chakraborty.2021}; 2019vcb, \cite{Quintin.2023}; Swift J023017.0+283603, \cite{Evans.2023, Guolo.2023}). The X-ray spectra during the quiescence in between QPEs can be well described by a multi-color disk model with an effective temperature of $kT\sim40-70~$eV. The X-ray spectra become much hotter ($kT\sim100-200~$eV) when being brighter during the eruptions \cite{Miniutti.2019, Arcodia.2021, Giustini.2020}.

The physical origin of QPEs is yet unclear. Some models have been proposed, ranging from accretion disk instabilities \cite{Sniegowska.2023, Kaur.2023, Pan-X.2022} to extreme mass-ratio inspirals---involving a MBH and a much smaller orbiting object \cite{Xian-J.2021, Lu_quataert.2023, Franchini.2023}. Interestingly, a repeating partial TDE is proposed to interpret the decades-long X-ray light curve of GSN 069, characterised by two long-lived repeating X-ray flares ~9 years apart \cite{Miniutti.2023}. This study highlights the potential connection between QPEs and TDEs. While the typical flare luminosities of QPEs will enable the detection of only the most nearby events with EP-WXT, the potential connection with TDEs provides a promising avenue for EP studies. Particularly, systematic late-time follow-up of TDEs with FXT may offer systematic assessment of the frequency with which QPEs occur in the wake of TDEs, which is crucial to shed new light on this phenomenon.

\subsection{Active Galactic Nuclei}
An active galactic nuclei is the bright center of a galaxy that is believed to be powered by a SMBH through the accretion process. There are currently hundreds of thousands of observed AGNs. 
The past several decades have seen enormous progresses made in our understanding of the AGN phenomena and their underlying physical processes.
However, there are still some questions, among many others, remain unanswered regarding the accretion process of SMBHs, such as the structure and instability of the accretion disk, the formation mechanisms and properties of the coronae and jets, and the connections among them.
%

In observations, AGNs show strong X-ray radiations with characteristic variations on timescales ranging from hundreds of seconds to years.
The X-rays originate from the vicinity of the BHs, providing crucial information directly related to the accretion disk, corona, jet, and the BH itself. 
Some of the information are encoded in the X-ray variability, which provides a powerful tool to probe the physical processes of accretion and jets close to the BHs \cite{Mushotzky1993, McHardy1985}. 
Previous studies along this line have mostly been focused on the short-term variability of AGNs using narrow-FoV, high-throughput telescopes, such as XMM-Newton and Chandra, leading to a great leap in the understanding of BH accretion in AGN. 
As a complementary approach, the study of the long-term evolution of AGN X-rays is crucial for understanding, e.g.  the accretion state transition, disk-corona structure and evolution. 
As of today, RXTE, Swift, and MAXI have monitored the long-term variations of dozens of AGNs (e.g. \cite{Rivers.2011, Rivers.2013, Matsuoka2009}); however, the sample size and/or sampling cadence are still limited.

\begin{figure}[H]%
\centering
\begin{tabular}{c}
\includegraphics[width=0.49\textwidth]{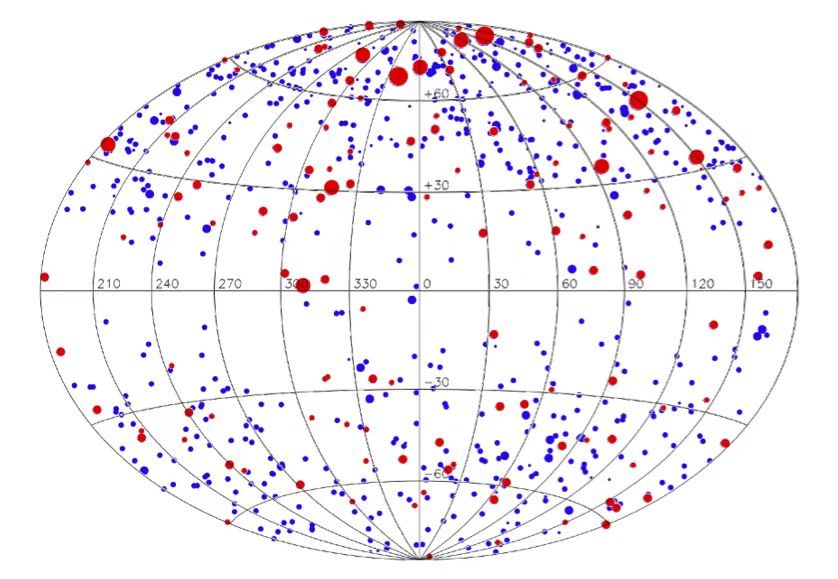}  \\
\end{tabular}
\caption{Distribution of about 800 known nearby bright AGNs (in the Galactic coordinates) that are likely to be observable by WXT with 5$\sigma$ detection per day (red, $\sim$140 AGNs) and per week (blue, $\sim$700 AGNs). The size of each symbol indicates the relative brightness as observed in the ROSAT all-sky survey. Credit: Jingwei Hu.}
\label{fig-epagn}
\end{figure}

The WXT's wide-FoV, sensitive sky surveys enable long-term  monitoring of hundreds of nearby X-ray bright AGNs (Figure\,\ref{fig-epagn}). It can also discover bright flares of AGNs in the more distant universe. Moreover, AGNs of particular interest can also be monitored on a long-term basis with FXT. 
Therefore, EP offers a great opportunity for studying the long-term X-ray variability of AGNs.
Here we discuss the extreme variability in AGNs, including changling-look AGN and flares in radio-quiet AGNs and blzars, and the long-term variability, including studies of AGN power spectral densities and state transitions.  

\subsubsection{Changing-look AGNs}
\label{sec-clagn}

Changing-look (CL) AGNs are a rare AGN phenomenon discovered over the past decades (e.g.  \cite{LaMassa.2015, MacLeod.2016, Graham.2017, Yang-Q.2018}). Their most distinctive feature is the dramatic variations in X-rays and optical/UV, including transitions from high-flux to low-flux state (turn-off) and vice versa (turn-on). These flux variations are also accompanied by significant changes in their spectra. The term ``changing-look'' was initially used to describe transitions between Compton thick and thin X-ray spectral states (e.g.  \cite{Guainazzi.2002, Matt.2003}), but later extended to describe significant changes in the optical/UV continuum and emission lines (e.g.  transitions between Types 1 and 2, see, e.g.  \cite{LaMassa.2015, MacLeod.2019, Trakhtenbrot.2019, komossa.2024}). Another important feature of CL AGNs is that these transitions typically occur on relatively short timescales, ranging from months to years. These timescales are significantly shorter than those predicted by the classical SMBH accretion disk theory, posing a significant challenge in explaining their underlying physical mechanisms.

Physically, CL AGNs can be broadly categorized into two types. One is caused by variations in the line-of-sight obscurer, known as ``changing-obscuration''. For example, clouds or outflowing gas can lead to significant changes in the line-of-sight column density $N_{\rm H}$, resulting in most CL events to be seen in X-rays and some in optical/UV (e.g.  \cite{Elitzur.2012, Gardner.2017}). The other type involves variations in the accretion flow itself, namely ``changing-state'', which is characterized by changes in the optical/UV continuum, emission lines, and X-ray spectra (e.g.  \cite{Noda.2018, Graham.2020}). For instance, the CL event of Mrk 1018 may be due to the instability of the accretion disk \cite{LaMassa.2017, Kim.2018}, while the CL event of 1ES 1927+654 may be caused by perturbations to the pre-existing AGN disk by a TDE \cite{Trakhtenbrot.2019, Ricci.2021}. To distinguish between these mechanisms, timely detection of CL events and the prompt triggering of high-precision follow-up observations are essential. Furthermore, the detection of a large CL AGN sample can help with categorizing and analyzing different CL phenomena.

\begin{figure}[H]%
\centering
\begin{tabular}{c}
\includegraphics[width=0.46\textwidth]{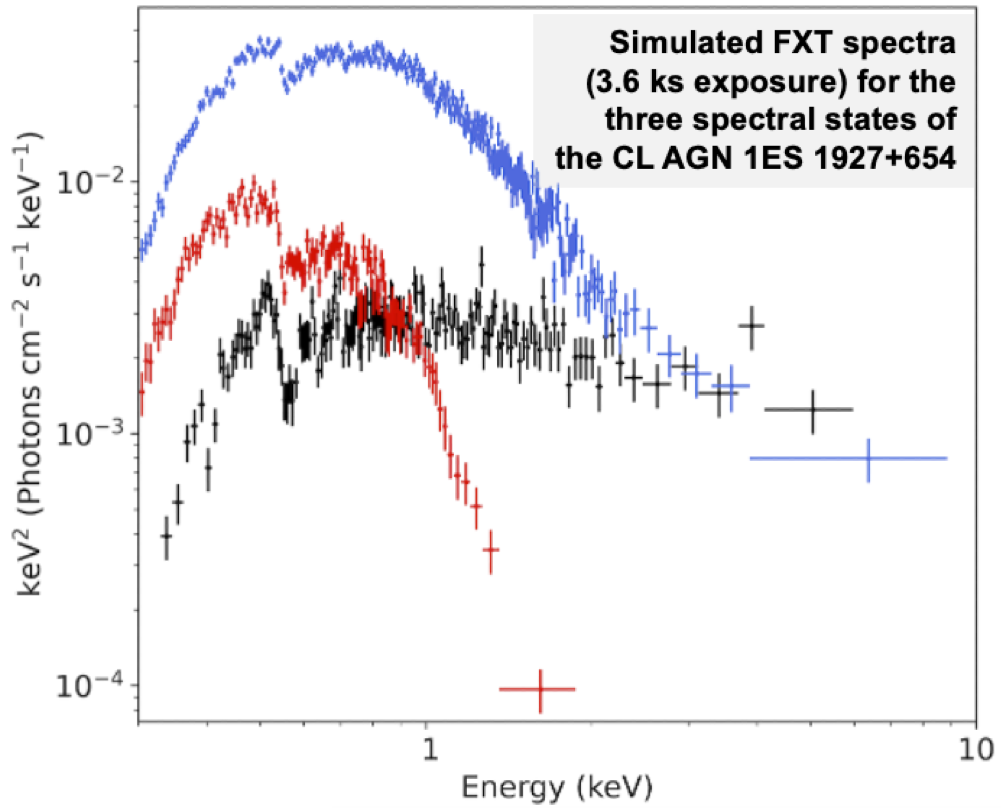}  \\
\end{tabular}
\caption{Simulated FXT spectra of 3.6 ks exposure time each (using thin filter) for the three distinct spectral states of the changing-look AGN 1ES 1927+654 observed in May 2011 (black), June 2018 (red) and December 2018 (blue), respectively. }
\label{fig-1es-simu}
\end{figure}

In recent years, an increasing number of CL AGNs have been discovered by various time-domain all-sky survey facilities. EP opens a new all-sky monitoring window in the soft X-ray band of 0.5--4 keV, likely leading to the discovery of a large CL AGN sample in X-rays. It can also perform long-term spectral monitoring with FXT. Figure\,\ref{fig-1es-simu} displays simulated X-ray spectra of the three typical spectral states of 1ES\,1927+654 \cite{Ricci.2021} as observed with FXT, showing the monitoring capability of FXT for this source. EP can also conduct long-term monitoring, possibly by synergies with other multi-wavelength facilities, of known CL AGNs to constrain the timescales of turn-on and turn-off events occurring alternately, thereby improving our understanding of their underlying physical mechanisms. 
In summary, EP holds promise for advance the studies of CL AGNs, such as the event rates, diversity, long-term evolution, and physical mechanisms underlying the CL AGN phenomenon.

\subsubsection{Flares from AGNs/blazars}
AGN flares have gained increasing attention in recent years as a distinctive phenomenon (e.g.  \cite{Lawrence.2016, Graham.2017}). These flares last for typically from days to years, and exhibit behaviors in multiple wavelengths that are not entirely consistent with typical TDEs. 
For example, AT\,2019cuk, a well-known CL AGN, exhibits day-long hard X-ray flares, suggesting a possible connection with coronal variations \cite{Masterson.2023}. IC\,3599, a Seyfert 1.9 galaxy \cite{komossa.1999}, showed soft X-ray flares different from those of typical TDEs, indicating a potential association with the instability of the inner disk region \cite{grupe.2015}. IC\,3599 is an excellent example of an extreme changing-look AGN, since the strong X-ray variability was accompanied by dramatic optical emission line variability as well. Similarly, the well-studied narrow-line Seyfert, Mrk\,335, has been observed to exhibit X-ray flares \cite{Wilkins.2015}, which are believed to result from dynamic changes in the X-ray corona at the base of the jet, and its variations are also caused by strong changes in a partially covering absorber \cite{komossa.2020}.

Blazars constitute a subset of radio-loud AGNs with kpc-sized jets closely aligned with our line of sight. 
Their broadband SEDs are characterized by a double-peaked structure. 
The low-energy peak mainly arises from synchrotron radiation of relativistic non-thermal electrons. If the jet is purely leptonic, the high-energy peak (above $\sim$100 keV), is thought to result from inverse Compton scattering of synchrotron or ambient photons (see, e.g.  \cite{Kirk.1998, Sikora.1994}) while in hadronic scenarios, it may be due to synchrotron emission from protons or from secondary decay products of charged pions (see, e.g.  \cite{Romero.2017}).
The ultra-relativistic conditions of the plasma propagate into the jets and the small viewing angles boost their broadband luminosities up to 
$10^{49}$\,ergs\,s$^{-1}$, making them the most extreme persistent objects over the full electromagnetic spectrum.
Blazars are often observed to show outbursts, typically 10 times or occasionally even 100 times brighter than in quiescence, which arise from dynamic variations in their relativistic jets.
Recent multi-wavelength and multi-messenger observations of blazars did put new constraints on the physical processes at play in these sources. In particular, it recently appeared that multi-component models better succeed in describing the blazar SEDs and short timescale variability than single-zone models (see, e.g.  \cite{MAGIC.2019,MAGIC.2020,Abeysekara.2020}).



WXT will catch AGN/blazar flares and ensure their monitoring over long timescales, enabling the studies of flare timescales, occurrence rate, diversity, and spectral evolution. 
This will also allow for a quick release of public alerts to trigger (quasi-) simultaneous multi-wavelength follow-up campaigns with other facilities. Flux changes at other wavelengths do not necessarily arise simultaneously, therefore time lags between different energy bands can be observed. 
Furthermore, FXT can perform high-cadence monitoring of new flares in their early phases, including flares detected by WXT and by other multi-wavelength facilities, enabling in-depth study of their spectral evolution.

\subsubsection{Power Spectral Density}


X-ray binaries (XRBs) show power spectral densities (PSDs) that are limited to certain frequency ranges, i.e., band-limited, with distinct breaks at high and low frequencies (e.g.  \cite{Belloni.1990, Belloni.2002}). Additional components can also emerge in the PSDs, such as various QPO signals (e.g.  \cite{Stella.1998, Remillard.2006, Ingram.2009}). The typical frequencies of these breaks and QPOs are linked to the BH masses and mass accretion rates. For example, the PSD shape of XRBs changes considerably with the accretion rate, reflecting changes in the accretion disk structure, such as the truncation radius of the inner disk (see \cite{Done.2007} for a review).

\begin{figure}[H]%
\centering
\begin{tabular}{c}
\includegraphics[width=0.46\textwidth]{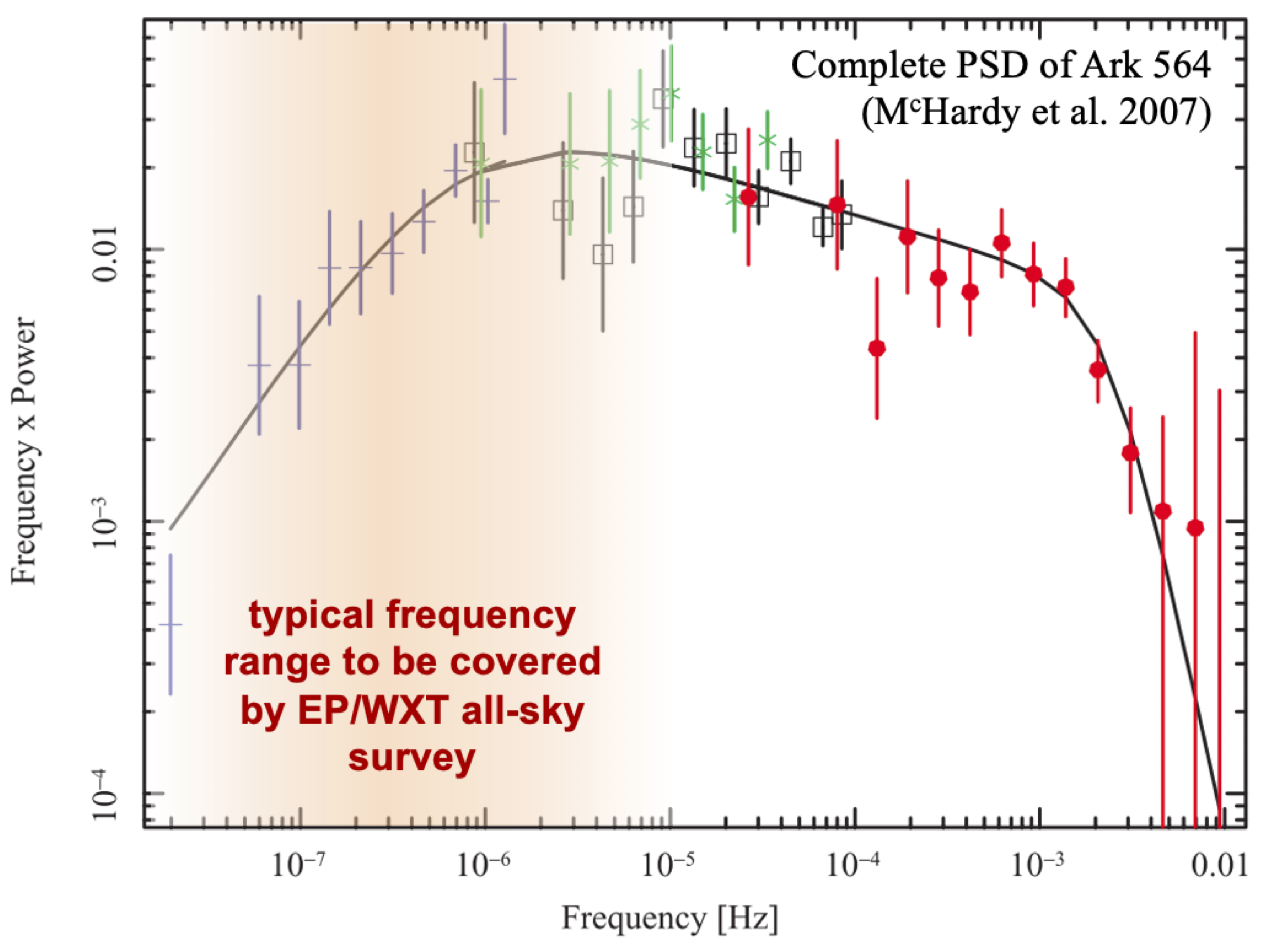}\\
\end{tabular}
\caption{The broadband 2-10\,keV PSD of Ark 564 (figure adopted from \cite{McHardy.2007}). The low-frequency range is revealed by RXTE's long-term monitoring, while the high-frequency range is revealed by XMM-Newton. The shaded range indicates the typical frequency range to be covered by the daily monitoring by the WXT's routine all-sky survey observations in 0.5-4\,keV.}
\label{fig-ark564-psd}
\end{figure}

AGNs are the scaled-up counterparts of XRBs and are expected to show similar behaviors in their X-ray PSDs, including the band-limited shape (e.g.  \cite{McHardy.2006}) and QPOs (e.g.  \cite{Gierlinski.2008, Jin.2020, Jin.2021}). However, studying the complete PSDs of AGNs requires long-term high-cadence monitoring to understand their variability properties below $10^{-5}$ Hz. This is challenging to achieve with traditional small field-of-view instruments due to the time-consuming and expensive nature of such monitoring. RXTE and Swift have monitored several dozen AGNs for multiple years, but with limited cadence (e.g.  \cite{Rivers.2011, Rivers.2013}). For example, RXTE's long-term monitoring successfully detected the low-frequency break of the AGN Ark 564 (Figure\,\ref{fig-ark564-psd}, \cite{McHardy.2007}), which is a rare occurrence. To systematically understand the similarities and differences between the AGN and XRB PSDs, it is necessary to monitor a larger sample of AGNs for multiple years with a cadence of days to weeks, combined with short-term variability, to obtain their complete PSD shapes.

\begin{figure}[H]%
\centering
\begin{tabular}{c}
\includegraphics[width=0.46\textwidth]{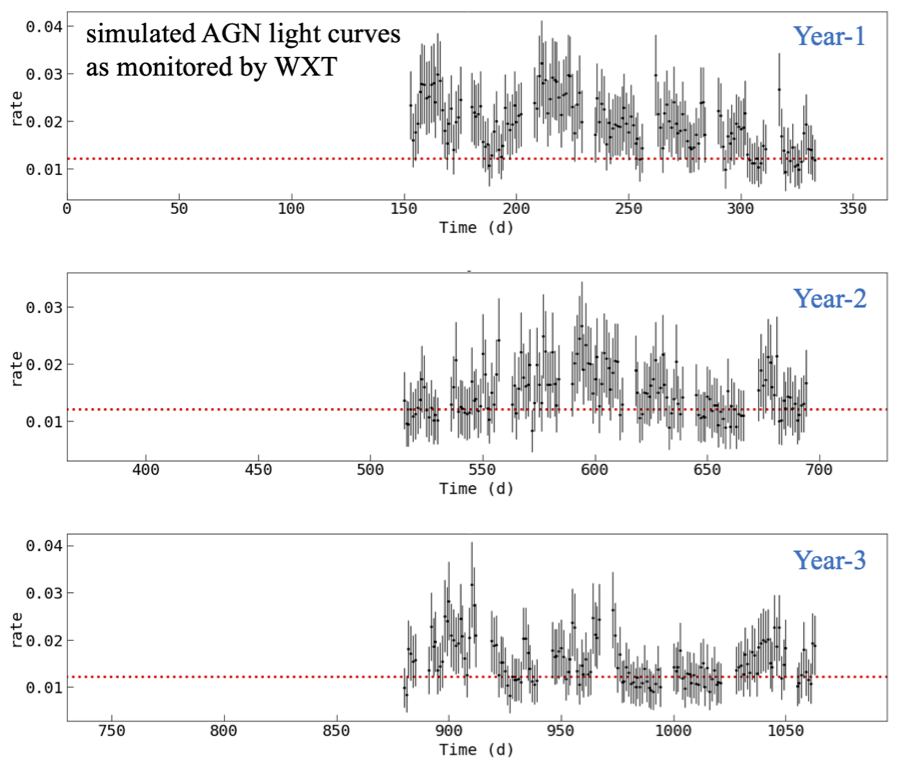}\\
\end{tabular}
\caption{Simulated AGN light curves as to be observed by the WXT, assuming five exposures of 1200 ks each per day for three years. The binning is one data point per day. The red line indicates the background level. Gaps in the light curves are due to pointing constraints, i.e., the source is not visible with EP.}
\label{fig-agnlc}
\end{figure}

WXT is detecting some $\sim 100$ bright nearby AGNs by stacking the daily data taken in the sky survey at a significance of at least 5\,$\sigma$. The sample size will increase to about several hundreds for weekly accumulative detections (Figure\,\ref{fig-epagn}). This allows for long-term monitoring of these AGNs over different timescales ranging from days to years (Figure\,\ref{fig-agnlc}), and will help constrain especially the low-frequency ranges of their PSDs and potentially detect the low-frequency breaks. Some of these AGNs (e.g.  3C 273, Mrk 335, NGC 4051) have already been well observed by satellites such as XMM-Newton, so their high-frequency PSD shapes are already known (e.g.  \cite{Gonzalez-martin.2012, Yang-H.2022}). By combining existing data, it is possible to extend their PSD frequencies to a much lower range. However, we note that the systematic uncertainty of WXT's flux measurement is about 10\%, so the intrinsic variability amplitudes of the monitored AGNs need to be larger than this level.

Through long-term monitoring of a large sample of AGNs, it is then possible to investigate potential systematic variances in PSD properties (e.g.  break frequencies, slopes, QPOs) across varying BH masses and accretion rates. This will enable us to conduct a comprehensive comparison with the PSD evolution in XRBs, thereby facilitating in-depth analyses of the similarities and differences between these two types of BH accretion systems.

\subsubsection{AGN State Transitions}
XRBs show a variety of spectral states that are thought to be caused by changes in the accretion rate. In AGNs, the search for corresponding state transitions is an important and long-standing topic (e.g.  \cite{Graham.2020, Yang-H.2022}). However, the variability timescales of SMBH accretion disks are expected to be $\gtrsim 10^5$ times longer than those of XRBs. Therefore, the detection of state transitions in individual AGNs is extremely difficult, if it occurs at all.

Previous studies have compared different AGN states to different XRB states. For example, the PSD of NGC 4051 is similar to that of Cyg X-1 in its high soft state \cite{McHardy.2004}, while that of Ark 564 is similar to that of Cyg X-1 in its very high state \cite{McHardy.2007}. One speculation is that AGNs may have states that correspond to all XRB accretion states. However, there is a key difference between AGNs and XRBs: AGN accretion states cannot be simply distinguished with X-ray spectra. This is because many factors can affect the X-ray spectra of AGNs, such as absorption by warm gas and torus, obscuration by a clumpy disk wind, and Compton scattering by diffuse gas. These factors can lead to significant variations in the X-ray spectral timing properties even when the accretion rate in the outer disk does not change (see, e.g.  1H 0707-495, RX J0134.2-4258 \cite{Jin.2023}).

\begin{figure}[H]%
\centering
\begin{tabular}{c}
\includegraphics[width=0.46\textwidth]{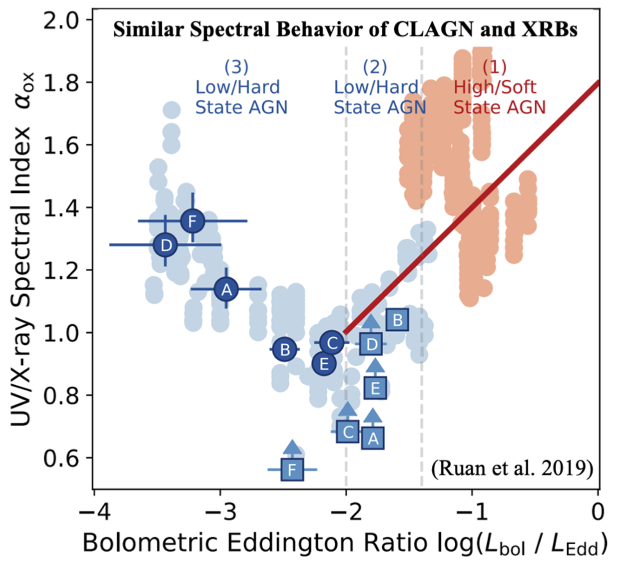}  \\
\end{tabular}
\caption{CL AGNs and XRBs can have similar spectral behavior for a wide range of mass accretion rate (figure adopted from \cite{Ruan.2019}). 
EP is expected to find more changing-look events of AGN to verify this spectral similarity in the comparison study of AGNs and XRBs.
}
\label{fig-state-trans}
\end{figure}

CL AGNs (see Section\,\ref{sec-clagn}) provide a new opportunity for this research topic. Although the physical mechanism of their changing-look phenomena is still unclear, the timescale of changing-look is relatively short, thereby allowing state transitions to be observed with regular monitoring. Interestingly, it has been found that the relationship between the spectral energy distribution (SED) and mass accretion rate of CL AGNs is similar to that of XRBs (see Figure\,\ref{fig-state-trans}, \cite{Ruan.2019}). 
%
As mentioned before, EP will monitor the long-term variability of hundreds of AGNs with high cadence, and will thus constrain the low-frequency end of the PSDs of these AGNs. It is then possible to identify which XRB accretion states these AGNs correspond to. Combining the PSD with more new CL AGNs discovered, EP may further verify and refine the relationship shown in Figure\,\ref{fig-state-trans}, and advance the study of state transition in SMBH accretion.

\subsection{Summary}

The large FoV and good sensitivity of EP-WXT enable early detection of new X-ray outbursts and special variability patterns of various SMBH systems, and allow for their long-term monitoring. This includes TDEs and variability of AGNs. Furthermore, SMBH systems typically exhibit multi-wavelength emission from radio to $\gamma$-rays, and potentially produce multi-messenger signals such as ultra high-energy cosmic rays, neutrinos, gravitational waves. Currently, optical survey facilities like ZTF and WFST are in operation, along with multi-messenger facilities such as LHAASO and IceCube, with future prospects including Rubin Observatory and Square Kilometer Array (SKA). EP can perform synergistic surveys with these facilities, conducting multi-wavelength, multi-messenger observations to better characterize various SMBH systems and their complex internal structures (e.g.  accretion disks, coronae, jets) and physical processes.

With the flexible scheduling of EP observations, FXT can rapidly follow up with high-precision observations and long-term monitoring of new transients discovered by WXT and other facilities. Additionally, FXT can synergize with other facilities to conduct deep multi-wavelength and long-term monitoring of specific SMBH systems. Given its high sensitivity as a follow-up telescope, it can also be used to monitor lower emission states of known sources.

To conclude, 
EP is expected to advance our understanding of SMBHs. Furthermore, AGNs and TDEs, as being among the highest-energy emitting sources in the Universe, also play a crucial role in multi-messenger astronomy; this is discussed in Section\,\ref{sec:mma}.


\section{Multi-messenger astronomy}
\label{sec:mma}


\subsection{Science objectives} 

Astronomy has entered the multi-messenger era. Besides the traditional electromagnetic channel, some astronomical objects with extreme physical conditions can be now observed with new messengers in GWs and neutrinos. 
Multi-messenger observations by combining two or more of these observational channels will reveal unprecedented clues to study the nature of these objects. EP is well-positioned to conduct X-ray observations of some of these multi-messenger targets. 

\subsubsection{X-ray counterparts to GW sources} 
\label{sec:X-rayCounterpartsGWsources}

Since the discovery of the first GW event GW150914 \cite{GW150914}, the LIGO/Virgo/KAGRA (LVK) collaboration has detected 90 GW events due to compact binary coalescences (CBCs) as of the 3rd operation run (O3) \cite{LIGOScientific:2021usb, LIGOScientific:2021djp}. 
More GW events are being detected with the ongoing O4 observations. The majority of GW events are binary black hole (BBH) mergers, which so far have not shown credible electromagnetic counterparts. Bright EM counterparts are expected from BNS mergers and some NS-BH mergers in which the NS is tidally disrupted outside the BH horizon. The former has been observationally confirmed by the observation of the GW170817/GRB 170817A event as well as its broad-band afterglow and kilonova detection \cite{GW170817,GW170817/GRB170817A}. Searches for EM counterparts for another BNS merger event, GW190425, and a few NS-BH merger events have not led to bright EM counterparts \cite{GW190425,GW190814}. 

\begin{figure*}
\centering
\includegraphics[width=18cm]{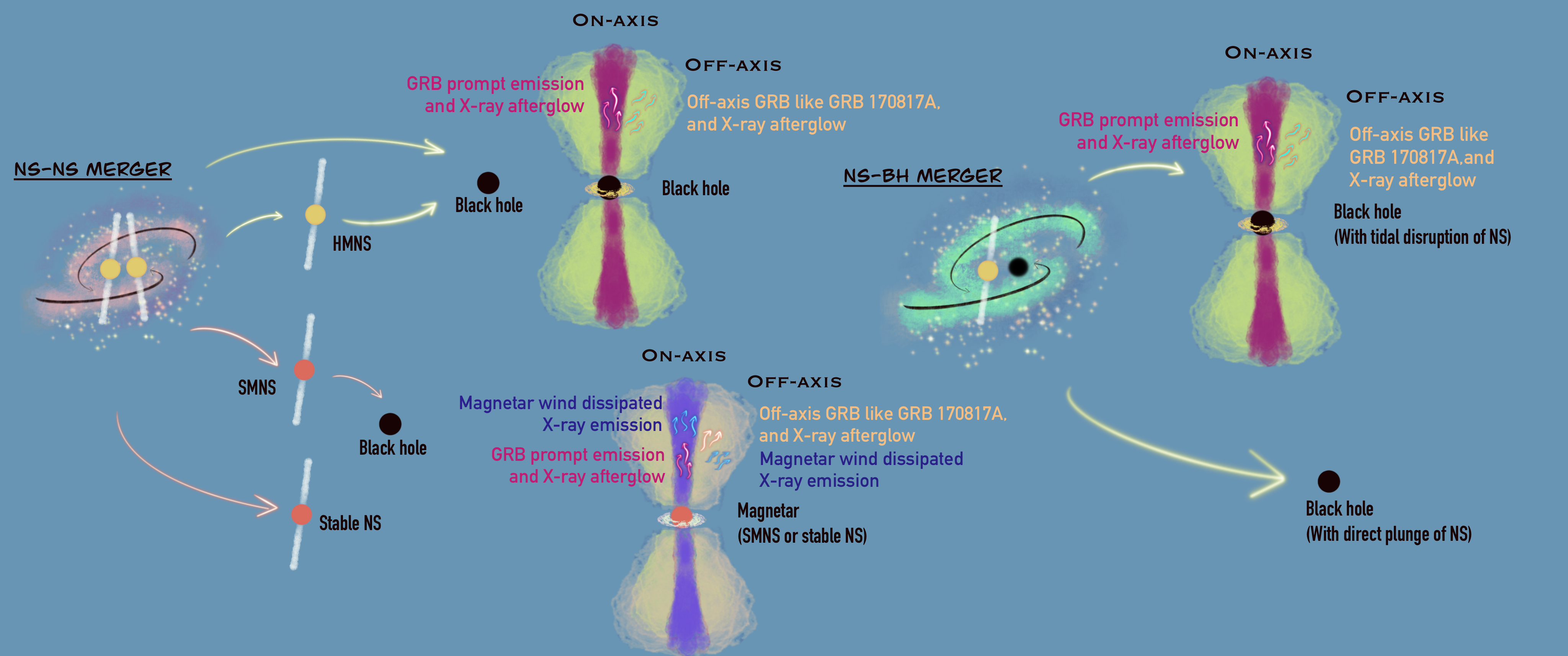}
\caption{A cartoon picture of various CBCs with different merger products and the geometry of various post-merger remnants. The EM counterparts as a function of viewing angle are also indicated. Credit: Hui Sun.}
\label{fig:sec4_figure1}
\end{figure*}

Bright X-ray emission is indeed expected from some CBCs, especially BNS mergers and some NS-BH mergers. This has been demonstrated by the association of GW170817 with a short GRB seen off-axis, with the detection of a hard X-ray/$\gamma$-ray prompt pulse followed by an X-ray afterglow \cite{Troja+17}. The properties of the signal depend on the merger remnant and viewing angle, which we summarize below (see Figure \ref{fig:sec4_figure1}). 

\begin{itemize}
\item {\bf On-axis BNS mergers: } These events are expected to be associated with bright short GRBs and their afterglows. The X-ray emission includes a prompt emission component and an afterglow component. For the former, the emission could be the spectral extension of $\gamma$-ray emission to the EP band, in the case of a BH engine, or a separate emission component independent of the prompt emission, in the case of a long-lasting central engine (likely a magnetar), as revealed in GRB 230307A detected by LEIA \cite{SunHui23}. For the afterglow emission (see Figure \ref{fig:sec4_figure2}), the X-ray emission could be a combination of internal dissipation emission, e.g.  the internal X-ray plateau emission \cite{Troja+07,YuYW2010,rowlinson13,lvhoujun15} and X-ray flares, and the standard external shock emission if the ambient density is high enough. This external shock afterglow component may give rise to a jet break at a later time \cite{Troja+16,Escorial2022}. At even later times, the light curve may show a flattening feature as the jet enters the Newtonian phase \cite{Ryan2024} and/or when the kilonova afterglow component emerges \cite{Troja2020, Hajela2022}, even though such late features are likely undetectable with EP unless the source is extremely nearby. The X-ray plateau emission could be the same component that appears during the prompt emission phase, as indicated in the case of GRB 230307A \cite{SunHui23}, but could be in principle a different component. 

\item {\bf Off-axis NS-NS mergers: } These sources are characterized by off-axis emission, as exemplified by the GW170817/GRB 170817A observations. The prompt emission would be characterized by a low-luminosity SGRB, as in the case of GRB 170817A, or, if the source is far away enough, characterized by an SGRB-less event because the predicted emission would fall below the sensitivity of the GRB detectors. The horizon of off-axis SGRB detection depends on the viewing angle, instrument sensitivity and energy range. For example, GRB 170817A would have been undetectable by the Fermi/GBM if it was further away than 65 Mpc \cite{zhangbb18}, but could be detected to up to 100 Mpc with Swift/BAT. 
As a reference, the peak flux of the prompt emission of GRB 170817A, extrapolated in the WXT energy band, is $1.73^{+2.97}_{-0.49}\times10^{-9}$\,ergs\,s$^{-1}$\,cm$^{-2}$.
At this flux level, it could only be marginally detected with the WXT at the luminosity distance of this burst, $\sim$40 Mpc (due largely to the short duration of the prompt emission), but may be detected at a further distance if it has a magnetar engine (see Section 3.2.4 for detailed simulations). 

Regardless of the central engine, a generic feature of these events is that the X-ray afterglow emission from an off-axis jet will gradually enter the line-of-sight, with the X-ray afterglow light curve steadily rising as a function of time \cite{ryan+20} (see Figure \ref{fig:sec4_figure3}). Following the peak, the afterglow will essentially behave as an on-axis GRB afterglow. The temporal evolution of the X-ray emission produced by the relativistic jet, combined with spectral information, brings key information about the energy of the burst, the jet geometry, the circumburst medium and the micro-physical parameters of the shocks. In addition, in the first seconds after the merger, different X-ray behaviours would arise depending on the merger product. If the merger product is a prompt BH or a BH shortly formed after a brief hypermassive neutron star (HMNS) phase, bright X-ray emission is not expected in the early phase. However, if the merger product is a long-lived NS, either supermassive neutron star (SMNS) or a stable NS, a bright plateau X-ray emission, possibly accompanied by X-ray flares, is expected \cite{zhangbing13} if the line of sight is not too far from the jet axis (e.g.  in the so-called ``free zone'' \cite{sunhui17}). Such signals, if detected, would be the smoking gun to probe a long-lived NS engine, and hence, give strong constraints on the NS equation-of-state (EoS) \cite{Gaohe16,LiAng16,Piro+19,AiShunke20,AiShunke23}.

\begin{figure*}
\centering
\includegraphics[width=16cm]{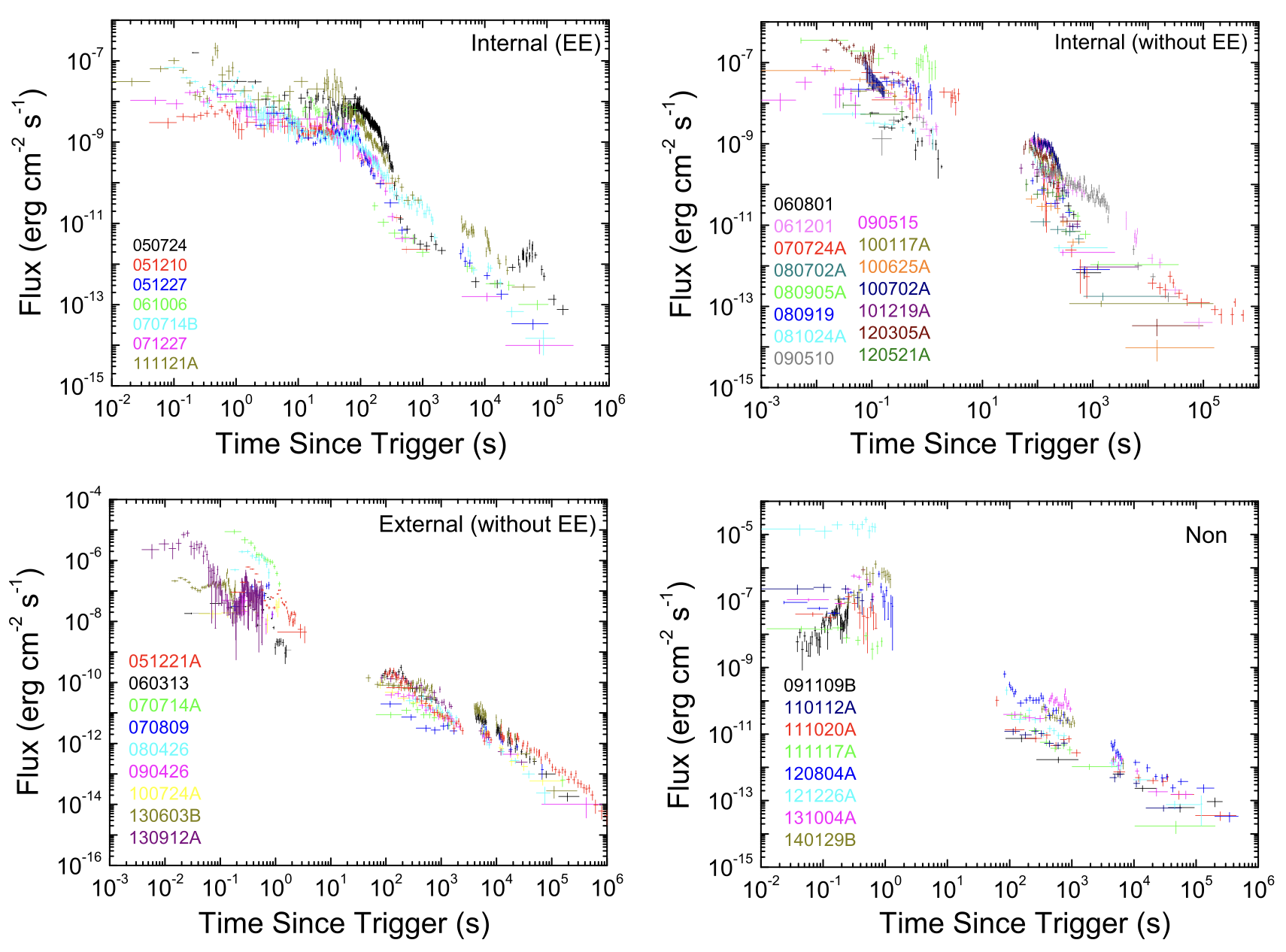}
\caption{0.3–10 keV X-ray afterglow light curves of various short GRBs with X-ray observations by Swift. The upper panels show the ones with an internal plateau, some of the bursts (upper left) also have an internal plateau showing up as extended emissions in the Swift-BAT band. The lower panels show the bursts with an external plateau (left) or no plateau (right). From \cite{lvhoujun15}. }
\label{fig:sec4_figure2}
\end{figure*}

\begin{figure}[H]
\centering
\includegraphics[width=0.49\textwidth]{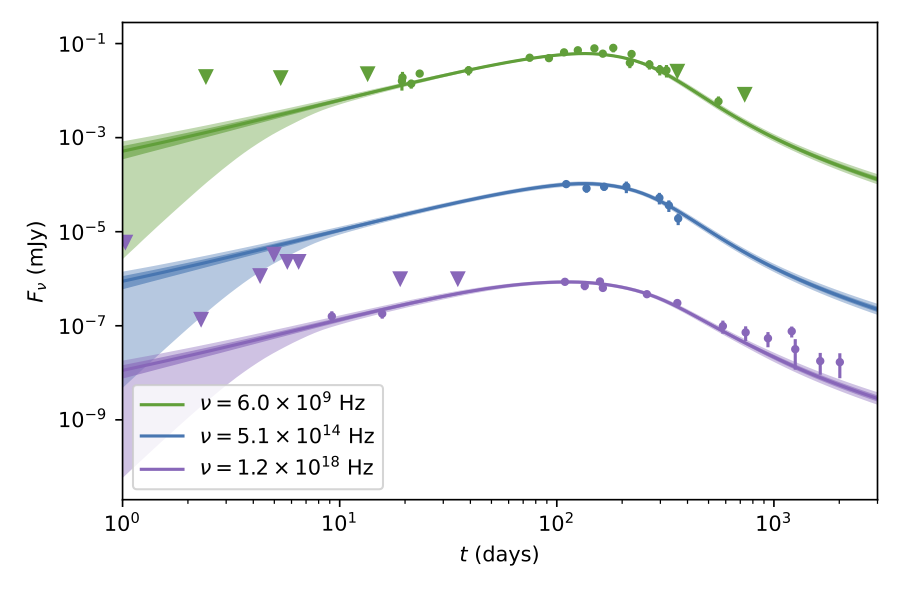}
\caption{The GW170817 light multi-band light curve. Radio, optical, and X-ray data in green, blue, and purple respectively. Triangles show 3$\sigma$ upper limits. Solid bands show the modelled afterglow emission after Markov Chain Monte Carlo (MCMC) fitting. The solid curve shows the median flux as a function of time, the pale bands show the 68$\%$ and 95$\%$ quantiles of the flux at each time. Figure from \cite{Ryan2024}.}
\label{fig:sec4_figure3}
\end{figure}

\item {\bf NS-BH mergers:} Such mergers can either lead to tidal disruption of the NS (in the case of large mass ratio $q=m_1/m_2$ with $m_1 > m_2$) or direct plunge of the NS into the BH. In the former case, a short GRB and a kilonova are likely produced. The discussion of the prompt emission and afterglow features for NS-NS mergers also applies, even though the central engine can only be a BH. The expected on-axis and off-axis light curves are similar to the case of NS-NS mergers (Figures \ref{fig:sec4_figure2} and \ref{fig:sec4_figure3}) except the bright X-ray plateau signature. For the plunging events, no bright X-ray counterparts are expected.

\item {\bf BBH mergers:} Usually no bright EM counterparts are expected from these events. Recent studies suggest that some BBH mergers may originate from AGN disks. A candidate EM counterpart for a BBH merger, presumably due to interaction between the kicked remnant and AGN disk, was reported \cite{graham20}, but not confirmed. So far no bright X-ray emission has been detected from any BBH merger events.

\item {\bf Tidal disruption events:} TDEs also produce gravitational waves. For typical TDEs around MBHs, the frequency of the GW at the closest approach is $f_{\rm GW}\sim10^{-4}$Hz, which falls within the detection band of space-bourne interferometers such as LISA \cite{Amaro-Seoane.2007}. However, the GW signal is weak, and LISA will only be able to detect extreme TDEs (in which the stellar mass $\gtrsim 60 M_\odot$) from very nearby galaxies \cite{Pfister.2022}. On the other hand, the disruption of WDs which approach MBHs during ``extreme mass-ratio inspirals'' (EMRIs)  is a more promising channel for producing detectable GWs \cite{Eracleous.2019, Toscani.2020}. In such events, the BH mass should be $\lesssim10^5 M_\odot$, and the gravitational wave signal will be emitted for many cycles (with $f_{\rm GW}\sim 0.01-10$~Hz) as the WD orbits the BH before the final disruption. The GW background produced by such events, coupled with its electromagnetic counterparts, will be valuable for probing IMBHs and their hosting galaxies/star clusters. 
\end{itemize}

Furthermore, while the GWs produced by the SMBH mergers are beyond the detection frequencies of the ground-based GW detectors like LIGO and Virgo, binary compact-star mergers can also happen in AGN accretion disks. These mergers can produce high-frequency GWs and also generate observable EM radiation \cite{Grobner.2020}.

A joint detection between a GW source and an X-ray counterpart will validate many physical scenarios as discussed above. Such a detection would also help to characterize the GW sources. By identifying the redshift of the GW source from an EM counterpart, one gives a direct measurement of the luminosity distance of the source, and hence, can more precisely constrain the inclination angle of the GW source \cite{2017PhRvL.119r1102F}. Joint constraints from the CBC member parameters and the X-ray plateau features may lead to constraining the initial spin frequency and ellipticity of the post-merger NS through modelling its spindown through magnetic dipole radiation and secular GW radiation \cite{Gaohe16,LiAng16,2023MNRAS.522.4294Y}. The strong interaction phase transition inside NSs may lead to changes in the merger remnant oscillation frequency as well as the moment of inertia and angular momentum evolution. Based on its dependence on the properties of the binary and phase transition parameters, the strong interaction phase transition can be studied based on the observed inspiral GW waveforms \cite{2023arXiv230508401M} and the radiation characteristics of the EM counterpart \cite{LiAng16,2021PhRvD.104h3004Z}.

\begin{figure}[H]
\centering
\includegraphics[width=0.5\textwidth]{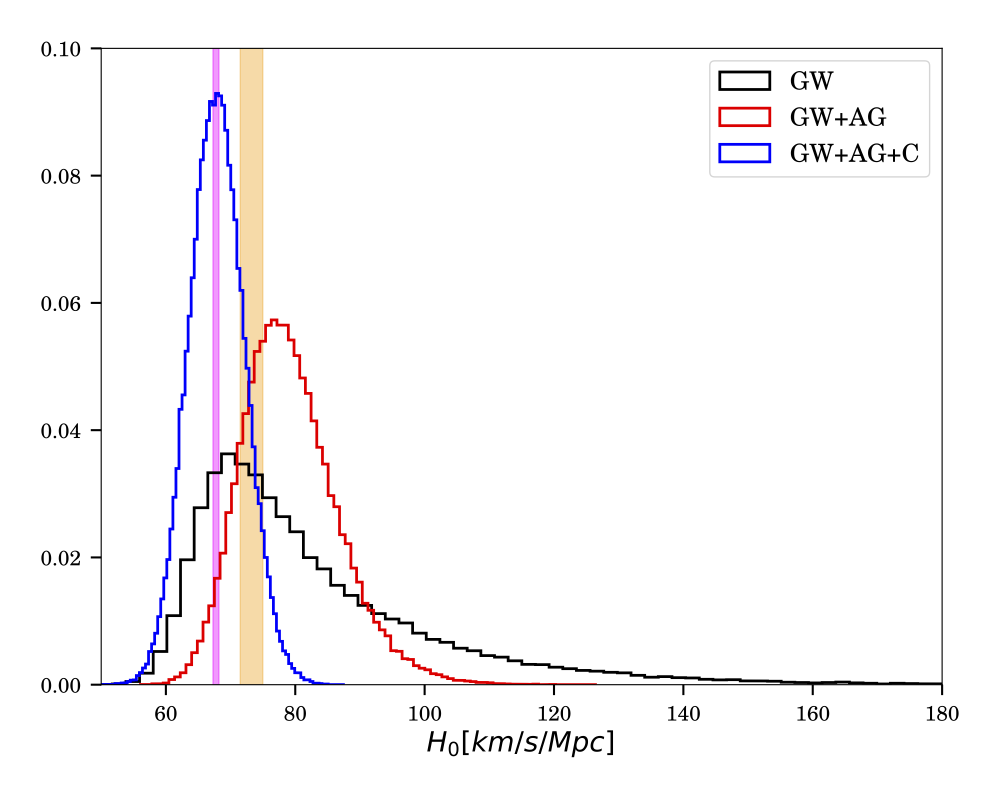}
\caption{Histograms of the Hubble constant $H_0$ posterior estimated for the event GW170817. The result of a GW-only analysis is represented in black, a GW and afterglow analysis (GW+AG) in red and a GW, afterglow and relativistic jet centroid motion analysis (GW+AG+C) in blue. The magenta and yellow shaded regions represent the 1$\sigma$ interval of the \textit{Planck} and SH0ES measurements respectively. The GW+AG analysis reduces the error on H$_0$ by a factor of 2 compared to GW-only, while including the jet centroid motion reduces it by a factor of 3. The Figure is adapted from \cite{Gianfagna2024}.}
\label{fig:sec4_figure4}
\end{figure}

The combination of a GW signal and EM information for mergers also provides a novel method to effectively measure the Hubble constant H$_0$ independently of distance scale estimators. This method should help solve the problem of the tension between the late and early time Universe measurements on H$_0$, once many events with GRB afterglow and GW detections become available. Applying this method to GW170817 already led to promising H$_0$ results, see Figure \ref{fig:sec4_figure4} \cite{Hotokezaka2018, Gianfagna2024}.

\subsubsection{X-ray counterparts to neutrino sources}\label{sec: xray_nu} 

High-energy neutrinos must be produced by energetic charged particles, namely cosmic-rays, so they are crucial messengers for studying the origin of cosmic rays. In astrophysical sources, such as AGNs, the energy in the jet or accretion disk is transformed into the acceleration of protons or heavier nuclei. These particles subsequently interact with radiation and/or ambient matter in the vicinity of the source to produce pions and other secondary particles that decay into neutrinos. 
In 2013, the IceCube Neutrino Observatory made the discovery of extraterrestrial high-energy neutrinos \cite{IceCube13}. Even though it has been very difficult to identify the sources of these neutrinos, progress has recently made through multi-messenger observations. The electromagnetic counterparts of these neutrino sources are typically X-ray emitters. EP is a valuable mission to study the counterparts of neutrino events and identify more neutrino sources. In the following we discuss several types of neutrino sources. 

\begin{itemize}
\item {\bf Blazars: } 
The IceCube Neutrino Observatory detected the first evidence of a neutrino source from the blazar TXS 0506+056 \cite{IceCube-a.2018}.
By combining $\gamma$-ray observations and neutrino data, the neutrino event IceCube-170922A is believed to be associated with a $\gamma$-ray flare from the blazar TXS 0506+056 \cite{IceCube-a.2018}. 
Most of the studies ascribe the high-energy neutrino emission of TXS 0506+056 to relativistic protons accelerated in the jet, through either photopion production with the radiation of the jet itself and the external radiation of the surrounding environment, or proton-proton collisions with matter of the jet and cloud/star entering the jet. 
Swift and NuSTAR detected X-ray emission from TXS 0506+056 \cite{Keivani_2018}. Swift monitored the X-ray flux from TXS0506+056 for 4 weeks, finding clear evidence for spectral variability before and after the neutrino alert. It was also found that an increase in the X-ray flux correlates well with the strong increase at VHE energies over several days after the alert (see panel C and D of Figure \ref{fig:sec4_figure5}). These findings suggest that X-ray activities may be correlated with neutrino events. In this context, X-ray observations have been crucial to test leptonic and hadronic models. In particular, the spectral energy distribution is mostly sensitive to hadronic acceleration processes over the 0.1--100 keV energy range (see, e.g.  \cite{2018ApJ...864...84K, 2019MNRAS.483L..12C}). Moreover, an excess of high-energy neutrino flux was found in 2014--2015 \cite{IceCube-b.2018}. Unfortunately, contemporaneous EM (in particular, X-ray) data are lacking over this period, which prevents the search for a possible EM flare associated to the neutrino one.

\begin{figure*}
\centering
\includegraphics[width=18cm]{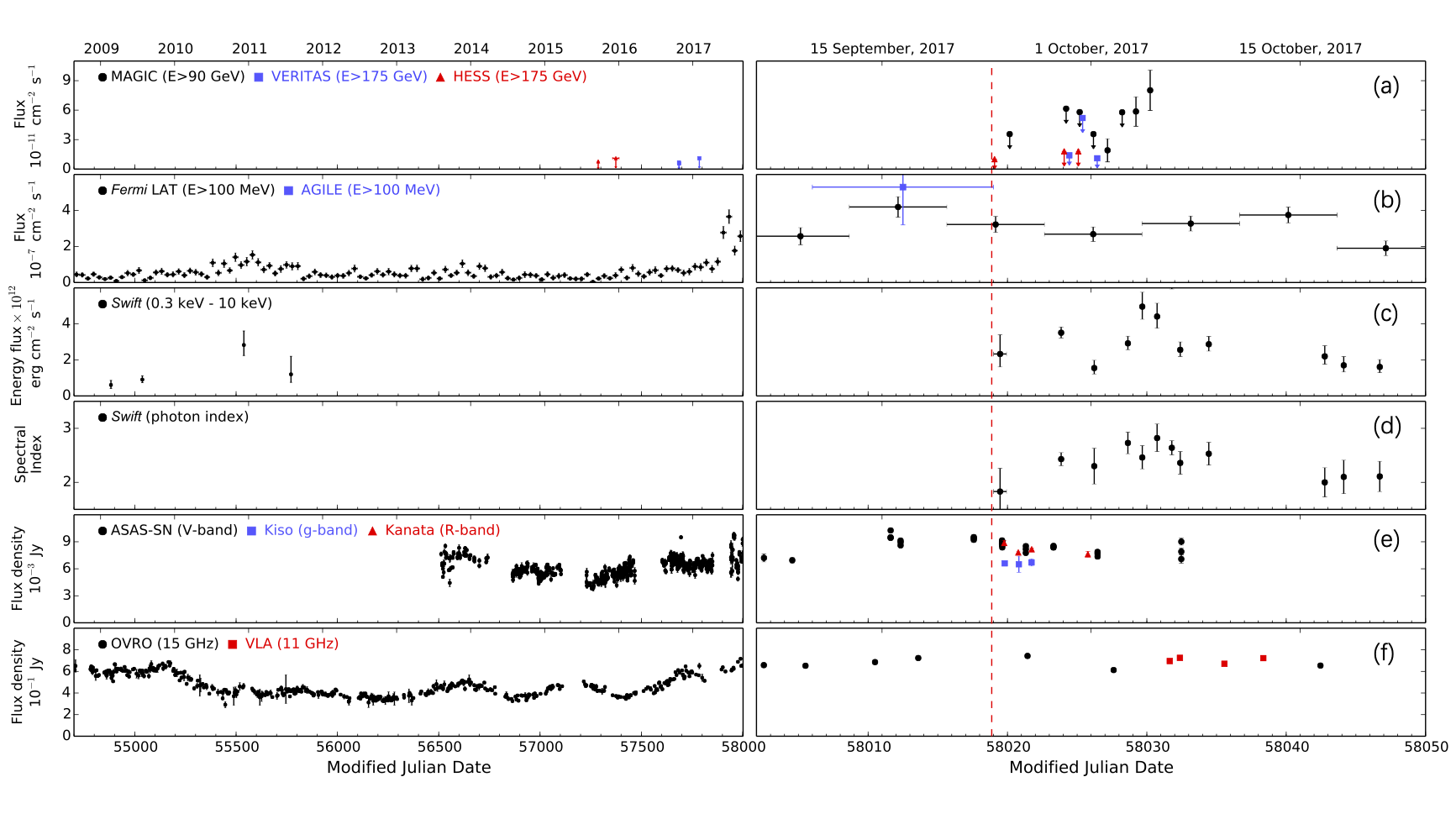}
\caption{Multi-wavelength light curve of TXS 0506+056 before and after the neutrino trigger. From top to bottom: (a) very high-energy $\gamma$-ray observations by MAGIC,
H.E.S.S. and VERITAS; (b) high-energy $\gamma$-ray observations by Fermi-LAT and AGILE; (c and d) X-ray observations by
Swift XRT; (e) optical observations by ASAS-SN, Kiso/KWFC, and Kanata/HONIR; and (f) radio observations by OVRO and VLA. The red dashed line marks the detection time of the neutrino IceCube-170922A. The figure is adapted from \cite{IceCube-a.2018}.}
\label{fig:sec4_figure5}
\end{figure*}

More recently, following the discovery of the neutrino event, IceCube-211208A, a multi-wavelength observing campaign was carried out and it was found that the blazar PKS 0735+178 lying just outside the 90\% error region of the neutrino event was in a flaring state in the optical, ultraviolet, X-ray, and GeV $\gamma$-ray wavebands \cite{2023ApJ...954...70A}. 
The possible association of neutrino emission with TXS 0506+056 and PKS 0735+178 strengthens the connection between blazars and high-energy neutrino events. Future multi-messenger monitoring of blazars is crucial to confirm this hypothesis and EP can play a major role in this context, in particular through a monitoring of these sources with the WXT.
X-ray observations are also important to understand the conditions and mechanisms for 
neutrino emission in blazars.


\item {\bf Seyfert galaxies:}
In 2022, the IceCube Collaboration reported an excess of neutrino events associated with NGC 1068, a nearby Type-2 Seyfert galaxy \cite{2022Sci...378..538I}, with a  significance of 4.2$\sigma$. In NGC 1068, an SMBH at the centre is highly obscured by thick gas and dust \cite{2022Natur.602..403G}. X-ray studies have suggested that NGC 1068 is among the brightest AGNs in intrinsic X-rays \cite{2015ApJ...812..116B}, which is generated through Comptonization of accretion-disk photons in a hot plasma above the disk, namely the coronae. Since the SMBH at the centre of NGC 1068 is obscured by gas and dust  and  is  also surrounded by strong  radiation,  efficient neutrino production  is  expected to  occur if cosmic rays are accelerated  \cite{2020PhRvL.125a1101M,2020ApJ...891L..33I}). Interestingly, the reported neutrino flux is higher than the GeV $\gamma$-ray flux, implying that $\gamma$-rays above 100 MeV are strongly attenuated by dense X-ray photons while neutrinos can escape. 

An important question is what differs NGC 1068 from other nearby Seyfert galaxies. NGC 1068 is not the brightest X-ray Seyfert, but  once the X-ray emission is corrected by accounting for absorption by the gas, NGC 1068 becomes  the intrinsically brightest Seyfert \cite{2015ApJ...812..116B,2016MNRAS.456L..94M}). Future identification of more Seyfert galaxies associated with neutrinos by X-ray observations will be useful to pin down the association between the neutrinos and highly obscured Seyfert galaxies and understand the underlying process of neutrino production in these sources. A few candidates for neutrino-emitting Seyfert galaxies have been proposed \cite{kheirandish21}.

\item {\bf Tidal disruption events:}


TDEs are promising sites for producing high-energy cosmic rays \cite{Farrar2009ApJ...693..329F} and neutrinos \cite{Wang2011PhRvD..84h1301W}. One natural production site is the relativistic jet, where the injected debris materials can gain large acceleration and interact with photons and other hadrons to produce ultra-high energy particles \cite{Farrar.2014, Wang-X.2016, Zhang-B.2017}. 
Neutrino emission is predicted to be produced in both successful jets \cite{Wang-X.2016,Dai2017MNRAS.469.1354D,Lunardini2017PhRvD..95l3001L,Senno2017ApJ...838....3S} and choked jets of TDEs \cite{Senno2017ApJ...838....3S}.  
On the other hand, it is suggested that the geometrically thick and possibly strongly magnetized accretion flows formed in TDEs (including their coronae) can also provide a good environment for accelerating particles to PeV energy level \cite{Dai-L.2017, Hayasaki.2019}. 

Recent optical observations of neutrino alerts have identified two flares from the centres of galaxies coincident with 100 TeV-scale neutrinos: AT2019dsg was possibly associated with the IceCube neutrino event IC191001A \cite{Stein.2021} and AT2019fdr possibly with IC200530A \cite{Reusch.2022}. 
The former is considered as TDEs from quiescent BHs, while the latter as a probable TDE in an AGN. 
In both cases, the neutrinos were detected $\sim 100-300$ days after the flare peak. 
Since these discoveries, various models have been proposed to explain the neutrino emission, including TDE jets \cite{2021NatAs...5..472W,Liu2020PhRvD.102h3028L}, corona and hidden wind \cite{Murase2020ApJ...902..108M}, as well as outflow-cloud interactions \cite{Wu2022MNRAS.514.4406W}.

In both of the TDE events, thermal soft X-rays were detected. In AT2019dsg, X-rays were detected starting from 37 days after discovery, and the X-ray flux faded extremely rapidly with an unprecedented decline rate in TDEs \cite{Stein.2021}. Temporal evolution of X-rays is also evident in the case of AT2019fdr. The X-rays are detected nearly 300 days after the neutrino detection. This shows that X-ray monitoring is important for further investigating the X-ray evolution associated with neutrino events from TDEs.


\end{itemize}



\subsection{Observing strategy}\label{sec:obs_strategy_mm}
EP is well-positioned to study multi-messenger sources with its supreme observational capabilities.

\subsubsection{Gravitational wave counterparts}

For GW events, we apply the following observing strategy for different types of events:

\begin{itemize}
\item {\bf Self-triggered events:} If a fast transient triggered WXT, which is spatially and temporarily associated with a GW event triggered by the LVK association, EP will perform the standard rapid observations using FXT with enhanced cadence to catch the entire lightcurve as much as possible. 
\item {\bf NS-NS mergers with a GRB counterpart:} Such events may be similar to GW170817/GRB 170817A. Based on the lessons learned to identify the counterparts of GW170817/GRB 170817A, the most efficient strategy is to use WXT to cover the error box provided by the smaller GW and GRB detectors. In the case of GRB 170817A, the GRB error box (derived from Fermi/GBM) is much smaller. Searching in the error box of GBM could have substantially shortened the discovery time of the EM counterpart. Based on the simulations using the discovery of the LEIA source associated with GRB 230307A, it is possible to perform two pointings during the first orbit. 
\item {\bf NS-NS mergers without a GRB counterpart:} WXT will promptly search for any bright X-ray counterpart within the GW error region. Depending on the distance derived from the GW data, it will be judged whether to perform one (if the expected X-ray plateau emission is faint) or two (if the expected X-ray plateau emission is bright enough to be detected in a time shorter than one observing cycle in the first orbit). If no detection is made after the first orbit, the observatory will make use of FXT to search for the counterpart based on the signal significance contours provided by the GW observatories. 
\item {\bf NS-BH mergers:} For such mergers with the NS tidally disrupted (marked as ``with a remnant'' in LVK alerts), an observing strategy similar to the NS-NS mergers will be carried out. In the case of a GRB detection, the GRB error circle will be first searched. The standard procedure applies for the case without a GRB detection, except that the priority of searching for bright X-ray plateaus is reduced. 
\item {\bf BBH mergers:} These targets will not be regularly observed unless special situations (e.g.  joint triggers in the $\gamma$-ray or neutrino channels). 
\end{itemize}

\begin{figure*}[ht!]
\centering
\includegraphics[width=0.99\textwidth]{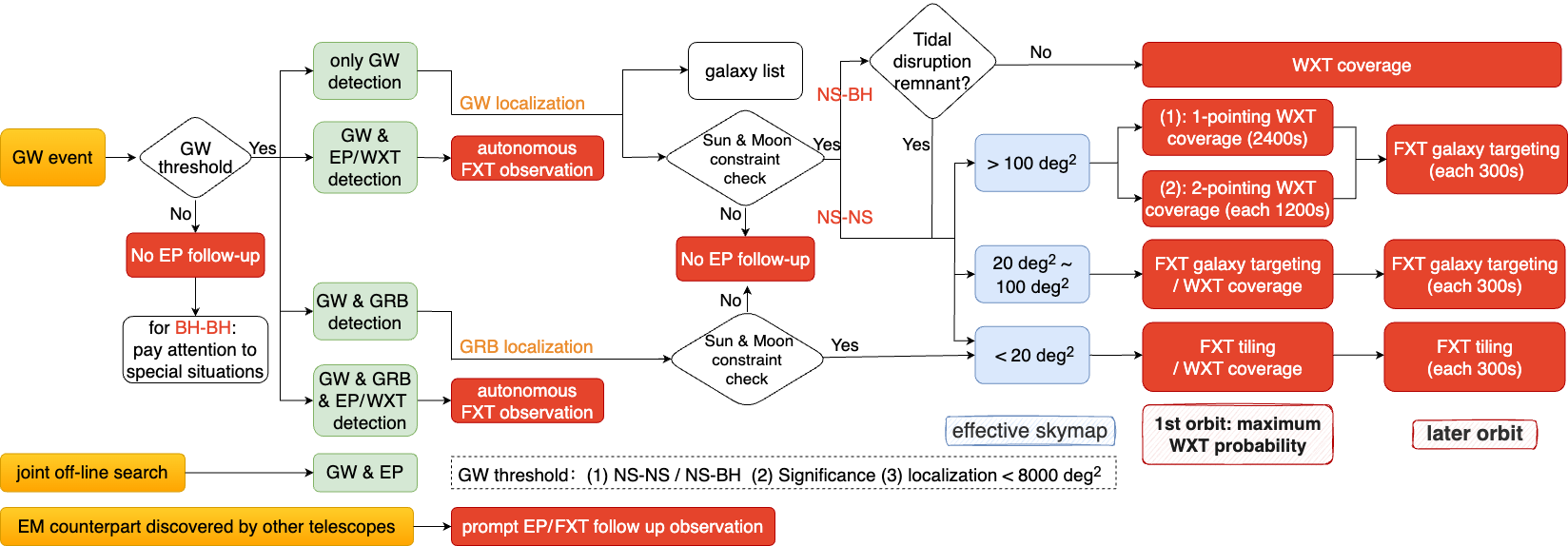}
\caption{EP pointing strategy of GW-related multi-messenger observations. Credit: EP science center.}
\label{fig:sec4_figure8}
\end{figure*}

A flow chart for the pointing strategy of GW-related multi-messenger observations is presented in Figure\,\ref{fig:sec4_figure8}.

For sources that match the GW threshold and the constraints on the multi-msessenger ToO observations, three observation strategies are applied by taking advantage of the WXT, the FXT and the galaxy list: 
(1) WXT coverage: the WXT FoV is adopted to cover the GW skymap. (2) FXT tiling: The area of the GW skymap is covered by multiple FXT FoV tiles. (3) FXT galaxy pointing: FXT is pointed to the selected galaxies with different significance, which are provided by GWOPS \cite{xu2020gwops}.
With the official start of the LVK collaboration O4 observation run, the WXT coverage observation mode has been successfully deployed on and used with LEIA \cite{ma2024enhancing}.
For sources with low significance or delayed alerts, a joint offline search by EP observations and GW event localization is adopted.

Three observation strategies are combined for different types of events and different situations. 
In the first orbit, the WXT coverage is considered first to take the benefits of the WXT FoV.
For the case of smaller effective GW skymap, the FXT galaxy targeting or the FXT tiling mode is considered in the meanwhile to calculate the overlapping sky probability of the WXT FoV and the GW skymap, and the final adopted observation mode in the first orbit is given to maximize the overlap of the GW probability region with the WXT FoV.
For the subsequent orbits, the FXT-related observation strategies are applied for better detection sensitivity.

Finally, in the case that an EM counterpart is already detected by other satellites in X-rays or other bands (e.g.  in optical), prompt FXT follow-up observations will be pursued.


\subsubsection{High-energy neutrino counterparts}

Based on what we know of the EM counterparts of high-energy neutrinos, we plan three observing strategies.

\begin{itemize}
 \item Regular monitoring of known or candidate neutrino sources: Known neutrino sources are AGNs, either blazars (such as TXS 0506+056) or Seyfert galaxies (such as NGC 1068). Regular monitoring of these sources is necessary to catch EM flaring phases. Stacked searches for enhanced neutrino signals from the enhanced EM-active episodes will help to understand the physical conditions to produce neutrinos from these sources.
  \item Study of archival WXT data in conjunction with archival neutrino datasets: The first strategy mentioned above can also be employed in retrospect: searching for an X-ray flare in the archival WXT data that coincides with a neutrino flare detected in archival neutrino telescope data. This approach was pursued following the detection of the TXS 0506+056 neutrino flare in 2014-2015. However, the lack of wide-field X-ray data during this time period prevented the construction of a sufficiently well-sampled lightcurve for the source, making it impossible to detect a potential X-ray flare associated with the blazar concurrent with the neutrino flare.
 \item Follow-up observations of triggered events: IceCube collaborations regularly release alerts of high-energy neutrino events. The ``Gold'' alerts provide a greater than 50\% chance probability of astrophysical origin, which could be followed up. Identification of TDEs or AGN flares in the error regions would lead to the identification of new neutrino counterparts. 
 \item Extensive observations of TDEs or other transients with possible neutrino associations are warranted in the case of a possible neutrino-transient association claimed by other teams. 
\end{itemize}

\begin{figure*}[ht!]
\centering
\includegraphics[width=0.9\textwidth]{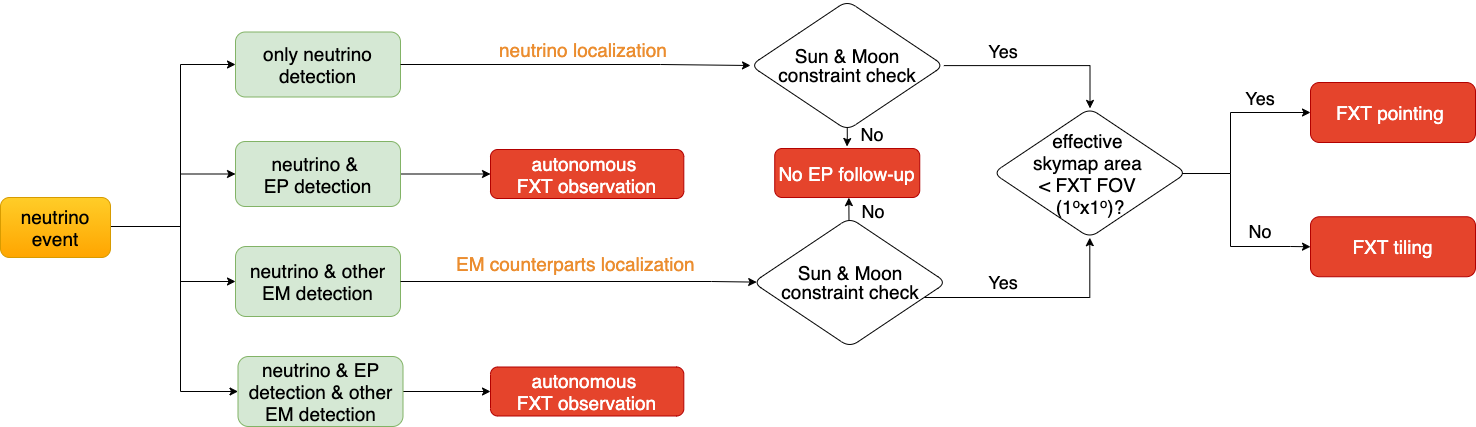}
\caption{Pointing strategy of Neutrino-related multi-messenger observations. Credit: EP science center.}
\label{fig:sec4_figure9}
\end{figure*}

A flow chart for the pointing strategy of neutrino-related multi-messenger observations is presented in Figure\,\ref{fig:sec4_figure9}.
Two observation strategies are applied: (1) single FXT pointing: the error region of the neutrino event will be observed by a single FXT pointing if it is smaller than the FoV of the FXT. (2) FXT tiling: The FXT FoV tiles are adapted to cover the neutrino skymap larger than the FoV of the FXT.

\subsection{Simulated observations of X-ray counterparts to GW sources}

Since there were no early X-ray observations of GW170817/GRB 170817A, it is not known what the brightness is of the early X-ray signals from NS-NS mergers. Nonetheless, there are two candidate X-ray sources that could be counterparts to NS-NS merger GW events, and one can, therefore, perform simulations based on the observations of these sources; these are presented here.

\subsubsection{CDF-S XT2-like events}

The Chandra-Deep-Field South X-ray-Transient \#2 (termed as CDF XT2) at a redshift $z=0.738$ had the signature of a new-born magnetar, and its host galaxy properties are consistent with those of SGRBs. So, it was regarded as a candidate X-ray transient from a NS-NS merger that left behind a long-lived magnetar \cite{Xue2019_XT2}, as predicted in \cite{zhangbing13} and \cite{sunhui17}. 

\begin{figure}[H]
\centering
\includegraphics[width=0.4\textwidth]{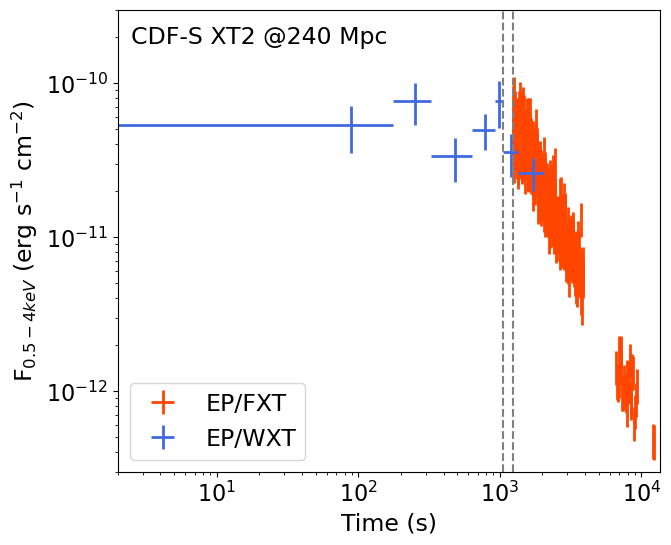}
\caption{
An example of simulated EP observations of CDF-XT2-like events at the distance of 240 Mpc. The blue and red data points represent the simulated WXT and FXT light curves, respectively. The left and right dashed vertical lines mark the time to trigger FXT follow-up and the start time of FXT observations after taking into account the slew maneuver, respectively. Credit: Jingwei Hu.}
\label{fig:sec4_figure10}
\end{figure}


 CDF-S XT2-like events can be simulated based on the given luminosity, X-ray spectrum, light curve, and absorption, and place it at different distances to evaluate the detection capability of such events by EP. An example of a simulated light curve is shown in Figure \ref{fig:sec4_figure10}. According to these simulations, the CDF-S XT2-like event are detectable by EP-WXT within 240 Mpc.

\subsubsection{LEIA counterpart of GRB 230307A}

 LEIA detected the X-ray counterpart \cite{SunHui23} to a nearby long-duration GRB 230307A with a compact star merger origin \cite{LevanA23,YangY23}. The X-ray light curve does not follow the behaviour seen at higher frequencies, which shows the emergence of a long-lived engine after the prompt $\gamma$-rays dim. The X-ray light curve is consistent with the spindown of a newborn magnetar \cite{SunHui23}. As a result, the LEIA source provides a very good candidate for being a source showing early X-ray emission of an NS-NS merger. 

We have simulated GRB 230307A-like events by using the same methods as for the CDF-S XT2-like events. According to these simulations, the GRB 230307A-like event can be detected by EP within a redshift of about 1.1.

\section{Compact stellar objects}\label{sec:sco}

\begin{figure}[H]
\centering
\includegraphics[width=0.48\textwidth]{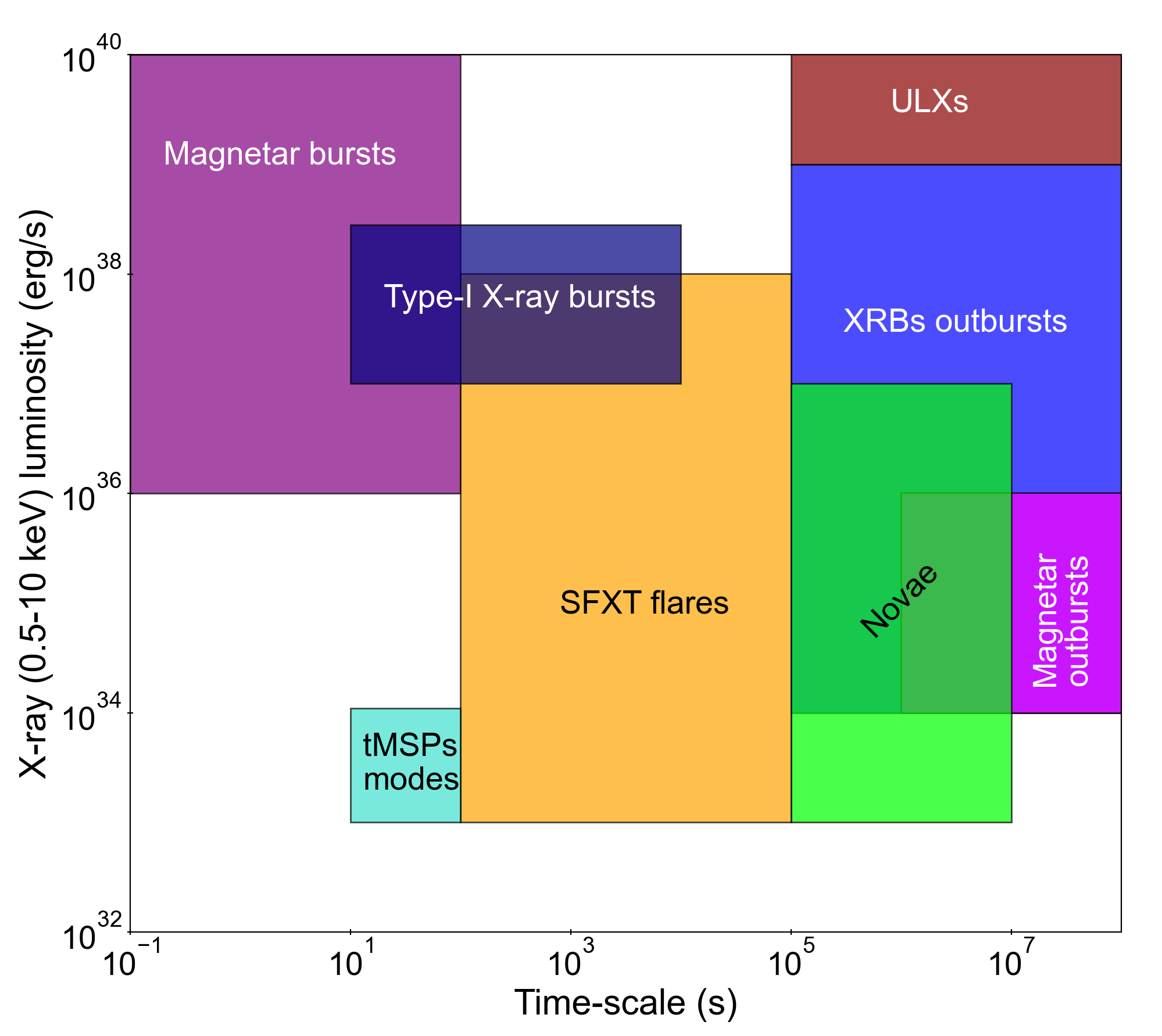}
\caption{X-ray luminosities and timescales of Galactic transient events related to compact objects.} 
\label{fig:bursting_classes}
\end{figure}

Most of the high-energy transients in our Galaxy and neighboring galaxies are associated with the presence of compact objects: WDs, NSs and BHs. Compact objects may produce bright X-ray emission via accretion in binary systems over different regimes, due to thermal surface emission, magnetic reconnection events, or acceleration mechanisms and shocks related to the interaction between plasma and magnetic fields over a wide range of states. By nature, most of these effects produce transient X-ray emission, as shown in Figure \ref{fig:bursting_classes}. EP provides an unprecedented view of the initial phases of these events, along with a unique opportunity for long-term monitoring of their evolution.
The X-ray emitting population of stellar-mass compact objects primarily consists of:\\
1) X-ray binaries (XRBs): these are systems where a NS or a BH orbits another object (low or high mass stars, a WD or even another NS). In these systems, the X-ray emission is mainly powered by mass accretion onto the compact object or, in some cases, by the thermal emission from the NS surface \cite{Negueruela2006, Reig2011, Walter2015, Bahramian2023, Kaaret2017, Papitto2022}.\\
2) Isolated NSs: these stars emit either predominantly through rotational energy loss or magnetic energy loss. In the former case, radiation is emitted via different acceleration mechanisms in the magnetosphere, as observed in typical radio, X-ray or $\gamma$-ray pulsars. In the latter case, magnetic energy loss results in thermal surface X-ray emission or, in more extreme cases like magnetars, resonant cyclotron scattering and possible magnetospheric reconnection events \cite{Kramer2005, Turolla2015, DeLuca2017, Kaspi2017, Esposito2021}. \\
3) Binary and isolated WDs: these sources are also very common in the X-ray sky, and emit either via mass accretion from a companion star (usually a relatively low-mass star) or via their surface thermal emission \cite{ChenWan1997, Ferrario2015, mukai17, Page2022}.

\subsection{X-ray binaries}
\label{sec:xrbs}
X-ray binaries come in two types, Low-Mass and High-Mass X-ray binaries (LMXBs and HMXBs), depending on whether the donor is an old, solar or sub-solar star or instead a young and highly massive $O$ or $B$ star. LMXBs are characterised by relatively short orbital periods, ranging from tens of minutes (in the extreme case of Ultra-Compact X-ray sources; see Section \ref{sec:ucxrb}) to days. On the other hand, HMXBs are wider systems with orbital periods of days to months. A further classification divides all types of XRBs into persistent and transient sources according to the long-term X-ray activity. Transient sources usually remain in a quiescent state with very low accretion rates and therefore low X-ray luminosity ($L_{X}\lesssim10^{33}$ ergs\,s$^{-1}$), which can last for several months or even decades \cite{Tanaka1996}. However, due to instabilities in the accretion disk \cite{Lasota2001,Hameury2020}, these transient sources occasionally exhibit outbursts, lasting for weeks to years. The peak luminosity during the outburst can vary a lot from source to source, going from ``faint outbursts'' at $L_{X}\sim10^{34}-10^{35}$ ergs\,s$^{-1}$ (see Sections \ref{sec:accNSs}, \ref{sec:tmsp}) up to the Eddington limit and beyond in the case of the Ultra-Luminous X-ray sources (section \ref{sec:ulxs}). 

\subsubsection{Accreting black-hole systems}\label{sec:accBHs}
The early identification of BH X-ray binaries (BHXRBs) dates back to the early 1960s. Currently, around 100 candidates have been discovered (e.g.  \cite{Corral-Santana2016}), with approximately 20 of them confirmed using dynamical methods \cite{Gandhi2019,zhao2023}. In fact, BHXRBs mostly remain in a quiescent state with X-ray luminosities typically ranging from $10^{29}$ to $10^{33}$~erg~s$^{-1}$. Quiescent BHXRBs can be detected using astrometric and radial-velocity techniques, while the transiently active ones can be identified through X-ray outbursts.

The observation of the initial phases of outbursts in BHXRBs poses a significant challenge due to the extremely low X-ray flux in their quiescent state and the limitations in sensitivity of X-ray instruments. While a few sources can be observed during quiescence using large space-borne X-ray telescopes like Chandra and XMM-Newton, the majority of new X-ray outbursts from BHXRBs are typically detected when they reach a sufficient brightness level to trigger X-ray all-sky monitors such as the RXTE/ASM, MAXI, and Swift/BAT. However, given the sensitivity of state-of-the-art X-ray monitors such as MAXI, this corresponds to a triggering luminosity of about 3$\times10^{36}$~erg~s$^{-1}$ for a source located at 8 kpc. As a consequence, most of the rising phase of X-ray outbursts gets missed. 

Currently, the early study of XRB outbursts primarily relies on continuous monitoring through ground-based optical telescopes. Over the past few years, about 50 known X-ray binaries have been monitored weekly in the optical band within the program X-ray Binary New Early Warning
System (XB-NEWS) \cite{Russell2019}, aiming to identify optical precursors of X-ray outbursts. This approach enhances the likelihood of capturing a new outburst during its early phase and triggering multi-band campaigns in time to provide adequate coverage. However, most of the known BHXRBs, and of course the ones that have yet to be discovered, are not covered by XB-NEWS. Therefore, it has been so far challenging for the X-ray telescopes to capture the short-term initial stages of outbursts, which has hindered the study of the entire outburst phase, especially the early temporal and spectral evolution, leaving some important questions unanswered. For example, when and where are the outbursts triggered in BHXRBs? What are the physical mechanisms responsible for BHXRB outbursts, and what factors influence the intervals between them? Moreover, capturing BHXRBs during the initial outburst stages increases the chances of detecting rare activities within accreting systems, specifically in relation to the early development and propagation of instabilities that culminate in complete outbursts. For instance, in the earliest reports about the recent outburst of the new BH binary candidate, Swift J1727.8--1613, some rare behaviours were observed unlike that seen in other BHXRBs, such as repeating bursts and rapid flaring activities \cite{connor23}, which became absent during later observations.

After the rising phase, a classical outburst typically conforms to the following evolution of the radiation state. During the initial outburst, BHXRBs are in a hard state, marked by the dominance of non-thermal components and the presence of strong low-frequency QPOs and band-limited noise. At this stage, the inner radius of the disk is generally believed to be truncated \cite{McClintock2001, Done2007}, although this is subject to debate (see below). As the outburst proceeds, BHXRBs enter the intermediate state, with the inner disk radius shifting inward. The intermediate state can be further subdivided into the hard and soft intermediate states, characterized by a softer spectrum in the soft intermediate state. Furthermore, the strong band-limited noise and QPOs observed in hard and hard intermediate states are replaced by relatively weaker broadband noise and QPO components in the soft intermediate state. Subsequently, BHXRBs may transform into the soft state as the thermal component progressively dominates the emission. In this state, a weak QPO and weak broadband noise may occasionally emerge alongside a stable inner disk radius. As the accretion rate decreases, BHXRBs return to intermediate, hard, and quiescent states with changes in their accretion geometry. 

During the soft state of a BHXRB outburst, it is widely accepted that the thin disk extends to the innermost stable circular orbit (ISCO). 
However, the truncation of the inner disk radius during the initial outburst phase has been a topic of extensive debate. While some studies suggest a truncation occurring at tens to hundreds of gravitational radii ($R_{\rm g}$) away from the BH, others, based on observations of the broad iron line \cite{Miller2006, Kara2019} and residual thermal disk emission \cite{Rykoff2007, 1999ApJ...527L..17L, 2000A&A...354L..67M, 2022iSci...25j3544L}, argue that it remains close to or at the ISCO. Furthermore, as BHXRBs undergo the subsequent soft-to-hard state transition and evolve back to the hard state, before returning back to quiescence, the evolution of the inner radius of the thin disk is still unclear. In the soft state, for most BHXRBs, it was found that the disk emission displays an exponential decay, indicating an exponential decrease in the mass accretion rate of the thin disk (e.g.  MAXI J1820+070 \cite{you2023}). Assuming that the mass accretion rate follows the same exponential law during the soft-to-hard transition state and the hard state, the inner disk radius can be estimated starting from the disk luminosity \cite{you2023}. It turns out that the inner radius increases with the decrease of the mass accretion rate, i.e., the thin disk recedes, which is consistent with the theoretical argument for the state transition \cite{esin1997}. However, how far away the thin disk in a BHXRB recedes before returning to the quiescent state cannot be determined, as the disk emission is no longer detected due to the sensitivity limitations of currently operating satellites.

The sensitivity of WXT, capable of reaching $\sim 1$\,mCrab ($2.4\times10^{-11}\,{\rm ergs\,cm^{-2}\,s^{-1}}$) with 1\,ks exposure time, surpasses the current generation of wide-field X-ray monitors, such as MAXI, by one order of magnitude. 
This improvement in sensitivity is expected to detect previously unknown BHXRBs, particularly faint ones, and to provide further insight into these mysterious celestial phenomena. Moreover, the unprecedented monitoring capability of WXT enables capturing the earlier phase of X-ray emission during the start of an outburst than the previous missions. 
WXT is capable of detecting a source of 8\,pc in distance whenever it gets brighter than about 2$\times10^{35}$~erg~s$^{-1}$. Once a new outburst is found, FXT can be triggered to observe the spectral and timing (i.e., aperiodic variability and QPOs) properties, and how they evolve during the rising phase (e.g.  \cite{Altamirano2012}). EP is expected to achieve more accurate determination of the outburst onset epoch in X-rays for a larger sample of BHXRBs, leading to estimation of the viscosity parameters of the hot disk via the time delays of the X-ray and optical flares and advancing our understanding of the short-timescale activities \cite{tucker18}. 

In addition, the low-energy bandpass of FXT down to $\sim$ 0.3 keV enables the investigation of the accretion flow geometry during the outburst, through combined observations with other high-energy satellites, such as HXMT-Insight. The optimal spectral and timing capabilities of FXT also enable detecting the disk emission and studying the accretion flow properties in both the decaying hard state and the quiescence state. 
Some nearby quiescent XRBs, such as HR~6819 (280~pc, \cite{Rivinius2020}) and V723~Mon (460~pc, \cite{Jayasinghe2021}), can also be monitored with FXT. This offers an opportunity to uncover the accretion characteristics during their quiescent states, bridging the existing gap between quiescent and transient states and providing new insights into the accretion flow geometry in this poorly studied yet critical evolutionary phase of the state transition of XRBs. 

\subsubsection{Accreting neutron stars with low-mass and high-mass companions}\label{sec:accNSs}

As a consequence of their different ages, NSs are found to be significantly less magnetized in LMXBs (B$\sim$10$^{8}$-10$^{9}$ G) than in HMXBs (B$\sim$10$^{12}$ G or higher). Most NS HMXBs have magnetic fields strong enough to channel the accreting matter onto the magnetic field lines and produce X-ray pulsations, while only in a fraction of NS LMXBs the accreting star is visible as an X-ray pulsar. A notable exception is represented by the family of Accreting Millisecond X-ray Pulsars (AMXPs, \cite{DiSalvo2021}), which host NSs spinning at hundreds of Hz and whose discovery allowed a first confirmation of the so-called recycling scenario (see Section \ref{sec:tmsp}). Additionally, NS HMXBs are usually divided into two subclasses, i.e., BeHMXBs which host a fast-rotating Be star and sgHMXBs which host a supergiant companion. Since the dawn of X-ray astronomy, several hundreds of NS XRBs have been discovered \cite{Avakyan2023} and firmly identified thanks to the detection of X-ray pulsations or type-I X-ray bursts (see below). 

Both NS LMXBs and HMXBs can be transient sources. NS LMXBs in outbursts evolve through different spectral-timing states, analogous to BHXRBs: the hard state, where the emission is dominated by a Comptonization component from a hot corona; the soft state, where the spectrum is mostly thermal, due to the contribution from the disk and sometimes from the NS surface. In addition, these states appear to be coded also by the appearance of QPOs at frequencies ranging from a few to hundreds of Hz (i.e., the so-called kHz QPOs, a unique feature of NS XRBs \cite{Mendez2021}). BeHXMBs usually display type-I outbursts coincident with their periastron passages, and sometimes giant and long-lasting type-II outbursts \cite{Reig2011,Kretschmar2019}. 

Over the past decades, these outbursts have been extensively investigated during their decay phases.
While the X-ray properties of these sources are relatively well understood during the outbursts, they are poorly studied in the initial outburst phases, as usually new outbursts are detected at a very late stage by X-ray all-sky monitors (see Section \ref{sec:xrbs}). In pulsating HMXBs, it remains uncertain whether the propeller effect (a sudden decrease in luminosity due to accretion suppression by the centrifugal barrier of the rotating magnetosphere \cite{Cui1997,Tsygankov2016}), which has only been observed during the fading phase of outbursts, also operates at the onset of outbursts. 
A high-cadence survey might also provide an opportunity to detect X-ray pulsations (which can also be a transient feature, e.g. \cite{Patruno2016}).

Along with bright and relatively long outbursts, NS XRBs can also display faint and/or short outbursts or forms of X-ray activity, which are typically challenging to catch. Rebrightening episodes, also called mini-outbursts or echo outbursts or simply reflares, are sometimes observed during the decay or even after the end of the main outburst in LMXBs \cite{Patruno2016}. Being often faint and short-lived ($L_{X}\sim10^{33}-10^{35}$ ergs\,s$^{-1}$), only a few of them have been studied in detail. Longer episodes of X-ray activity with peak X-ray luminosity below $L_{X}\sim10^{36}$ ergs\,s$^{-1}$, typically defined as ``faint'' X-ray outbursts, have also been observed \cite{Wijnands2015}. Some systems can show both faint and bright outbursts, see e.g.  \cite{Marino2019b, Ng2021}, while others have only been observed in the faint accretion regime, the so-called Very Faint X-ray Transients (VFXTs) \cite{Muno2005,Bahramian2021}. In a few of these sources, the so-called "burst-only" systems, the existence of an ongoing sub-luminous accretion regime was implied by the detection of type-I X-ray bursts (see below) during periods when they were too faint to be detectable by the X-ray all-sky monitors \cite{Cornelisse2002}. 

Some sgHMXBs exhibit dramatic X-ray flares, known as Supergiant Fast X-ray Transients (SFXTs). 
They spend most of the time in the quiescent state and experience sporadic outbursts lasting up to a few days. These outbursts are marked by bright flares, which can last hundreds of seconds and reach a luminosity of $\sim$$10^{38}$\,ergs\,s$^{-1}$ \cite{Romano2014,Walter2015,Romano2023}. Recently, in the NS LMXB Cen X-4 an optical-to-X-ray brightening phase during quiescence was observed, which, however, did not lead to the onset of a full ``bright'' outburst \cite{Baglio2022}. This wide range of transient and dim X-ray episodes of activity is hard to reconcile within the standard Disk Instability Model (DIM, \cite{Lasota2001}), which is commonly accepted to explain the outbursts-quiescence cycle in transient XRBs. However, these faint outbursts are typically too faint to trigger the X-ray All-Sky X-ray Monitors, making this exotic accretion regime poorly studied so far.

In the light curves of NS LMXBs sampled with a binsize of $\sim$ seconds, spikes often emerge with peak fluxes up to the Eddington luminosity and timescales from seconds to minutes. These events are termed type-I X-ray bursts, or thermonuclear X-ray bursts. They are characterized by specific behaviors, including softening of the spectrum during the decay phase and an 'alpha' parameter (i.e., the ratio of the burst's fluence to that of the outburst) that falls within the range expected  for accretion and thermonuclear burning. 
Most bursts are single-peaked, except for the brightest ones which show photospheric radius expansion (PRE; due to radiation pressure) (for a review, see \cite{Galloway2020}).
Since the first discovery of thermonuclear bursts in the Galactic XRB 3A~1820--30 in 1975, a total of 118 such binaries have been observed to exhibit this phenomenon\footnote{https://personal.sron.nl/$\sim$jeanz/bursterlist.html}. 
Observations of the bursts have been one of the key objectives of the X-ray missions such as RXTE, NICER, and Insight-HXMT. 
This is because the properties of the bursts and their interactions with the surrounding environment provide insights into the nature of the NS and the accretion process. For instance, the oscillation during a burst can reveal the NS's spin, while the flux at touchdown time can be used to estimate the Eddington luminosity, thereby constraining the mass and radius of the NS. 
In addition, observations of changes in the persistent emission, such as enhancements in the soft X-ray band, deficits in the hard X-ray band and a bump peaking at 20--40 keV or discrete emission features, can shed light on the influence of the burst emission on the accretion process.
The surrounding accretion flow could also have impacts on the burst emission, e.g.  the obscuration of the burst emission by the disk, Comptonization of the burst emission by the corona/boundary-layer (e.g.  \cite{Worpel2015, Degenaar2018}).
In addition, we note that current observations are mostly concerned with regular bursts during outbursts. Some rare events, i.e., super-bursts produced by carbon burning and bursts under a very low accretion rate, are yet to be explored.

For sources at $\sim$8\,kpc, any new X-ray activity with a luminosity above $2\times10^{35}$ ergs\,s$^{-1}$ in 0.5--10\,keV will trigger WXT. 
EP is thus expected to be able to catch the onset of an outburst during its rise, allowing for multi-wavelength campaigns already at this early activity stage.  Follow-up observations with FXT in the early stage of outbursts will enable (i) the study of the pulse profiles, spin and spectral evolution of pulsating HMXBs; (ii) monitoring the emergence of low- and high-frequency QPOs and search for X-ray pulsations in LMXBs. 
EP can not only catch new outbursts earlier than previous observations, but also detect fainter outbursts than before from all types of NS XRBs, particularly VFXTs. 
EP is expected to discover new SFXTs and study their activities, especially relatively faint ones. 
The combination of the WXT's survey capability and the FXT's fast response and other capabilities (particularly the fast timing mode at a resolution of 42\,$\mu$s) makes EP well-suited for investigating the demography of the bursting behaviours in XRBs.

\subsubsection{Transitional Millisecond Pulsars}\label{sec:tmsp}
Radio millisecond pulsars (MSPs) are NSs with weak magnetic fields  ($B\simeq10^8-10^9$\,G), spinning hundreds of times per second.
Currently, about 430 of these MSPs have been discovered in the field of the Galaxy\footnote{\url{http://astro.phys.wvu.edu/GalacticMSPs/GalacticMSPs.txt}.}, while 305 have been found in 40 globular clusters\footnote{\url{https://www3.mpifr-bonn.mpg.de/staff/pfreire/GCpsr.html}.}. Some 30 of these MSPs,  termed ``redbacks'' \cite{Roberts2013}, reside in binary systems with orbital periods shorter than a day and have a donor star with a mass ranging from 0.1 to 0.4\,$M_{\odot}$. 
Their radio signals exhibit significant eclipses due to intense interaction between the pulsar particle wind and matter from the donor star. Such interaction produces synchrotron radiation which is observable at X-ray and $\gamma$-ray energies. In fact, all the redbacks known so far \cite{Koljonen2023}
have been detected in the GeV domain by the Fermi satellite. Considering that over 2200 sources in the latest Fermi LAT catalog remain unidentified \cite{Ballet2023}, some of these might be redbacks, whose X-rays may be too faint to be detected with the current instruments. Thus the actual number of redbacks could be significantly larger than the current value.

For decades, the formation of MSPs in binaries was understood within the context of the ``pulsar recycling scenario'' \cite{Alpar1982}. This scenario posits that radio MSPs are formed following a prolonged phase where matter is transferred from a low-mass donor star to an NS within a compact binary system. Throughout this phase, the accreting matter spins up the NS to spin periods of the order of a millisecond, and the system is detected as a bright LMXB. Once this phase ceases, the pulsar is anticipated to turn on as a radio pulsar. This scenario was challenged a decade ago by the discovery of three redbacks swinging between a radio pulsar state and an extremely peculiar active X-ray state with a luminosity of 10$^{33}$--10$^{34}$ ergs\,s$^{-1}$ (i.e., 2--3 orders of magnitude fainter than persistently bright accreting NSs \cite{Papitto2022}). These systems are referred to as ``transitional'' MSPs \cite{Papitto2022}). The state transitions in transitional MSPs can take place over timescales as short as days, but the exact mechanisms that trigger these transitions and power the multiband emissions in the active X-ray state are currently poorly understood. Understanding these phenomena has proven difficult so far, primarily because of the very limited sample size. As a matter of fact, their active X-ray states have eluded detection by past and present X-ray wide-field monitors such as RXTE/ASM, Swift/BAT and MAXI. Instead, these systems were identified through dedicated extensive multi-waveband observing campaigns.

EP is expected to detect more transitional MSPs and thus deepen our understanding of their emission mechanisms. Notably, the sensitivity of the WXT in the soft X-ray energy range is at least one order of magnitude larger than that of the current above-mentioned X-ray wide-field monitors. In just a single day of data taking, the WXT in the survey mode can detect X-ray sources within 4\,kpc with luminosities as low as $10^{34}$ ergs\,s$^{-1}$ in the soft X-ray energy range. This is comparable to the luminosities of transitional MSPs in their active X-ray state. Therefore, any transition of known or candidate transitional MSPs from the radio pulsar state to the active state will possibly manifest themselves as an X-ray variable/transient detectable by the WXT. It is worth noting that most of the currently identified redbacks are found within a distance of 4\,kpc \cite{Koljonen2023}.
In summary, with the unprecedented sensitivity of the WXT, EP has the potential to enable, for the first time, witness of transitions to an active X-ray state in known and candidate redbacks in ``real time'' (i.e., within days at most).

\subsubsection{Ultraluminous X-ray Sources}\label{sec:ulxs}
Ultraluminous X-ray Sources (ULXs) are non-nuclear accreting compact objects with an apparent luminosity above the Eddington limit of stellar mass BHs (see \cite{Kaaret2017} for a review). Identifications of high-velocity outflows \cite{Pinto2016}, shock-ionized surrounding bubble nebulae \cite{Pakull2002}, and NS accretors \cite{Bachetti2014}, suggest that the majority of them are powered by supercritical accretion, i.e., the mass accretion rate is much greater than that just needed to power the Eddington limit. 

The nature of supercritical accretion is still poorly understood. Recent numerical simulations revealed that supercritical accretion will lead to a turbulent, thick inflow with strong winds \cite{Ohsuga2011,Jiang2014,Sadowski2016,Kitaki2021}. In this picture, the central accretion disk is confined within a funnel like geometry filled with high velocity winds. When the X-rays from the central disk propagate in the funnel, they will undergo Comptonization and some degree of mild, geometric beaming effect \cite{King2023}. The optically thick part of the outflow as well as the outer accretion disk may produce soft thermal X-ray emission \cite{Gu2016,Qiu2021}. Therefore, the ULX spectrum may emerge as a combination of a hard Comptonization component plus a soft thermal component \cite{Sutton2013,Middleton2015}.

This physical picture is reminiscent of the unification model for AGNs, i.e., obscuration by the optically thick outflow and the viewing angle together determine the observed energy spectrum. ULXs viewed at low inclinations are hard and highly variable, while those viewed close to edge-on may appear to be soft or even supersoft \cite{Middleton2015,Feng2016,Urquhart2016,Zhou2019}. Along with the variation of accretion rate and consequently the opening angle of the central funnel, hard and soft ULXs may transition between each other. The best approach to test the scenario is to monitor variable ULXs that transition between different spectral regimes \cite{Pinto2017}. Several examples have been found, but the sample size is still small. 

The long-term monitoring program also allows us to reveal possible super-orbital modulations in ULXs (such as in NGC 5907 ULX-1 \cite{Walton2016}, NGC 300 ULX1 \cite{Carpano2018,Vasilopoulos2019}, M82 X-2 \cite{Brightman2019}, M51 ULX-7 \cite{Vasilopoulos2020}, and NGC 7793 P13 \cite{Fuerst2021}). Collecting more data enables a direct comparison with similar behaviours seen in Galactic XRBs like accreting X-ray pulsars (e.g.  Her X-1, LMC X-4, and SMC X-1). Compared to Galactic sources, modulation in ULXs seems to be more prevalent. This needs to be further confirmed observationally.  

A major difference between extragalactic ULXs and super-Eddington accreting XRBs in the Milky Way (such as V404 Cyg \cite{Kimura2016} and Swift J0243.6+6124 \cite{Doroshenko2020} at the peak of their outbursts) is that many of the former seem to be in outburst for decades while the latter are transient sources, perhaps due to their distinct binary properties and mass-transfer mechanisms. However, transient ULXs like CXOU~J203451.1+601043 \cite{Wang2019} have been discovered, but are very rare. The monitoring may disclose more candidates and shed light on the history of binary evolution of supercritically accreting compact objects. 

The fraction of ULXs containing a NS accretor is one of the hot topics, and also concerns the history of binary evolution \cite{Shao2015,Middleton2017}. The monitoring program may detect pulsating ULXs via the propeller effect. With proper modelling, this may allow us to constrain the fraction of ULXs with a NS, or at least to establish a lower limit.

The FXT is an ideal instrument to perform long-term monitoring of a good number of ULXs with short snapshots, for a comprehensive understanding of ULX spectral behaviors and the physical nature of supercritical accretion. A FXT observation with an exposure of 1~ks can reach a limiting luminosity of $3 \times 10^{39}$~ergs~s$^{-1}$ for a source at a distance of 10~Mpc. Therefore, with the FXT, we may monitor a fair number of ULXs beyond the local group.


\subsubsection{Ultra-Compact X-ray Binaries}\label{sec:ucxrb}
Ultra-compact X-ray binaries (UCXBs) are binary star systems with orbital periods of less than 60 minutes, where both components are compact objects, such as the cores of evolved giants, WDs, NSs or BHs \cite{ref3,ref4,ref6}. UCXB systems typically consist of a WD donor and an accretor that can be a NS or a BH \cite{ref3}. 
UCXBs are guaranteed sources of GWs in the low-frequency range ${10}^{-4}$ to ${10}^{-1}$ Hz \cite{ref5,ref6,ref7}. 
To date, there are 20 confirmed UCXBs, of which 19 have NS accretors, except for 47 Tuc X-9, whose nature is still uncertain. Among the NS accretors, 10 are AMXPs \cite{ref4}. Meanwhile, according to a recent study based on multi-wavelength phenomenology, there are 25 potential UCXBs that warrant further investigation \cite{ref4}. 
Multi-wavelength observations revealed the transfer of chemical substances during mass accretion. They can be used to test binary evolution theories, especially the common-envelope phase \cite{ref6}. 

When the accretor draws matter from the donor star through Roche lobe overflow, the matter forms a hot accretion disk that emits X-rays \cite{ref7}. The typical peak X-ray flux of UCXBs ranges from ${10}^{-12}$ to $ \rm {10}^{-8}\, ergs\, s^{-1}\,cm^{-2}$, and only a few candidates have a peak flux of the order of $\rm {10}^{-13}\, ergs\, s^{-1}\,cm^{-2}$ \cite{ref4}. Furthermore, how chemical composition affects the formation rate and the evolutionary process is an active topic \cite{ref9,ref10}. X-ray  observations with XMM-Newton and Chandra have revealed emission lines of Fe or other elements in the spectra \cite{ref10,ref11}. 

It is therefore important to enlarge the UCXB sample by finding new UCXBs and verifying true UCXBs from the currently known candidates. EP is a good facility for such studies. 
On the one hand, the WXT has the potential to detect more candidate UCXBs. On the other hand, 
the FXT can be used to explore the known UCXBs and to observe extensively the candidates to verify their nature, as well as to study the chemical composition of the donor star. 

\subsubsection{Population studies and detection of new XRBs}

A probe into the distribution of sources in luminosity and their evolutionary stages spreading over a wide range of astronomical timescales ($\sim$Myr to Gyr) can be achieved by population studies. EP is an ideal facility for the population studies of XRBs, which contribute the bulk of the X-ray emission of our Galaxy. Population studies can shed light on the origin and evolution of XRBs, and help understand how their properties vary in a range of environments, from actively star-forming galaxies to older stellar systems with different metallicities. Furthermore, the monitoring of nearby galaxies has uncovered unique objects such as ULXs, exotic transients like quasi-periodic eruptions, and stellar mass BHs with higher masses than those found in our Galaxy \cite{silver08}. 
HMXBs, which comprise massive donor stars, provide a calibration of the star formation rate. LMXBs, on the other hand, are shaped by the cumulative effect of the star formation episodes experienced by the host galaxy throughout its lifetime, and is proportional to its total stellar mass (see \cite{grimm2002}).

The non-uniform spatial distribution of XRBs, coupled with the non-uniform X-ray absorption and unknown distances, make the population study of XRBs difficult within the Milky Way galaxy. 
Nevertheless, the WXT, with its improved sensitivity and spatial resolution, enables a census of the Galactic XRB population and their activities at much lower X-ray flux levels than those achieved by the previous and other operating X-ray all-sky monitors. 
In fact, the neighboring galaxies in the Local Group provide even more ideal sites for the population studies, as well as the detection of new transient XRBs.
The Magellanic Clouds (MCs), the nearest star-forming galaxies, host a large population of BeHMXBs that go into outbursts either at the periastron passage (type I) or during large outbursts (type II) associated with disk instability or warping where the luminosities can reach about the Eddington limit \cite{haberl2016}. 
M31, with global properties similar to our Galaxy, hosts interesting classes of objects, such as super-soft sources including post outburst optical novae, and BHXRBs (e.g.  \cite{barnard2014,orio2017}). In particular, regions extending some tens of arcmin away from the center of M31, have not been properly monitored by the previous and current X-ray missions.
The situation is similar for other galaxies in the Local Group.

EP is instrumental in studying the population and evolution of variable sources and transients in X-ray in the nearby Universe.
The large Grasp of the WXT enables the monitoring of the entire MCs and nearby galaxies with an unprecedented combination of the sampling cadence and sensitivity during its all-sky surveys. 
Moreover, the FXT, with a relatively large FoV of 1 squire degree (among Wolter-I telescopes), high  sensitivity and spatial resolution, is capable of monitoring the entire, or a large part of, galaxies in the neighboring universe. 
During sky surveys with WXT, the FXT points to targets selected from a sample of nearby galaxies within a few tens of Mpc, which are monitored at a range of cadences from a week to months.   
The outskirts (i.e. some tens of arcmin away from the center) of M31 can also be monitored.
Hence, with both the WXT and FXT, the variability of many known X-ray sources will be monitored and studied in detail over long time scales; new transients with (relativly) high X-ray luminosities are also expected to be detected, since the brightest part of the XRB population is above the sensitivity of WXT/FXT.
Any outbursts or new sources found will further be followed up quickly with the FXT, and will be studied and classified by triggering multi-wavelength follow-up monitoring campaigns.
We expect a significant fraction of these transients to be outbursts or flares originating from compact stellar objects in binaries, WD, NS or BH, which may be revealed from their temporal and spectral properties.

\subsection{Intermediate-mass black holes}
\label{sec:imbhs}

IMBHs (see Section\,\ref{sec:IMBH-TDE}) are of particular interest in the BH astrophysics research today. The outskirts of early-type galaxies are an ideal hunting ground for IMBHs.
This is because partly of their assembly history, which involves many mergers with satellite dwarfs, and partly of less dust obscuration and less contamination by other bright sources of X-ray emission typical of star-forming regions (i.e. stellar-mass super-Eddington ULXs, supernova remnants). Moreover, the location in a low-density stellar halo makes it easier to resolve and identify multiwaveband counterparts. Monitoring early-type galaxies in clusters and fossil groups will be a key strategy for such a search.

The (very few) strong IMBH candidates reported to-date are indeed transient X-ray sources in old stellar halos (perhaps inside a leftover star cluster), reaching luminosities above $\approx$10$^{42}$ ergs s$^{-1}$.
They include, for example, HLX-1 near the S0 galaxy ESO 243-49 in the Phoenix Cluster \cite{farrell09} and J2150$-$0551 inside a globular cluster at the edge of another S0 galaxy \cite{lin18}. 
The serendipitous nature of those initial discoveries (later confirmed by systematic follow-up multiband studies), and the occasional reported detections of other unidentified, ultraluminous X-ray flares near early-type galaxies \cite{irwin16,peng19} suggest that the IMBHs found are likely the tip of the iceberg. 
There is a great prospective for the discovery of new transient IMBHs with EP. The good sensitivity of the WXT in the soft X-ray band is particularly suited for the discovery of IMBHs in the disk-dominated, thermal high/soft state ($L_{\rm X} \approx 0.03$-1 $L_{\rm Edd}$). A $\sim$10$^{4}$ $M_{\odot}$ IMBH with a luminosity of $10^{42}$ ergs s$^{-1}$ will likely have a soft X-ray spectrum, which can be detected with the WXT at $\sim$5$\,\sigma$ out to a distance of $\sim$50\,Mpc with an exposure of 10\,ks.

\subsection{Cataclysmic Variables (CVs)}
\label{sec:cv}
Accreting WDs are typical soft X-ray sources and have been traditionally discovered in various past X-ray surveys. The survey strategy and the high sensitivity of the WXT are instrumental for systematic searches for variable accreting WDs, increasing the number of known accreting WDs of all classes. The high-sensitivity time-domain survey of the WXT  would yield an invaluable dataset about the long-term variability behaviors of these objects with higher cadence than previous missions. In addition, accreting WDs could be discovered serendipitously during FXT observations, similar to the case of other focusing X-ray telescopes such as XMM-Newton \cite{rosen16} and Swift \cite{evans14}. With a larger FoV and higher effective area, the FXT is expected to discover significantly more accreting WDs than, for example, Swift/XRT. Moreover, the quick response and real-time transient search in FXT observation fields make it promising to lead multi-wavelength studies. Accreting WD binaries have been traditionally widely studied in the optical and UV bands. Thus, the study of such sources provide good opportunities for synergies between EP and various optical observing facilities.

\subsubsection{Novae}
Novae are transient outbursts of accreting WDs (see, e.g.  \cite{bode_classical_2008} and \cite{chomiuk_new_2021} for reviews). The WD in the cataclysmic variable system accretes from its companion. The accreted matter accumulates on the surface of the WD until a thermonuclear runaway is triggered if the accreted material is hydrogen-rich and the mass accretion rate is below a critical rate. The thermonuclear runaway leads to the expansion and subsequent ejection of the envelope, resulting in an outburst lasting for years whose emission covers a broad electromagnetic band from radio to $\gamma-$ray. X-ray emission has been detected in different phases of the nova outburst. It has long been predicted that a short-duration (a few hours long) soft X-ray flash in the very early ``fireball'' phase of the nova outburst occurs when the expanding envelope reaches the WD photosphere, before the emergence of the optical emission (e.g.  \cite{starrfield_soft_1990} and \cite{hillman_nova_2014}). But it was not until 2021 that such an X-ray flash was detected \cite{konig_x-ray_2022}, and this remains to be the only case. Due to lack of observations, theoretical studies of the fireball phase are still quite limited.

X-ray monitoring of novae also revealed long-term hard X-ray radiation that peaked after the optical maximum \cite{mukai_novae_2008}. This hard X-ray component can be fit using an optically-thin plasma model with a temperature of $1-10 ~\rm keV$, and has an X-ray luminosity in the range of $\sim 10^{33}-10^{35}~\rm ergs~s^{-1}$. This emission is believed to originate from the shock due to the ejecta interacting with the wind from the companion or due to slower ejecta colliding with fast ejecta \cite{sokoloski_flows_2017}. The detections of radio and $\gamma-$ray emission \cite{abdo_gamma-ray_2010,ackermann_fermi_2014} from novae confirm the existence of shocks in novae. 

X-ray observations also revealed supersoft (a few tens of eV) emission from novae that is presumably emitted by the nuclear burning of hydrogen on the WD surface \cite{ness_chandra_2003,nelson_x-ray_2008,ness_xmm-newton_2011}. This supersoft component is obscured by the ejecta during the initial stage of the outburst, and becomes visible to the observer after the ejecta becomes optically thin. 

The WXT is an efficient instrument to detect the X-ray flash from the fireball phase, given its large FoV and soft X-ray coverage. Based on the FoV and observing efficiency of EP and the Galactic nova rate, we roughly estimate that the WXT will detect 1-3 fireball phase flashes of Galactic novae. Thus EP will greatly increase the sample of nova fireballs. Furthermore, the X-ray luminosity of the hard X-ray component, i.e., $\sim 10^{33}-10^{35}~\rm ergs~s^{-1}$ at soft X-rays, is above the FXT sensitivity for Galactic novae. FXT monitoring observations therefore help to understand the particle acceleration processes in novae. Finally, new insights on the nature of the supersoft novae emission will also be provided by the FXT. Taking a typical value of $60~\rm eV$ for the effective temperature \cite{wolf_hydrogen_2013}, the supersoft emission (at $\sim$ the Eddington luminosity) of a Galactic nova at $10~\rm kpc$ will have a flux of $\sim 10^{-12}~\rm ergs~cm^{-2}~s^{-1}$, well above the FXT sensitivity.

\subsubsection{Polars}
Polars, also known as AM Herculis stars, are magnetic CVs where the strong magnetic field (10--230 MG) of the WD synchronizes its spin period with the binary orbit \cite{warn95}. Accretion occurs directly along the magnetic field lines without an intervening accretion disk. The X-ray emission from polars arises from the stand-off accretion shock near the WD surface \cite{knigge2011}. The post-shock plasma cools via thermal bremsstrahlung and cyclotron emission, producing a hard X-ray spectrum below 10 keV \cite{cropperPolars1990}.

There are approximately 100 known and suspected polars, detected in the hard X-ray surveys such as Swift/BAT and INTEGRAL. 
Notably, three of the four known asynchronous polars (BY Cam, V1432 Aql, and CD Ind) are included among the 14 BAT-detected systems. An unresolved question is: are all asynchronous polars the result of recent nova eruptions, or is there an alternative pathway to asynchronism \cite{mukai17}? Interpreting the complex light curves of polars poses a challenge. While multiple emission peaks may suggest distinct emission regions resulting from a complex magnetic field geometry \cite{masonXrayEmissionCataclysmic1985}, it is important to note that the presence of multiple regions alone does not necessarily imply field complexity, as is supported by observations of two polar systems, QQ Vul and V834 Cen, made with the EXOSAT satellite \cite{osborne1987}. Initially, both systems exhibited complex light curves, but subsequent observations revealed drastically different curves \cite{ambrunaTransitionXRayLight1994}. Since the magnetic field configuration of a polar should not change within a few years, the changing light curves observed in these systems can be attributed to other physical parameters, such as variations in accretion rate, rather than changes in the magnetic field geometry.

Mechanisms causing asynchronism, and the cause of soft X-ray variations require further in-depth studies. The WXT is well-suited for long-term monitoring studies of known polars, as well as to expand the sample size of polars. Due to their synchronized rotation, polars display strictly periodic X-ray modulations at the orbital period from viewing angle-dependent absorption and emission. WXT could systematically monitor a large sample of polars to uncover long-term variability, outbursts and changes in their periodic signals. Follow-up observations of polars discovered in WXT surveys or exhibiting interesting behavior could be performed with the FXT. The broad energy coverage of the FXT in the 0.3-10 keV band allows for a detailed analysis of X-ray light curves across soft and hard bands, enabling a more precise determination of whether the changing light curves result from evolving magnetic fields or variations in physical parameters.

\subsubsection{Intermediate Polars}
In Intermediate Polars (IPs, also called the DQ Her-type stars), the magnetic fields are not strong enough to synchronize the WD spin with the binary orbit, and accretion occurs via a truncated disk. IPs are commonly considered to be more luminous than polars in the hard X-rays \cite{mukai17}, so that only broadband X-ray spectral and timing analyses can help gain insights into the magnetic field strength and the WD mass in this class \cite{verme23}. IPs are mostly below the current detection limit of all-sky X-ray monitors such as MAXI and Swift/BAT. However, they are known to have a rich variety of behaviours in their X-ray light curves on a wide range of time scales, including outbursts, QPOs, eclipses and dips, orbital and spin modulations \cite{warn95}. 

In addition to traditional CVs, AM CVn objects, in which the WD accretes from a degenerate hydrogen-deficient companion star, are receiving increasing attention during recent years as strong low-frequency Galactic GW sources. They have compact orbits with typical orbital periods between 5 and 65 minutes. 
Some members of the known population are weak X-ray emitters, and currently the two with shortest periods, V407 Vul \cite{motch15, ramsay00} and HM Cnc \cite{israel99} were both discovered in the past ROSAT survey. AM CVn objects are still a relatively poorly studied class of accreting WDs, thus increasing the number of X-ray detected AM CVn objects to form a systematic sample would be vital for any further studies aiming to investigate their physical properties \cite{solh10}.

EP provides the opportunity to discover new candidate IPs, either via monitoring with the WXT or serendipitous detections by the FXT. Joint observations of EP with satellites with advanced capabilities in the hard X-ray band, such as NuSTAR and HXMT-Insight, will be instrumental to confirm their IP nature so as to gain a broad X-ray band perspective. EP enables systematic, long-term, high-cadence monitoring of a substantial number of these sources in the future. Such a detailed monitoring has not been previously achieved and is important, as it provides observational insights (duty cycle, variations of the orbital and rotation period, etc.) into the physical properties of these objects. Furthermore, by combining the X-ray survey of the WXT with followup observations by the FXT, it is possible to (i) detect more and more AM CVn objects during their outburst phases, (ii) search for modulations in their X-ray light curves and (iii) analyse their energy spectra or determine the nature of these sources.

\subsection{Isolated Neutron Stars}
\label{sec:ins}
The isolated NS population encompasses very diversely emitting classes of objects powered by both thermal and non-thermal processes, shining from radio to $\gamma$-rays. Each NS class offers unique insights into the extreme physics of these objects and their interactions with their environments. 

\subsubsection{Rotation-powered Pulsars}
The most common class of NSs are the radio pulsars. The energy reservoir of all those pulsars is well established to be their rapid rotation. A key ingredient to activate the radio emission is the acceleration of charged particles, which are extracted from the star’s surface by an electrical voltage gap \cite{Goldreich1969, Ruderman1975}. All isolated pulsar rotation periods increase with time, implying a decay in their rotation frequencies. The evolution of pulsars necessarily goes from being born with fast rotation and high magnetic fields to ending their lives with a slower rotation and a decayed magnetic field. The exact trajectory and timescale are different from pulsar to pulsar, depending on their birth properties: in particular it depends on their spin-down energy loss, as well as the magnetic field strength at birth and its evolution in time. Besides the canonical radio pulsars, other types of NSs have also been discovered in the past couple of decades, with their observational properties that, until few years ago, were thought to be at variance with those of the rotation-powered pulsars. 

Young isolated pulsars are very bright X-ray emitters, usually steady in flux (but see also \cite{Wilson-Hodge2011}). However, phase-rotation variabilities and timing anomalies have been observed mainly in the radio band. EP is able to monitor the timing stability of the brightest X-ray pulsars, detecting glitches and possible flux changes either in terms of flickering (as observed in the Crab pulsar) or possible X-ray bursts for the most magnetic ones, in line with magnetars.

\subsubsection{Magnetars}
Among the most puzzling manifestations of NSs are the magnetars, the strongest magnetized pulsars. Discovered in the 70s due to their emission of powerful flares from their intense magnetic energy, they were initially misidentified as GRBs. Subsequently, they were recognized as powerful steady high-energy emitters ($L\sim10^{34-36}$ ergs\,s$^{-1}$), with rotational periods ranging between 0.3-12s despite their young age. Their dipolar magnetic fields, inferred from spin-down energy loss, are estimated to be on the order of $\sim10^{14-15}$ Gauss. The known magnetar population includes about 30 X-ray pulsars whose emission is very hardly explained by the common scenario for pulsars. In fact, the very strong steady X-ray emission of these objects is too high and variable to be fed by the rotational energy alone (as for radio pulsars), and there is no evidence suggesting accretion from a companion star. Their strong magnetic fields are believed to form either through dynamo action in a rapidly rotating proto-NS, magneto-rotational instabilities within the highly convective stellar core, or as fossil fields remnants from a highly magnetic massive star ($\sim$1k\,Gauss). 

Given these strong magnetic fields, the emission of magnetars is thought to be powered by the decay and the instability of their strong fields \cite{Duncan1992, Thompson1993, Thompson2002}. This powerful X-ray output is usually well modelled by a thermal emission from the NS hot surface (about $3-5\times 10^{6}$\,K) reprocessed in a twisted magnetosphere through resonant cyclotron scattering \cite{Thompson2002, Nobili2008, Rea2008}, a process favored only under these extreme magnetic conditions. On top of their persistent X-ray emission, magnetars emit very peculiar bursts and outbursts on several timescales (from a fraction of seconds to years; see \cite{Rea2011} and \cite{Turolla2015} for recent reviews) emitting large amount of energy ($10^{40}-10^{46}$\,ergs; the most energetic Galactic events after the supernova explosions). These bursts are caused by large-scale rearrangements of their twisted magnetic field, either accompanied or triggered by fractures of the NS crust (similar to stellar quakes). 

\begin{figure}[H]
\centering
\includegraphics[width=0.48\textwidth]{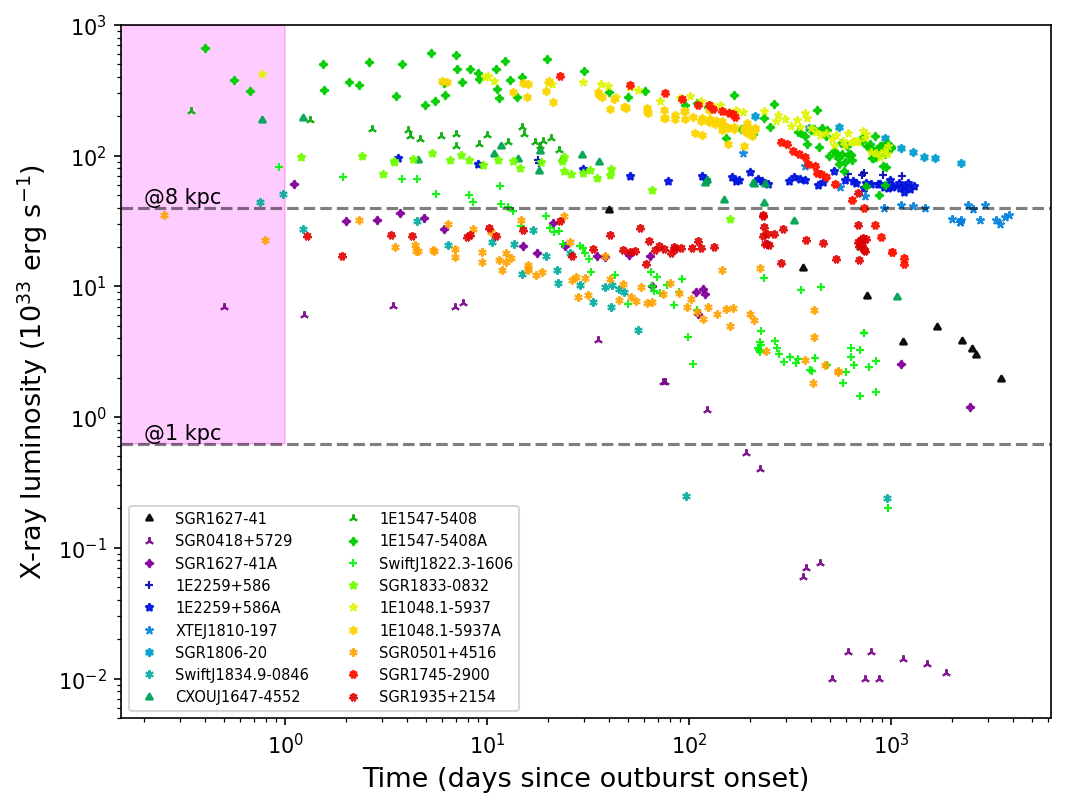}
\caption{Evolution of the X-ray luminosity for all magnetar outbursts observed from 1997 to 2018 (see the Magnetar Outburst Online Catalog at \url{http://magnetars.ice.csic.es/}). The two horizontal gray lines indicate the anticipated X-ray luminosity sensitivity that can be achieved with the WXT when observing for one day, assuming a distance for the magnetar of either 1\,kpc or 8\,kpc (i.e., at the Galactic Centre). The magenta shaded area marks the region corresponding to the earliest outburst phases that is expected to be populated by the WXT. Credit: F. Coti Zelati.}
\label{fig:magnetar_outbursts}
\end{figure}

EP will be a valuable asset in discovering and characterizing new magnetar outbursts. It is expected that we might identify approximately one new outburst per year during EP's lifetime, provided these episodes exhibit similar X-ray fluxes at their peak as previous events. However, this mostly likely represents only a lower limit to the anticipated outburst rate: EP will also explore new regimes by detecting faint and subtle flux enhancements from magnetars (see Figure\,\ref{fig:magnetar_outbursts}), as well as those outbursts that are not necessarily accompanied by short bursts. This capability is vital as it allows for the study of outlier events.

EP will also improve our ability to measure the rise timescale of magnetar outbursts, a critical factor that, so far, has only been constrained to less than four days in the case of one event, the 2006 outburst from CXOU\,J164710.2--455216 \cite{Muno2007}. This will not only provide better constraints of the actual peak luminosity and the energy and decay timescales of outbursts, but it also offers an opportunity to validate and refine models by comparing observed rise times with those anticipated by these models (e.g.  \cite{Pons2012, DeGrandis2022}). Ultimately, constraining the rise timescale will be paramount for advancing our comprehension of the triggering mechanisms of magnetar outbursts and how energy propagates from the crust to the surface layers, shedding light on the structure of NSs themselves.

EP observations will also enable the study and validation of correlations between various parameters (e.g.  X-ray luminosity at the outburst peak, energy released and decay timescale, magnetic field strength and characteristic age) across a broader range of values than previously possible (see \cite{CotiZelati2018}). This is again crucial to test the predictions of models for magnetar outbursts. For instance, the internal crustal heating model predicts the existence of an anticorrelation between the quiescent luminosity level and the maximum increase in luminosity in magnetar outbursts. According to this model, once the energy injected into a magnetar exceeds approximately 10$^{43}$\,erg, the photon luminosity emitted from the surface reaches a maximum of around 10$^{36}$ ergs\,s$^{-1}$, since the crust becomes so hot that it emits most of the energy in the form of neutrinos \cite{Pons2012}. While this anticorrelation has indeed been observed in the sample of outbursts studied so far \cite{CotiZelati2018}, investigating it across the broadest possible range of parameter values is vital for better understanding the neutrino emissivity's self-regulating effects and the overall mechanisms at work in the interior of a magnetar.

EP is set to advance our understanding of magnetars by not only detecting more outbursts from these sources, but also by simultaneously tracking the evolution of their timing and spectral properties (at least for the brighter and closer sources). Recent studies have revealed that magnetars can alternate between distinct, stable states and that their thermal emission may be largely driven by the dissipation of magnetospheric currents (e.g.  \cite{CotiZelati2020}). Hence, there is growing evidence suggesting that magnetar emission can show remarkable variability even outside of outburst periods. Besides, a few magnetars were observed to exhibit timing anomalies that did not necessarily coincide with changes in their emitted radiation, suggesting that such anomalies may originate within the NS interior (e.g.  \cite{Younes2020}). However, the lack of high-cadence, targeted observations has so far precluded a thorough investigation of these flux changes and timing anomalies. 

EP will systematically detect and monitor both the flux changes over a wide range of timescales as well as timing anomalies, possibly linking them to changes in radiative properties. Moreover, EP enables us to measure or tightly constrain the spin-down rate of these sources during phases when the torques imparted by magnetospheric reconfigurations are substantially weaker than those during outbursts. Obtaining precise measurements of such “secular” spin-down rate is vital to better constrain key parameters such as the strength of the dipolar magnetic field at the star surface as well as the characteristic age and the spin-down luminosity of these stars. The determination of these parameters, in turn, is crucial to inform magneto-thermal evolution models that predict how magnetars cool over time (e.g.  \cite{Vigano2013, Dehman2023}).

\subsubsection{X-ray Dim Isolated Neutron Stars}
The X-ray Dim Isolated NSs (XDINSs) are a small group of seven nearby (hundreds of parsecs), thermally emitting, isolated NSs. They are radio quiet while relatively bright in the X-ray band, and five out of the seven have spin periods similar to the magnetars. Moreover, their magnetic fields seem to be comparable to the electron critical field. Unlike most X-ray emitting pulsars, XDINSs display X-ray spectra that closely resemble perfect black bodies, typically with an effective  temperature of $kT\sim0.1$\,keV, and sometimes include broad lines. This characteristic makes them key objects for a large branch of research on the NS equation of state, and the cooling properties of the NS crust and core. 

EP enables long-term monitoring of XDINSs and their timing and spectral behaviour. Long-term variations of XDINSs are expected given their high magnetic fields. In particular, in the scenario of XDINSs being old magnetars, we can expect an X-ray outburst to occur, albeit at a significantly lower rate. The WXT daily monitoring has the potential to detect such magnetar-like outbursts for the first time.

\subsubsection{Central Compact Objects}
The group of Central Compact Objects (CCOs) consists of radio quiet thermally emitting X-ray sources ($kT$$\sim$0.2--0.5\,keV) located near the centre of luminous shell-like SNRs. The association with SNRs implies ages of a few tens of kyrs at most. Three CCOs show X-ray pulsations with periods in the 0.1–0.4s range, with small spin-down rates. The rotational energy loss of the three pulsed CCOs is certainly too small to give a detectable contribution to their observed X-rays. Their extremely low inferred dipolar magnetic field at the surface ($B\sim10^{10-11}$\,G) is puzzling given their young ages and high surface temperatures. Furthermore, the spin-down characteristic ages of these three CCOs exceed the ages of their associated SNRs by orders of magnitude, indicating that they were either born with spin periods and magnetic fields very close to their current values, or underwent strong accretion during the fallback phase soon after the supernova explosion. This accretion might have buried a strong magnetic field within their crust. This scenario is further invoked to explain the large spin-modulated X-ray emission (i.e., $\sim$60\% in the SNR Kes\,79) and the presence of bright, small (only a few kilometers in radius) thermally emitting hot spots in most of these objects, which are challenging to reconcile with a scenario involving a weakly magnetised NS.

By using the FXT window mode (with a readout period of 2ms), young pulsars, including CCOs, can be searched in SNRs.

\subsubsection{Pulsar Wind Nebulae around pulsars}
Pulsars release their rotational energy through relativistic winds, which generate bright pulsar wind nebulae (PWNe). These nebulae can be observed across the electromagnetic spectrum, emitting synchrotron and inverse Compton radiation, as well as optical emission lines when they interact with the surrounding medium. These phenomena provide valuable insights into relativistic shocks, particle acceleration, and the properties of interstellar gas \cite{2006ARA&A..44...17G}. We gained insight into PWNs thanks to the unprecedented high-resolution X-ray imaging capabilities of Chandra \cite{2000SPIE.4012....2W}. Recently, the space-resolved polarized results of the PWN revealed that the magnetic field is highly uniform \cite{2022Natur.612..658X,2023NatAs...7..602B}. Furthermore, PWNe are important objects of TeV-PeV sources \cite{2021Natur.594...33C,cao23}.

The Crab Nebula is the protypical PWN powered by the energetic Crab pulsar PSR B0531+21
($\dot E=4.5\times10^{38}$\,ergs\,s$^{-1}$). It is among the brightest objects in the X-ray and $\gamma$-ray sky, and has been used as the standard candle for the calibration of numerous 
high-energy satellite instruments. Strong flares in the GeV band have been observed with AGILE and Fermi/LAT \cite{2011Sci...331..736T,2011Sci...331..739A}. A search in 11 yr of Fermi data revealed 17 significant flares, implying an average rate of 1.5 per year \cite{2021ApJ...908...65H}.
The flares have a short rise time of $\sim$1\,day and last for a few days to a few weeks, with an integrated $\gamma$-ray flux above 100\,MeV increased by a few to a few ten of times compared
to the quiescent value. While the flares are believed to be synchrotron emission from particles accelerated in the inner PWN, the exact mechanism producing such high-energy emission is still unclear. It could be due to Doppler boosting or magnetic reconnection \cite{2012ApJ...749...26B}.
Short Chandra X-ray observations during the $\gamma$-ray flares did not find any obvious X-ray variability \cite{2013ApJ...765...56W}. The good sensitivity of EP allows for much deeper searches in future, for the Crab Nebula and other PWNe that show flares.

The young pulsar PSR B0540-69 is the first to exhibit a brightening of its PWN following a change in its spin-down rate. This provides a crucial insight into the relationship between the central engine and variations in the PWN \cite{2019NatAs...3.1122G}. So, the connection between young pulsars and its surrounding PWN is important to study the braking mechanism of pulsars as studied by \cite{2019NatAs...3.1122G,2020ApJ...900L...7G}. It is suitable to monitor the spin evolution of the pulsar and the flux of PWN considering the sensitivity of the FXT.

\section{X-ray flaring stars}
\label{sec:flarestar}

Since the era of Einstein Observatory, it has been established that stars of virtually all spectral types are X-ray emitters \cite{Rosner1985}, and mostly stronger than the Sun. 
The variability in the soft X-ray sky is dominated by stellar flares, mainly from late-type stars in the vicinity of the solar system.  
The study of flaring stars has been an active interdisciplinary field of stellar physics and solar physics. 
Previous studies of stellar X-ray flares were based mostly on heterogeneous samples acquired by serendipitous detections with small-FoV telescopes (e.g. \cite{Pye2015}).  
The previous and the other operating wide-field monitors, on the other hand, having hard X-ray passband and/or limited sensitivity (e.g. Swift/BAT, MAXI), are biased to detect extreme or the nearest flaring events \cite{Osten2016, Tsuboi2016}. 
Hence there has been a long lack of systematic, sensitive surveys of stellar flares in the soft X-ray band, in which the spectral peaks of the majority of the flares lie. 
With a combination of the good sensitivity and large FoV of the WXT, as well as the quick follow-up capability of the FXT, EP performs an unprecedented survey of stellar flares and is expected to revolutionize the reserch in this field.
This section discusses about X-ray flares from late-type stars, particularly in light of their potential influence on exoplanets that orbiting them, and coronal activities from contact binary systems.


\subsection{Flares and coronal mass ejections on late-type stars}

Late-type stars are known for their high-level magnetic activities and associated energetic coronal eruptions. Such magnetically driven explosive phenomena manefest themselves in the form of stellar flares, coronal mass ejections (CMEs) and energetic particles.
Understanding the nature and origin of coronal eruptions on late-type stars is an active topic of current research in stellar physics.
This understanding also provides an indispensable basis for the assessment of the impact of stellar magnetic activity on the habitability of exoplanets around the host stars, on which our current knowledge is very limited. 
In the subsection, the prospects of EP observations in the research on stellar flares, CMEs and their potential impact on exoplanets are discussed, respectively. 

\subsubsection{X-ray flares on late-type stars}
Observations in X-ray, UV and optical bands show that late-type main-sequence stars can produce stellar superflares with an energy level of 10$^{33}$ ergs to 10$^{38}$ ergs, which is 10-10000 larger than that of the strongest flare ever observed on the Sun \cite{2012Natur.485..478M}. These high-energy stellar flares can produce excess electromagnetic radiation and energetic particles, which may gradually erode and alter the atmospheres of close-in planets around host stars \cite{2007AsBio...7..167K}. 
At present, stellar flares on late-type stars are speculated to be analogs of solar flares, although key processes of stellar flares occurring in the hot stellar coronae are far from well understood \cite{2010SSRv..157..211G,2022ApJ...933...92C}.

With an unprecedented combination of sensitivity, cadence and soft X-ray passband, the survey of the EP-WXT will detect X-ray flares from a large number of late-type stars.
Moreover, follow-up and ToO observations of the EP-FXT will enable detailed X-ray observations of high-temperature emission of stellar flares on late-type stars. By comparing soft X-ray data with the EP-WXT and harder X-ray observations with FXT and/or other X-ray telescopes, one can study the ``Neupert effect" (the observed temporal correlation of the hard X-ray flux with the time derivative of the soft X-ray flux in solar flares) and the underlying flare energy dissipation process in stellar flares in detail \cite{1996ApJ...471.1002G}. Moreover, joint observations of EP and TESS will offer a comprehensive understanding of the emission nature of stellar flares across multiple wavelengths. This approach will not only provide insights into the enigmatic origin of the long-lived white-light emission often detected in optical stellar flares, but also shed new light on the poorly understood processes of chromospheric evaporation and condensation in stellar post-flare loops \cite{2022ApJ...933...92C,2022ApJ...928..180W}. In addition, quasi-periodic pulsations (QPPs) are commonly observed in solar and stellar flares during X-ray observations \cite{2020STP.....6a...3K,2021SSRv..217...66Z}. These phenomena are believed to be closely associated with the fine evolutionary nature of astrophysical energy release processes, including possible modulation of MHD oscillations (e.g.  \cite{2016ApJ...823L..16T,2023SoPh..298...40K}) or intermittent magnetic reconnection (e.g.  \cite{2016ApJ...823..150T}). With X-ray observations by the WXT and FXT, investigation of QPPs is enabled for a large sample of stellar flares, which would shed light on the origin of QPPs.

\subsubsection{Stellar coronal mass ejections} 
Coronal mass ejections are characterized as spectacular expulsions of magnetized plasma from late-type stars into the heliosphere/astrospheres, which can directly affect the atmospheric and magnetic environments of orbiting planets.
The known relationship between flares and CMEs on the Sun \cite{2009IAUS..257..233Y,2011SoPh..268..195A} suggests that superflares on late-type stars are likely to be accompanied by violent CMEs reaching energy levels up to 10$^{33}$--10$^{38}$ ergs. If such violent CMEs frequently occur on a host star, they would undoubtedly become one of the decisive factors that affect the habitability of its orbiting planets \cite{2007AsBio...7..167K,2020IJAsB..19..136A,2022SerAJ.205....1L,2023huitian}. Meanwhile, stellar CMEs can significantly contribute to the loss of the stellar mass and angular momentum, which is important for stellar evolution \cite{2013ApJ...764..170D}. 

On the Sun, CMEs are routinely detected from coronagraph images. 
However, this approach is currently impossible for other stars due to the impossibility to spatially resolve the extended stellar coronae \cite{2019ApJ...877..105M,2022SerAJ.205....1L}. Many authors have attempted to detect stellar CMEs mainly through the analysis of plasma motions revealed by Doppler shifts in time-resolved spectroscopy (e.g.  \cite{2022ApJ...928..180W,2022ApJ...933...92C}), but only a handful of stellar CME candidates have been detected using this technique \cite{1990A&A...238..249H,2016A&A...590A..11V,2019NatAs...3..742A,2022ApJ...933...92C,2022NatAs...6..241N,2022A&A...663A.140L}. Instead, recent research in Solar physics has revealed that the CME-induced coronal dimming, which refers to transient darkening of coronal emission on the solar disk caused by the mass loss of coronal plasma \cite{2001ApJ...561L.215H,2012ApJ...748..106T}, can be used to differentiate powerful solar flares with CMEs from those without \cite{2016SoPh..291.1761H}. Interestingly, the dimming can be observed as a noticeable decrease lasting for seven hours on average EUV spectral line intensities obtained with sun-as-a-star spectral lines (around 0.5$\%$-6$\%$)  \cite{2019ApJS..244...13M,2022ApJ...931...76X}. 

For more active late-type stars, similar coronal dimming signals may also appear in shorter wavelength soft X-ray spectral lines \cite{2020IAUS..354..426J}. 
Recently, searches for coronal dimming signals associated with stellar flares by utilizing the observational data at the extreme ultraviolet and soft X-ray wavelengths from the Chandra, XMM-Newton, and Extreme Ultraviolet Explorer (EUVE) satellites have been performed \cite{2021NatAs...5..697V}. In total, 21 possible coronal dimming events that could be attributed to CMEs on 13 stars were identified. Generally the soft X-ray flux during the coronal dimming phase exhibits a significant decrease of more than 30$\%$ compared to the pre-flare flux level.
The coronal dimming signal thus can be used as a novel and promising proxy for detecting stellar CMEs in X-ray observations \cite{2020IAUS..354..426J,2023huitian}.

Furthermore, when CME is propagating towards the line of sight of an observer, its internal material (particularly the cooler, partially ionized plasma) can temporarily obstruct the radiation emitted from the stellar atmosphere or even the flaring region \cite{2023huitian}. This results in photoelectric absorption of the continuum spectrum radiated by the star.  
Possible signatures of such absorption may have been found in the X-ray spectra of stars during giant flares \cite{1999A&A...350..900F}, which are considered as an indirect evidence for stellar CMEs (e.g.  \cite{2017ApJ...850..191M,2023RAA....23i5019W}). 
Combining the WXT survey with follow-up/ToO observations with the FXT, EP is well suited for hunting the X-ray signals of mass-loss-induced coronal dimming and the X-ray absorption during stellar flares, thereby offering an unprecedented opportunity to find and characterize CMEs from stars.


\subsubsection{Effects of stellar flares on exoplanets}

Exoplanet science is beginning a new phase that will focus on defining exoplanet atmospheres and discovering possibly habitable worlds.
Potentially habitable exoplanets are believed to mostly exist within the habitable zones of G-K-M dwarf stars, especially M-type dwarfs \cite{2019AJ....157..113B,2019AJ....158...75H}.  
Stars and the planets that orbit around them engage in intricate interactions driven by various physical phenomena \cite{Poppenhaeger2019}.
Giant flares and CMEs on these stars cause high-level X-ray and extreme ultraviolet radiations (X-EUV) as well as intense particle fluxes, which will have profound effect on the nature of global chemical and climatic environments of the exoplanets that are close to their host stars, and even pose a threat to the habitability of exoplanets 
\cite{Gudel2002,2007AsBio...7..167K,Segura2010,Lammer2011,2019LNP...955.....L,2020IJAsB..19..136A,2022huitian,2023huitian}. 
Thus, planetary atmospheric evolution is linked to XUV emission \cite{Watson1981, Lammer2003, Baraffe2004, Poppenhaeger2021}. 
Many studies have concluded that the XUV radiation emitted by host stars are potentially harmful to life \cite{Heath1999, Tarter2007, Lammer2009, Shields2016, Meadows2018}. 
On the other hand, the stellar XUV radiation is also essential for catalyzing prebiotic chemistry \cite{Ranjan_Sasselav2016}.

As one specific approach to study star-planet interaction, open clusters provide an ideal site to study the high-energy diversity of the host stars of exoplanets. 
In a study by \cite{Fujii2019}, it was determined that around 7\% of planets with distances ranging from 1 to 10 astronomical units (AU) are displaced due to interactions with stars in clustered environments. Such scenarios offer a valuable testing ground for refining theories about planet formation and development, especially concerning planets orbiting young stars. The stars within a given star cluster exhibit similarities in terms of their metallicity, helium abundance, and age, as they originate from the same primordial cloud. This uniform information is advantageous for refining the model of exoplanetary atmospheres. The early evolutionary path shapes the demographic composition of the planetary population in significant ways. 

With its expansive FoV, EP is particularly advantageous for observing open clusters.
The FXT has the capability to observe specific host stars of exoplanets, while the WXT simultaneously observes the remaining stars within the cluster and tracks their stellar activities, including phenomena like flares. Our anticipation is that the combined accumulation of exposure time, reaching around $10^5$ seconds, should prove adequate to attain a sensitivity level of $\rm \mathrm{10^{-12} \, ergs\, s^{-1}\, cm^{-2}}$ for the WXT.
With these observations, we hope to gain insights into the questions as to how the characteristics of stars influence the nature of the planets that emerge, and how we can use these star properties to learn more about the origin of planets.

\subsection{Coronal activities from contact binaries}

Contact binaries are commonly found in the Galaxy. A contact binary consists of two stars positioned very closely together, and both components are enclosed within their Roche lobes. These stars share a common envelope between them, giving rise to phenomena such as angular momentum loss and mass loss. To a certain extent, these physical processes can influence the evolution of contact binaries and trigger X-ray radiation \cite{2013MNRAS.430.2029Y, 2020A&A...635A..89M}.

Investigating the X-ray coronal activity of contact binary systems can offer deeper insights into stellar magnetic activity and coronal heating. In a physical model of coronal radiation, the density of radiation in the extreme ultraviolet region is as high as \(N_e = 2 \times 10^{13} \, \text{cm}^{-3}\) \cite{1998ApJ...502..918B}. Currently, numerous known W UMa binary systems have been observed as X-ray radiation sources. 57 of these sources have been detected in the ROSAT all-sky survey, with X-ray luminosities of \(\sim\) \(1 \times 10^{30} \, \text{ergs}\,\text{s}^{-1}\) \cite{2001A&A...370..157S}. This X-ray radiation is believed to be associated with chromospheric and solar corona activities. 

Studying the X-ray coronal activity of contact binaries still presents several challenges and obstacles. Due to a phenomenon known as `supersaturation', X-ray activity observed in W UMa stars is typically weaker than in other rapidly rotating stars \cite{2023AJ....166...51P}. X-ray emission was observed from W-type systems, with five out of the eight A-type systems in the sample also showing detectable X-ray flux. The measured X-ray luminosities ranged from $\sim$ \(8 \times 10^{28} \, \text{ergs}\,\text{s}^{-1}\) to $\sim$ \(2 \times 10^{30} \, \text{ergs}\,\text{s}^{-1}\) \cite{1984ApJ...277..263C}. Only a few number of contact binaries have been observed with the necessary sensitivity. Moreover, there are exceedingly few examples where orbital modulation in X-rays has been observed within the eclipsing binary systems. The X-ray-emitting contact binary system 2MASS J1120034-220134 was observed with Swift, and during the entire observation, only $\sim 30$ X-ray photons were detected \cite{2016AJ....151..170H}. The observed X-ray flux between 0.3 and 10 keV was found to be $10^{-14}$~erg~cm$^{-2}$~s$^{-1}$; significant X-ray orbital modulation could not be observed. 

Coronal activities within binary systems often manifest as irregular and swiftly fluctuating phenomena. Hence, it is important to capture the intricate time-evolving and periodic fluctuations inherent in coronal activities \cite{2022ApJ...934...20S}. 
The EP-WXT enables observations of the coronal activities in soft X-rays for a large population of stars, while the FXT can be used to monitor stellar systems at relatively high cadences to capture the above fluctuation signatures in coronal activities.

\section{Observatory Science}
\label{sec:obssci}

Although the key science objectives of EP are concerned with time-domain and multi-messenger astronomy, the mission is also well suited for the general science in X-ray astronomy. As a space X-ray observatory, EP has a wide-range of observing targets and scientific goals, from planets in our solar system to quasars and clusters of galaxies in the distant universe.  
Compared to other X-ray satellites, the unique feature of EP is the combination of two focusing telescopes, one of wide-FoV and one of relatively large light collecting power and good spatial resolution. This is further complemented with the fast two-way satellite-ground communications and rapid response of the spacecraft.
These features also render EP the potential to make unexpectedly discoveries. 
This section discusses the expected contributions of EP to several selected topics in the observatory science domain, including galaxy clusters and groups, supernova remnants and the search for possible X-ray counterparts of PeV sources as detected recently by LHAASO. 

\subsection{Clusters and groups of galaxies}
\subsubsection{Mergers and cluster outskirts}
Galaxy clusters represent the most massive relaxed systems in the universe and are built by the hierarchical merger and accretion of smaller systems (for a review see \cite{Kratsov2012}).
Most of the baryonic matter in these systems is in the form of a hot plasma atmosphere, the intracluster medium (ICM), although the majority of the matter is dark.
Due to the high temperature of the ICM (typically several $10^7$~K for clusters), the emission is in the X-ray band.
The 100s to 1000s of galaxies within the cluster exist within the atmosphere and are gravitationally bound.

The centers of a large fraction of these systems are  relaxed, with dense and cool central cores where the intracluster medium shows short mean radiative cooling times \cite{Fabian2012}.
However, around 1/3 of the systems are undergoing merging activity (e.g. \cite{Ghirardini2022}).
Around the virial radius, the material in a cluster transitions from the virialised body to an unrelaxed accreting regime.
Clusters are surrounded by substructure such as infalling groups and filaments (e.g. \cite{Reiprich2021}).

As they are X-ray bright, the central parts of clusters are relatively well characterised by existing X-ray observatories, such as Chandra and XMM-Newton.
It is in the outskirts of clusters where our knowledge of the ICM becomes much poorer, due to the declining X-ray surface brightness (the X-ray brightness is proportional to the density-squared, integrated along the line of sight).
Understanding the outskirts of clusters is critical for understanding several problems in astrophysics, such as how they grow and interact with the cosmic web \cite{WalkerLau2022}.

Merging clusters are also excellent locations to study the growth of clusters (e.g. \cite{Molnar2016}).
In these systems the non-equilibrium physics of the ICM can be observed, such as shocks and contact discontinuities (also known as cold fronts \cite{Markevitch2007}). 
In these systems we can study the impact on mergers on AGN and star formation.
It is also possible to directly observe the presence of dark matter within the system \cite{2006Clowe}.



The combined on-axis effective area of the FXT telescopes is about 600 cm$^2$  at 1\,keV, which is higher than that of Chandra at present and smaller than that of XMM-Newton (1900\,cm$^2$) and eROSITA (2200\,cm$^2$) \cite{Predehl2021}. Since EP is in a low orbit, its particle background is low (predicted to be around 5 times lower than that of eROSITA; \cite{zhangj22}). High signal-to-noise images can be obtained by combining data from repeated visits. Exposure times of 8\,ks should produce comparable data with the eROSITA final survey data (2.2\,ks).
In the harder X-ray band its sensitivity may be better than eROSITA due to its low particle background.

FXT is able to provide valuable first-hand data for the aforementioned studies related to the merging clusters, infalling groups and large-scale filament structures. Combined with the optical observations of the member galaxies, we can also construct samples of cluster member galaxies in different gravitational environments and at different stages of dynamical evolution, and study the correlation between the physical properties of the member galaxies, and the dynamical evolution of the host galaxy cluster and the local environment of the galaxy, so as to reveal the environmental effects during the evolution of the galaxies.

Nearby clusters of galaxies often exceed the observational fields of view of the XMM-Newton and Chandra telescopes. The complete images can only be obtained by mosaicing together multiple observations.
Examples include the X-ray brightest cluster of the sky, Perseus, and the nearest cluster, Virgo.
Other merging clusters like A399/A401 \cite{a399}, A3391/3395\cite{2018ApJ...858...44A, 2022A&A...661A..46V}, etc., are also not suitable for Chandra and XMM-Newton due to their large extent.
The large 1-deg wide FoV of FXT is ideal to observe these nearby galaxy clusters effectively.
eROSITA has demonstrated the advantage of a wide FoV for observing clusters and their environment (e.g. \cite{2022A&A...661A..46V,2022Sanders}).

In the outskirt of clusters, there are numerous small groups around galaxy clusters that will eventually merge with the main system. Their faint X-ray radiation can trace dark matter halos around galaxy clusters. Such X-ray substructures have been found in a number of nearby galaxy clusters, such as A2142 \cite{2014Eckert}, Coma \cite{2016Sasaki} and A780 \cite{2016Grandi}.
Based on the analysis of a much wider field of optical member galaxies and weak gravitational lensing, such infalling groups are prevalent in the radial range of 1--2 times $R_{200}$ (an overdensity of 200 of the critical value).
At these radii, it is difficult to observe nearby clusters due to the small FoV of Chandra and XMM-Newton relative to the size of the cluster.
The variable background of XMM-Newton also makes such analyses challenging.
The Suzaku satellite also carried out many observations targeting the outskirts of galaxy clusters 
due to its low particle background \cite{2009Bautz,2010Hoshino,2011Akamatsu,2012Humphrey,2020Nugent,2020Sarkar}. However, its low spatial resolution (1.8 arcmin HPD) limited such studies, as it is critical to mask background AGN to reduce the background level.
Similar to eROSITA, the FXT's has the advantage of a larger FoV than Chandra and XMM-Newton, and better spatial resolution than Suzaku.

Besides infalling groups, there are also numerous filaments connecting with clusters in the outskirt. Existing telescopes have seen some faint filament signals around some galaxy clusters, such as A85 \cite{2003Durret}, A222/223 \cite{2012Dietrich}, A1750 \cite{Bulbul2016}, A2744 \cite{2015Eckert}, A3017 \cite{2017Parekh}, A133 \cite{2018Connor}.
The FXT allows for deep observations in neighboring clusters of galaxies, revealing the distribution of material in the outskirts of the clusters.
The filamentary material around clusters is part of the so-called mission baryons (e.g. \cite{Cen1999}).
eROSITA has shown FXT-like telescopes are able to detect this filamentary emission \cite{Reiprich2021}.

Clusters of galaxies do not only emit in the X-ray band.
One important tracer of cluster astrophysics is synchrotron radiation in the radio band.
The main forms are diffuse radio halos, radio relics and mini-halos \cite{2012Feretti,2019Weeren}. They are thought to be associated with the shock and turbulence generated during merging. However, it is difficult to understand under what scenarios these structures are formed and how they can be maintained.
The X-ray morphology of clusters (e.g. \cite{Ghirardini2022}) can be used to understand the merging history and kinematic state of clusters.
By combining this information with the radio structure, we can help understand the merger history and current dynamic state of clusters using the FXT.

\subsubsection{Superclusters}
The largest building blocks of the cosmic web are galaxy superclusters, known as `clusters of clusters'. 
Despite being a important large scale structures, they do not have a commonly accepted definition (see e.g. \cite{Liu2024}).
Superclusters comprise a few to dozens of rich galaxy clusters and a large number of galaxy groups.
They are detected using several different techniques, such as the galaxy density field (e.g. \cite{Einasto2007}) or by finding friend-of-friends, either using galaxies (e.g. \cite{Sankhyayan2023} or clusters (e.g. \cite{Bohringer2021,Liu2024}).

Superclusters are important for several reasons.
Firstly, the formation and evolution of superclusters are governed by dark matter and dark energy, and they can be used to discriminate between different cosmological models \cite{Einasto2021}.
Secondly, superclusters have complex inner structures from dense clusters to low-density bridges and filaments (e.g. \cite{Reiprich2021}), they are ideal laboratories to study the physical processes influencing the evolution of member galaxies and clusters.
Thirdly, superclusters are excellent places to detect the warm-hot intergalactic medium (WHIM) locked in the filaments (e.g. \cite{Ghirardini2021}), because in these crowded systems, the intercluster medium may also be compressed, which would enhance its emission to the level accessible to current telescopes, and can reveal important clues regarding the missing baryon problem. 

Superclusters can be identified based on existing cluster catalogs like Abell clusters, or from overdensity regions based on optical, X-ray, and SZ surveys (e.g. \cite{Einasto2007,Planck2013,2022Liu,Sankhyayan2023,Liu2024}). 
The FXT, with its large FoV and sufficient sensitivity, is an excellent instrument for studying superclusters.
With FXT, we can map the superclusters in X-ray to study the distribution and properties of baryonic matter within them. As the most massive systems, superclusters act as great attractors, growing through mergers and accretion of matter from surroundings, we can study the dynamics of member clusters and groups. We can also examine the bridges and filaments between clusters, and compare the properties of member galaxies in different environments. 

\subsubsection{eROSITA follow-up}
The eROSITA X-ray telescope detected more than 10$^4$ galaxy clusters and groups in the first All-Sky Survey (eRASS1) \cite{2022Liu,2022Bulbul,Bulbul2024}.
Although 2400 eRASS1 clusters have temperature measurements, the limited depth of the eRASS survey ($\sim$ 150~s for eRASS1 and $\sim$ 600~s for eRASS:4),
means that temperatures are not measured in a substantial fraction of systems.
It is therefore necessary to perform deeper X-ray follow-up observations on a representative subsample of eRASS clusters.
Although XMM-Newton is suited for measuring temperatures of clusters, the lower particle background and large field of view of FXT make it an excellent instrument to do follow up, in particular thanks to its nearly identical response with respect to eROSITA. 

The immediate goal of following eROSITA clusters with EP can be as simple as acquiring high-significance measurements of the overall ICM temperature for a large sample, and possibly ICM temperature profiles for a smaller sample. 

The realization of this goal will provide us with a comprehensive understanding of the temperature discrepancy between the current X-ray instruments. A fundamental issue faced by X-ray astronomers is the cross-calibration of different instruments. In particular, current X-ray telescopes such as XMM-Newton, Chandra, eROSITA, and NuSTAR are found to deliver inconsistent ICM temperature measurements, where the discrepancies range from a few to $>50$\%\ (e.g. \cite{2015Schellenberger, 2022Wallbank, 2023Liu}). Such a discrepancy possibly originates from inaccurate calibration of the instruments, or the absence of multi-phase gas in the ICM, or both. Solving this issue is of great importance in the study of clusters, as the gas temperature is a fundamental property of a cluster, and the derivation of many quantities such as hydrostatic mass, relies on precise temperature measurements. In the coming years, $\sim 10^5$ new clusters and groups will be discovered in eRASS.
Using FXT in combination with other telescopes for representative samples of clusters will help us resolve the relative temperature differences due to the different instrument calibrations.
This is particularly important to use this legacy sample for further astrophysical and cosmology studies. There are several hundred clusters commonly observed by Chandra, XMM-Newton, and in eRASS. Among them, a sample of clusters covering a wide range of temperature, mass, redshift, and dynamical status will be selected, and intermediate deep observations with FXT to obtain precise temperature measurements be performed. The product of such work will help to understand and solve the temperature discrepancy. With the $T_{\rm true}-T_{\rm eRO}$ relation established, one can further perform a hydrostatic study for the clusters where we have temperature profiles with good quality, and investigate the hydrostatic bias through the comparison with weak lensing masses.

\subsection{Supernova remnants}

As important energy and chemical sources in galaxies, supernovae play a crucial role in regulating the evolution of the galactic ecosystem.
The SN is the endpoint of a massive star or WD. A SN produces a huge amount of kinetic energy and heavy metals, which are gradually transferred to the interstellar medium in the form of a SN remnant. The shock waves of young and middle-aged SNRs are typically faster than a few hundred km~s$^{-1}$, which can heat the gas to $\sim 10^6$--$10^7$~K, a temperature for radiating thermal X-ray emission \cite{vink20}. Some SNRs resulting from massive stars contain pulsar wind nebulae that generate non-thermal X-ray emission. 

Observations in the past decades have shown that X-ray measurements are among the most efficient ways to unveil the properties of SNRs \cite{vink12}. Despite decades of study, a few crucial questions in the SNR fields are still waiting to be answered with future observations: What is the population of SNRs? Are SNR and PWNe factories of cosmic rays with PeV energy? What detailed physical processes are happening at the SNR shock? EP provides a new opportunity to observe SNRs in the X-ray band and help us to study the above questions.

\subsubsection{A more complete view of SNR population}

The X-ray all-sky surveys by ROSAT and eROSITA have shown that SNRs are among the brightest extended X-ray sources. Over 300 SNRs have been found in our Galaxy, but only tens of them have been well observed with large X-ray telescopes \cite{ferrand12}.
There has still not been a complete X-ray view of the SNR population in our Galaxy. Furthermore, the observed sample is highly biased. Very extended SNRs with large angular sizes (over a degree) are typically not the preferred targets for XMM-Newton or Chandra due to the relatively small FoV of these telescopes. These large remnants may be either the nearest, old, very energetic, or evolving in the tenuous medium in the high Galactic latitude. These groups are of great importance since they are the extreme targets but have not been well studied. 

EP can enlarge the SNR sample, especially the very extended SNRs. The eROSITA all-sky survey has successfully discovered a few new, extended SNRs with $\lesssim 1$~ks exposure \cite{beker21,churazov21,khaibullin23}.
Given the long exposure of the WXT all-sky survey, there is a good chance to identify new SNRs at high Galactic latitudes.
Follow-up observations with the FXT or other X-ray instruments enables not only to constrain the SNR properties, but also to obtain the density and chemical abundances to understand the gas properties above the Galactic halo.

FXT allows to obtain a complete view of the very extended SNRs, which were detected by ROSAT but not fully covered by previous XMM-Newton/Chandra/Suzaku observations. With FXT, one can perform the first spatially resolved spectroscopic analysis of these peculiar (nearby, old, or evolving in low-density medium) remnants, and obtain their evolution stage, environment impact, and possible SN type. This will provide important insight into the SNR population in our Galaxy.

\subsubsection{Identification of the counterparts of PeV sources}

SNRs have been established as a factory of cosmic rays (CRs) \cite{ackermann13}. 
An unsolved and important question is whether SNRs are the ``main'' factories of CRs up to the PeV energy. 
To answer this question, one needs observations to test if some SNRs can generate PeV CR protons.  Identification and investigation of the counterparts of the PeV sources is a 
direct way to resolve the PeVastron factory problem.

LHAASO \cite{CaoZhen2019} recently released the first source catalog \cite{cao23}, which includes 43 sources with emissions energies above 100~TeV. 
The majority of detected Galactic TeV sources are
likely to be associated with SNRs and PWNe \cite{hess18}.
It is reasonable to explore if some of the PeV sources are SNRs or PWNe.
Unfortunately, a large part of the LHAASO sources lack X-ray observations to constrain their source types, making it difficult to uncover the origin of the very high-energy CRs.

X-ray observations are an efficient way to identify the high-energy sources. 
SNRs and PWNe are both extended X-ray sources which can be distinguished by imaging and spectral analysis.
SNRs usually emit thermal X-ray emission, although the young SNRs with fast shock velocities may contain non-thermal emission.
PWNe, a sub-type of SNRs in broad terms, are powered by pulsars and pure non-thermal sources.
X-ray analysis efficiently diagnoses the physical parameters of the sources, such as the evolution stage, energy, or density (if thermal), and identifies the PeV counterparts. 
The information is crucial for constraining the radiative mechanisms of the PeV sources, distinguishing if CRs are leptonic or hadronic, and finally, understanding the CR acceleration mechanisms.

\subsubsection{Very extended SNRs}

Very extended SNRs provide good targets for studying the SNR evolution and radiative mechanisms with good spatial detail.
Some very extended SNRs are evolving in a tenuous, relatively uniform ambient medium. X-ray observations of these SNRs can provide the effective temperature, density, and chemical properties and allow a good comparison with the standard Sedov-Taylor evolution model with or without pre-SN mass losses. This can help to constrain the evolution and progenitor properties of SNRs.

X-ray observations of the very extended SNRs can also help to diagnose different radiative processes, such as charge exchange (CX), and spatially resolve the structures.
In middle-age SNRs, CX and resonant scattering processes may play an important role in producing or modifying strong line emissions.
CX processes occur in the interface of a neutral medium and ionized gas, where the collision leads to the electron transfer from the atom to the ion. The highly-charged ion, such as OVII, that captures the electron can result in X-ray emission.
For O VII K$\alpha$ triplet, the CX emission can be distinguished from the thermal emission due to collisional excitation, by showing a large G-ratio (between forbidden+inter-combination to resonance transitions).
Another process that can produce a large G-ratio is the resonant-line scattering as the optical depth is enhanced, but the radial profile across the SNR and line properties are different from the CX results.
Both processes can be identified even using X-ray CCD data, since they can cause a line shift to the soft energy band \cite{roberts15}.
Although the K$\alpha$ triplet can hardly be resolved in a CCD spectrum, we can still interpret the G-ratio through its centroid energy, with a redshifted line centroid indicative of a high G-ratio. 
Ignoring these processes may result in an incorrect determination of the chemical abundance or temperature of the hot plasma.
The FXT is able to explore the CX and resonant scattering processes in extended SNRs (not only in the Cygnus loop).

\section{Summary}\label{sec:sum}

The driving science objectives of the Einstein Probe are to discover and characterise cosmic X-ray transients and to monitor the X-ray variability of known sources, so as to reveal their properties and gain insight into their nature and underlying physics.  
To this end, EP achieves an unprecedented combination of detecting sensitivity and monitoring cadence that are not accessible to the other wide-field X-ray monitors, complemented by powerful follow-up capability. 
As an interdisciplinary mission of time-domain and X-ray astronomy, EP has a wide range of scientific topics, from the nearby to the high-redshift Universe, some of which are among the most cutting-edge research in contemporary astrophysics. 
By expanding the monitoring horizon beyond the Milky Way, EP starts to enable a systematic census of extragalactic transients and variables fainter than typical GRBs in the soft X-ray regime. 

In recent years, some extragalactic fast X-ray transients of heterogeneous origins, mostly found in serendipitous surveys or from archival data, have excited great interests. However, their nature remains mysterious except for a few well-studied cases (e.g. SN SBO 2008D). 
EP may help begin to unveil their demography and nature, by catching more of them on the fly. 
Particularly, shock-breakouts generated from core-collapse SNe (mostly of II-P type) are among the most promising classes predicted to be detectable with WXT, which may shed light on the SN physics and SN progenitors.
New data from EP may test and constrain models for some FXRTs, particularly models invoking a fast-spinning magnetar as an aftermath of NS-NS mergers, which has profound implications for the detection of the electro-masgnetic counterparts of gravitational-wave events.
EP may revitalise the GRB research by extending the detecting bandpass to soft X-rays for prompt emission and early afterglows. It is also expected that EP will provide new insight into the understanding of the variants of GRBs, such as X-ray flashes, low-luminosity GRBs and ultra-long GRBs, as well as the physics of GRB jets and afterglows.
The legendary discoveries of high-redshift GRB ($z>6$) by Swift may be continued by EP and SVOM, although their identification and redshift measurement pose a great challenge to the follow-up observations of such sources. 

EP is well-positioned to search for X-ray counterparts of some of the multi-messenger events. For on-axis short GRBs originating from NS-NS or NS-BH mergers, EP may detect their prompt and afterglow emission in the soft X-ray band, along with the several GRB monitors operating at higher energies. If the merger product is a long-lived magnetar, X-ray transient emission may be produced within a large solid angle as predicted by some models, which is predicted to be detectable with WXT, even for off-axis GRB. EP is an ideal observatory to test such models and to provide possible constraints to the NS equation-of-state. EP may also detect flaring X-ray emission associated with (transient) neutrino sources, thanks to the high-cadence monitoring of a large part of the sky with WXT. 

EP will advance our understanding of SMBH, particularly their demography and accretion physics, by detecting and characterising a sample of nearby, X-ray bright TDEs and highly variable AGNs, especially at an early phase of their X-ray flares. This will complement the studies of faint TDEs by X-ray surveys such as eROSITA and of optical TDEs by ground optical telescopes. EP will also provide legacy data of long-term monitoring of various AGN samples, such as changing-look AGN and blazars, at a range of timescales.   

Looking back to our own Milky Way and neighbouring galaxies, EP also provides a new perspective to the dynamic X-ray views of compact stellar objects therein, i.e., BHs, NSs and WDs, either isolated or in binary systems.
With the improved monitoring sensitivity of WXT, EP will detect new X-ray bursts or state transitions with relatively lower peak fluxes, or at an earlier phase of the flaring process, than those detected in previous studies. These observations would provide valuable data to in-depth understanding of the physical processes involved in these systems. EP may also discover new compact objects by catching their flaring X-rays, especially objects that are possibly less bright at their emission peaks than those bright, well-studied systems. The improved sensitivity and soft X-ray passband make EP possibly a well-suited and unique instrument to study stellar flares, by carrying out a systematic census of stellar activities in the X-ray band. EP is expected to detect a large number of stellar flares of various types, and to help answer some of the interesting questions in stellar physics such as coronal mass ejection and their possible effects on the atmosphere of exoplanets orbiting the host stars.  

EP's capability as an X-ray observatory and its unique features also enables a wide range of other studies than time-domain astrophysics. These include a wide range of targets, such as supernova remnants, cluster of galaxies and solar system objects. The unique wide-field, spectroscopic-imaging capability in the soft X-ray band provide a valuable tool to study diffuse X-ray emission on large scales. 

During it's two-year operations, the EP-WXT pathfinder LEIA has detected tens of transients, including the prompt emission of the very bright GRB\,230307A possibly resulting from a NS-NS merger \cite{Sun.2023}, and the longest-lasting and most energetic stellar X-ray flare ever detected \cite{mao.2024}. 
During the commissioning and early operations, EP has detected about 80 X-ray transients with high signal to noise (S/N) ratios, and several hundreds with low S/N (marginal detections). A significant fraction of the EP FXRTs have no counterparts as GRBs, signifying that the soft X-ray band is as an important novel window for the GRB and fast extragalactic transient research. 
One of the EP detected GRBs, EP240315a, was found at a redshift of 4.859, which demonstrates EP's potential to detect GRBs at even higher redshifts \cite{LiuY2024arXiv}. 
The first EP transient, EP240219a, turned out to be an X-ray-rich GRB, presenting both challenges and opportunities for studying the physical origins of X-ray flashes, XRRs, and classical GRBs \cite{Yin.2024}. 
EP has also detected some new X-ray sources undergoing X-ray flaring and X-ray outbursts from some known AGNs and X-ray binaries. These include the finding of an intermediate polar, EP240309a/EP\,J115415.8-501810 \cite{Potter2024}, an outburst of a Be-WD binary in the SMC \cite{Marino2024}. The peculiar X-ray transient, EP240408a, whose nature remains an enigma, may represent a new type of transients with intermediate timescales of the order of more than 10 days \cite{zhang.2025}. 
A large number of X-ray stellar flares have also been detected with EP so far.

The success of LEIA and the early operations of EP have demonstrated the performance of EP and its promise for discovering a large sample of various kinds of X-ray transients and flares. 
It can thus be expected that the science objectives of EP, as presented in this paper, could be largely accomplished in the following years to come of EP operations.

\Acknowledgements{
We would like to thank a large number of colleagues, many of whom are not listed as co-authors of this paper, for stimulating discussions on the scienceobjectives and science cases of the EP mission over the past ten years. 
EP is a space mission supported by Strategic Priority Program on Space Science of Chinese Academy of Sciences, in collaboration with ESA, MPE and CNES (Grant No.XDA15310000, No.XDA15052100). 
This work was supported by the National Natural Science Foundation of China (Grant Nos. 61234003, 61434004, 61504141) and CAS Interdisciplinary Project (Grant No. KJZD-EW-L11-04).
We gratefully acknowledge the China National Astronomical Data Center (NADC), the Astronomical Data Center of the Chinese Academy of Sciences, and the Chinese Virtual Observatory (China-VO) for providing data resources and technical support.}

\InterestConflict{The authors declare that they have no conflict of interest.}


\bibliographystyle{scpma}
\bibliography{ref} 

\end{multicols}
\end{document}